\documentclass[aps,twocolumn,prd,tightenlines,nofootinbib,superscriptaddress,floatfix]{revtex4}
\usepackage{placeins} 
\usepackage{amssymb,amsbsy,amsmath,amsfonts,amstext}
\usepackage{xcolor}
\usepackage{dcolumn}
\usepackage{bm}
\usepackage{lipsum}
\usepackage{siunitx}
\usepackage[T1]{fontenc}
\usepackage[utf8]{inputenc}
\usepackage{graphicx} % Required for inserting images
\usepackage[english]{babel}
\usepackage{dsfont}
\usepackage{slashed}
\usepackage{bbold}
\usepackage{booktabs}
\usepackage{comment}
\usepackage{esint}
\usepackage{nicefrac}
\usepackage{braket}
\usepackage{multirow}
\usepackage{wrapfig}
\usepackage{blindtext}
\usepackage{float}
\usepackage{sidecap}
\usepackage{tikz}
\usepackage{tikz-feynman}
\usepackage{feynmp-auto}
\usetikzlibrary{decorations.pathmorphing}
\usepackage{csquotes}
\usepackage[most]{tcolorbox}
\usepackage{xfrac}
\usepackage{soul}
\usepackage{ragged2e}
\usepackage{etoolbox}

\usepackage{ulem}
\usepackage{cancel}

\newcommand{\blue}[1]{\textcolor{blue}{#1}}

\def\qt0{\tilde{q}_0}

\def\bea{\begin{eqnarray}}
\def\eea{\end{eqnarray}}

\definecolor{myGreen}{rgb}{0.2,0.72,0.2}
\definecolor{myWhite}{rgb}{0.98,0.98,0.98}
\definecolor{myGray}{rgb}{0.7,0.7,0.75}
\definecolor{myGold}{rgb}{0.8,0.64,0.24}
\definecolor{myPurple}{rgb}{0.71,0,0.71}

\catcode`,\active

\catcode`\,12

\allowdisplaybreaks

\usepackage[colorlinks=true,linkcolor=blue,citecolor=blue]{hyperref}
\begin{document}

\title{Subtracted Dispersion Relations for Virtual Compton Scattering off the Proton}

\author{Igor Danilkin}

\affiliation{Institut f\"ur Kernphysik and PRISMA\textsuperscript{++} Cluster of Excellence, Johannes Gutenberg-Universit\"at, Mainz, D-55099, Germany}

\author{Barbara Pasquini}

\affiliation{Dipartimento di Fisica ``A. Volta", Universit\`a degli Studi di Pavia, I-27100 Pavia, Italy}
\affiliation{Istituto Nazionale di Fisica Nucleare, Sezione di Pavia, I-27100 Pavia, Italy}

\author{Matteo Ronchi}

\affiliation{Institut f\"ur Kernphysik and PRISMA\textsuperscript{++} Cluster of Excellence, Johannes Gutenberg-Universit\"at, Mainz, D-55099, Germany}

\affiliation{Dipartimento di Fisica ``A. Volta", Universit\`a degli Studi di Pavia, I-27100 Pavia, Italy}

\author{Marc Vanderhaeghen}

\affiliation{Institut f\"ur Kernphysik and PRISMA\textsuperscript{++} Cluster of Excellence, Johannes Gutenberg-Universit\"at, Mainz, D-55099, Germany}

\date{\today}

\begin{abstract}
We present a once-subtracted dispersion relation (DR) formalism for the virtual Compton scattering (VCS) process from threshold up to the $\Delta(1232)$ energy region. The formalism aims at extracting the nucleon's electric and magnetic generalized polarizabilities from the $e^- p \to e^- \gamma p$ process, in view of the precision goals of the present and near-future experiments at Jefferson Lab. The present work improves upon the existing unsubtracted DR formalism in several ways. The required $s$- and $t$-channel discontinuities in the subtracted dispersion integrals are obtained in a largely data-driven manner in this energy region. The $s$-channel dispersive integrals are reconstructed from $\gamma^\ast p \to \pi N \to \gamma p$ using pion photo- and electro-production data, while the $t$-channel dispersion integrals are evaluated from $\gamma^\ast \gamma \to \pi \pi \to N \bar N$ using recent dispersive analyses of both the $\gamma^\ast \gamma \to \pi \pi$ and $\pi \pi \to N \bar N$ processes. We compare our results to VCS data and show the sensitivity of the observables to the nucleon's scalar generalized polarizabilities, which enter the present formalism as subtraction constants.   
    
\end{abstract}

\maketitle

\section{Introduction}

The electromagnetic polarizabilities of the nucleon are elementary structure constants that encode its response to slowly varying external electric and magnetic fields, and provide stringent tests of our understanding of low-energy QCD. In real Compton scattering (RCS), the leading scalar polarizabilities (electric $\alpha_{E1}$ and magnetic $\beta_{M1}$) and their spin counterparts can be accessed through precision measurements of photon-nucleon scattering; see~\cite{A2CollaborationatMAMI:2021vfy,Li:2022vnz} for the most recent experiments and \cite{Mornacchi:2022cln} for the first simultaneous extraction of the six leading-order proton polarizabilities. The extraction of polarizabilities has long served as a benchmark for effective field theory and dispersion-theoretical descriptions of hadron structure; see~\cite{Drechsel:2002ar,Griesshammer:2012we,Hagelstein:2015egb,Pasquini:2018wbl} for reviews. The nucleon polarizabilities also enter the hadronic corrections to precision Hydrogen spectroscopy, where the polarizability contribution corresponds to the two-photon exchange process between the lepton and the proton in the Hydrogen atom. These two-photon exchange corrections are currently a dominant source of theoretical uncertainty in the Standard Model prediction of the Lamb shift and Hyperfine splitting of muonic Hydrogen and light muonic atoms; see~\cite{Antognini:2022xoo,Pachucki:2022tgl} for recent reviews.  An improved understanding of these polarizability corrections is timely and necessary in view of the ongoing spectroscopy experiments, which aim to extract nucleon and nuclear radii with improved precision. Virtual Compton scattering (VCS), in which the incoming real photon is replaced by a space-like virtual photon, extends this program to generalized polarizabilities (GPs) \cite{Guichon:1995pu,Drechsel:1997xv}. The GPs, which depend on the photon virtuality $Q^2$, allow us to map the spatial distributions of the induced polarization and magnetization densities of the nucleon~\cite{Gorchtein:2009qq}, therefore providing spatially resolved information on the nucleon's response to external electromagnetic fields. 
A substantial experimental program has established VCS as a precision tool, with measurements performed at the Mainz Microtron (MAMI)~\cite{MAMI1,A1:2006cdk,MAMI3,MAMI4,Blomberg:2019caf,MAMI6,A1:2020nof}, MIT-Bates~\cite{MIT-Bates1,MIT-Bates2}, and Jefferson Lab (JLab)~\cite{JLab1,JLab2,Li:2022sqgnature}; see~\cite{Guichon:1998xv,Drechsel:2002ar,Fonvieille:2019eyf} for reviews.

Dispersion relations (DRs) provide a particularly powerful tool for extracting polarizabilities and GPs from RCS and VCS data, respectively, because they connect these low-energy structure parameters to the excitation spectrum of the nucleon and to photo- and electro-production data through analyticity and unitarity. Their application to the Compton process has a long history, dating back to the works of Gell-Mann, Goldberger, and Thirring~\cite{Gell-Mann:1954ttj} and Goldberger~\cite{PhysRev.97.508}, who derived DRs for forward Compton scattering. These pioneering works laid the foundation for numerous sum rules for the forward RCS amplitude, obtained by combining DRs with low-energy theorems~\cite{Gell-Mann:1954wra,Low:1954kd}; see~\cite{Drechsel:2002ar,Pasquini:2018wbl} for more detailed discussions. More recently, DRs for non-forward RCS have become a practical tool for extracting information on the nucleon's static polarizabilities from RCS observables from low energies up to the $\Delta(1232)$ resonance region. The formalism has been developed and applied to data using both unsubtracted DRs~\cite{Lvov:1996rmi} and subtracted DRs~\cite{Drechsel:1999rf,Pasquini:2007hf}. The unsubtracted DR formalism has also been extended to VCS in~\cite{pasquini2000,Pasquini:2001yy} and used to extract GPs from VCS data, as recently reviewed in \cite{Fonvieille:2019eyf}.

In the unsubtracted VCS DR formalism of~\cite{pasquini2000,Pasquini:2001yy}, the twelve VCS amplitudes are separated into Born contributions, corresponding to a nucleon intermediate state in the Compton process, and non-Born contributions, corresponding to all intermediate excited states. These non-Born contributions are represented by unsubtracted fixed-$t$ dispersion integrals over their $s$-channel discontinuities, which are saturated at moderate energies by $\pi N$ intermediate states. For three of the twelve VCS amplitudes, these $s$-channel integrals are supplemented at higher energies by energy-independent asymptotic contributions. The asymptotic contributions correspond physically to the $t$-channel exchange of the $\pi^0$ state and the $\pi \pi$ state in the $f_0(500)$ channel, as well as $s$-channel dispersive contributions beyond the $\pi N$ state. Although this approach provides a good description of existing VCS data up to the center of mass energy $\simeq 1.25~\text{GeV}$, it also highlights a limitation of the unsubtracted formalism: the amplitudes corresponding to the $\pi^0$ and $\pi \pi$ $t$-channel exchanges do not decrease sufficiently fast at high energies for the corresponding integrals to converge. In the unsubtracted DR approaches for RCS~\cite{Lvov:1996rmi} and VCS~\cite{Pasquini:2001yy}, this issue was addressed by evaluating the $s$-channel dispersion integrals up to a finite cutoff and approximating the remainder by effective $t$-channel poles. The associated model dependence of this unsubtracted finite-energy sum rule scheme is difficult to quantify reliably, especially for the description of the $\pi \pi$ $t$-channel exchange through an effective $f_0(500)$ pole exchange, and becomes a limiting factor. To improve on this point, a fixed-$t$ once-subtracted DR formalism was developed for RCS~\cite{Drechsel:1999rf,Drechsel:2002ar}. Due to the extra subtraction in the $s$-channel dispersive integrals, the corresponding amplitudes then exhibit fast convergence at the expense of introducing two $t$-dependent subtraction functions. The latter are calculated for zero value of the crossing-symmetric energy (corresponding to the Mandelstam variable condition $s = u$) by once-subtracted DRs in $t$ for the crossed $\gamma\gamma \to N \bar N$ process. In the RCS work of~\cite{Drechsel:1999rf,Drechsel:2002ar}, one of these subtraction functions was saturated by the $\pi^0$-pole, while the other was calculated through the $\gamma \gamma \to \pi \pi \to N \bar N$ process using a dispersive amplitude for the $\gamma \gamma \to \pi \pi$ process, constrained by data for this reaction and convoluted with the $\pi \pi \to N \bar N$ amplitude taken from~\cite{hohler1983pion}. The leading left-hand cut contributions to the subtraction functions were modeled by $\Delta(1232)$ exchange in the $s$- and $u$-channels. In the subtracted DR formalism, the subtraction constants at $s = u$ and $t = 0$ can be written directly in terms of the leading scalar and spin polarizabilities, which can then be obtained from a fit to RCS data.

Since the development of the subtracted DR formalism for RCS nearly three decades ago, major progress has been achieved in the dispersive description of the relevant $t$-channel dynamics. The amplitudes for the real and virtual photon-photon fusion processes $\gamma\gamma \to \pi\pi$ and $\gamma \gamma^\ast \to \pi \pi$ have been analyzed within single- and coupled-channel dispersive frameworks that include $\pi\pi$ rescattering in $s$- and $d$-waves and allow for a data–driven description of scalar and tensor isoscalar resonances~\cite{Garcia-Martin:2010kyn,Moussallam:2013una,Colangelo:2017fiz,Danilkin:2018qfn,Hoferichter:2019nlq,Danilkin:2019opj}. In parallel, the $\pi\pi \to N\bar{N}$ partial waves that enter the $t$-channel discontinuity of the Compton amplitudes have been reconstructed with high precision from the Roy-Steiner equations for $\pi N$ scattering~\cite{Hoferichter:2015hva}, providing a modern alternative to the older H\"ohler parameterization~\cite{hohler1983pion}, extending the range in $t$ up to around 4~GeV$^2$. These developments allow us to extend the subtracted DR formalism to the VCS process and suggest that a more constrained and quantitatively controlled dispersive treatment, where the dominant $t$-channel contributions are anchored to experimental information, is now within reach. The aim of the present work is to mirror the successful subtracted DR approach for RCS~\cite{Drechsel:1999rf,Drechsel:2002ar} and to construct a largely data-driven subtracted DR formalism for VCS from threshold up to the $\Delta(1232)$-resonance region, in view of the precision goals of present and near-future VCS experiments at Jefferson Lab (JLab)~\cite{VCS-II:2023jhu,VCS-BSA}.

The paper is organized as follows. In Section~II we briefly review the VCS process and the definition of generalized polarizabilities. Section~III presents the 
dispersive formalism for VCS. We first briefly summarize the unsubtracted approach with asymptotic contributions and then develop the once-subtracted fixed-$t$ DRs together with the $t$-channel DRs for the subtraction functions. In Section~IV we provide numerical results for the VCS amplitudes and illustrate their impact on the latest VCS observables from JLab, including comparisons with the unsubtracted framework and studies of sensitivity to the $t$-channel input. Section~V contains our conclusions and outlook. Further technical details of the formalism are given in three Appendices.

\section{Virtual Compton Scattering and Generalized Polarizabilities}

\begin{figure}
	\centering
	 \includegraphics[width=0.45\textwidth]{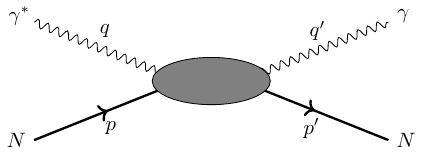}
	\caption{The virtual Compton scattering process on a nucleon.}
	\label{fig:vcs}
\end{figure}

Virtual Compton Scattering (VCS) describes the subprocess 
\begin{equation}
\gamma^\ast (q) + N (p) \rightarrow \gamma(q') + N(p'), 
\end{equation}
where a space-like virtual photon with four-momentum $q$ scatters off a nucleon with mass $M_N$ and four-momentum $p$, producing a real photon and a nucleon in the final state with four-momenta $q'$ and $p'$, respectively; see Fig.~\ref{fig:vcs}. We denote the initial photon's space-like virtuality by $q^2 = - Q^2 < 0$. 

\begin{figure}
	\centering
	 \includegraphics[width=0.48\textwidth]{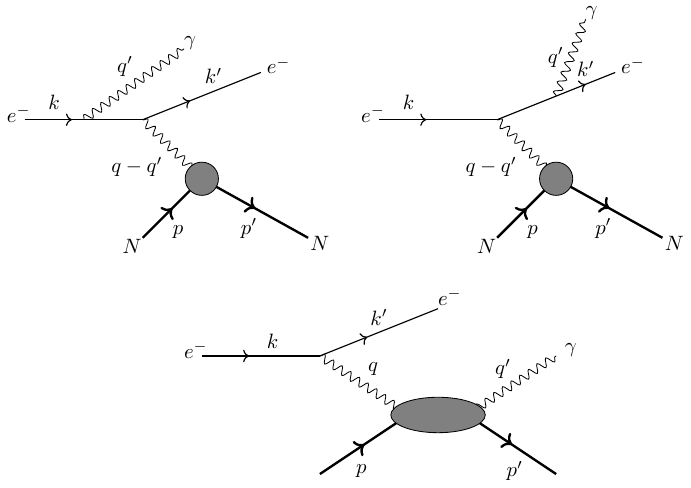}
	\caption{The $e N \to e \gamma N$ process: Bethe-Heitler process (upper diagrams) and VCS process (lower diagram).}
	\label{fig:epepgamma}
\end{figure}

The Mandelstam invariants for the VCS process are given by 
\begin{equation}
 s = (p + q)^2, \quad t = (q - q')^2, \quad u = (p - q')^2,    
\end{equation}
with the constraint $s + t + u = 2 M_N^2 - Q^2$.  
In describing the VCS process, it is useful to introduce the crossing energy variable $\nu$ as
\begin{equation}
\nu = \frac{s - u}{4 M_N}. 
\end{equation}
Furthermore, to describe VCS observables, it is common to use the $\gamma^\ast N $ center-of-mass (c.m.) frame. In this frame, the c.m. energy $W$ and the angle $\theta$ of the outgoing photon momentum direction $\hat{q}^{\, \prime}$ relative to the virtual photon momentum direction $\hat{q}$, can be expressed in terms of Lorentz  invariants as
\begin{eqnarray}
W &\equiv& \sqrt{s},  \\
\cos \theta &\equiv& \hat{q} \cdot \hat{q}^{\, \prime} = 
\frac{1}{\lambda^{1/2}(W^2,M_N^2,-Q^2)} \nonumber \\
&&\hspace{0.85cm}\times \left\{ W^2 - M_N^2 - Q^2 + \frac{2 W^2 (t + Q^2)}{W^2 - M_N^2} \right\}, \nonumber 
\end{eqnarray}
with the K\"all\'en function,
\begin{eqnarray}
\lambda(x,y,z) = x^2 + y^2 + z^2 - 2 x y - 2 x z - 2 yz.
\end{eqnarray}
Experimentally, the VCS process can be accessed through the $e N \to e \gamma N$ reaction, as shown in Fig.~\ref{fig:epepgamma}. At leading order ${\cal O}(e^3)$ in the electromagnetic coupling, the $e N \to e \gamma N$ reaction  contains both the Bethe-Heitler (BH) subprocess and the VCS subprocess. In the BH subprocess, the real photon is emitted from the lepton line, and the virtual photon has four-momentum $q - q'$, with spacelike virtuality $t$. It can be calculated exactly in terms of QED and the nucleon elastic form factors~\cite{Guichon:1998xv}. 
The VCS subprocess is described by the invariant amplitude
\begin{eqnarray}
{\cal M}_{\rm{VCS}} &=& \bar u(k',h')\, (-i e q_l \gamma_\mu)\, u(k, h) \left(\frac{-i}{q^2}  \right) \varepsilon^\ast_\nu(q',\lambda') \nonumber \\
&\times & (-i e^2)\, \bar N(p', s')\, {\cal M}^{\mu \nu} N(p,s),   
\label{eq:vcs}
\end{eqnarray}
where $u(k,h)$  ($\bar u(k',h')$) are the initial (final) electron spinors with four-momenta $k (k')$ and helicities $h$ $ (h')$ respectively,  and
$N(p,s)$ ($\bar N(p',s')$) are the initial (final) nucleon spinors with four-momenta $p$ $ (p')$ and helicities $s $ $(s')$, respectively; and 
$\varepsilon^\ast_\nu(q',\lambda')$ is the final photon polarization vector with four-momentum $q'$ and helicity $\lambda'$. Furthermore, in Eq.~(\ref{eq:vcs}) we introduced a lepton charge factor $q_l = -1$ ($q_l = +1$) for an electron (positron) beam, with $e > 0$ denoting the magnitude of the electric charge unit. 

The VCS tensor ${\cal M}^{\mu \nu}$ in Eq.~(\ref{eq:vcs}) can be decomposed into a Born (B) part, corresponding to a nucleon intermediate state, and a non-Born (NB) part, as discussed in~\cite{Guichon:1995pu}. 
Following the early work of~\cite{rhobasis}, Refs~\cite{Drechsel:1997xv,Drechsel:1998zm} showed how ${\cal M}^{\mu \nu}$ can be expressed in a gauge-invariant tensor basis
\begin{eqnarray}
{\cal M}^{\mu \nu} =\sum_{i=1}^{12} f_{i}(Q^2,\nu,t)\, \rho_i^{\mu \nu},
\label{eq:vcsbasis}
\end{eqnarray}
where the 12 independent gauge-invariant tensors $\rho_i^{\mu \nu}$ are given in Appendix A of \cite{Pasquini:2001yy}.  The basis is chosen such that the non-Born parts of the invariant amplitudes $f_i(Q^2, \nu, t)$ are free of kinematic singularities and constraints~\cite{Drechsel:1997xv,Drechsel:1998zm}. 
The Born terms for the amplitudes are collected in Appendix B of ~\cite{Pasquini:2001yy}. 

The invariant amplitudes $f_i$ have the following behavior under nucleon crossing combined with charge conjugation ($\nu \to - \nu$)~\cite{Pasquini:2001yy}
\begin{eqnarray}
f_i(Q^2, \nu, t) &=& f_i(Q^2, -\nu, t), \quad (i = 1, 2, 5, 6, 7, 9, 11, 12), 
\nonumber \\
f_i(Q^2, \nu, t) &=& - f_i(Q^2, -\nu, t), \quad (i = 3, 4, 8, 10). 
\end{eqnarray}
Furthermore, in~\cite{Pasquini:2001yy} the amplitudes $F_i(Q^2, \nu, t)$ were conveniently introduced, which are all even functions in $\nu$,
\begin{eqnarray}
F_i(Q^2, \nu, t) &\equiv& f_i(Q^2, \nu, t), \quad (i = 1, 2, 5, 6, 7, 9, 11, 12), 
\nonumber \\
F_i(Q^2, \nu, t) &\equiv& \frac{1}{\nu} f_i(Q^2, \nu, t), \quad (i = 3, 4, 8, 10). 
\end{eqnarray}
Note that, since the odd $f_i$ vanish at $\nu=0$, the corresponding $F_{i=3,4,8,10}$ are regular there.

For finite initial photon virtuality $Q^2$, the limit of a low outgoing photon four-momentum, $q' \to 0$, corresponds to the kinematic point $\nu = 0, t = -Q^2$. In this limit, the non-Born VCS amplitudes $F_i^{\mathrm{NB}}$ can be parameterized in terms of six generalized polarizabilities (GPs)~\cite{Guichon:1995pu,Drechsel:1997xv}. A convenient choice and multipole notation were proposed in~\cite{Guichon:1998xv}.  The relations between these six GPs and the invariant amplitudes are 
\begin{align}
  P^{(L1,L1)0}(Q^2)&= \sqrt{\frac{2}{3}} N_q 
   \bigg[ \bar F_1 
   + 4 M_N^2 (1 + \tau) \bar F_2  \nonumber\\
   &+ 4 M_N^2 \tau \bigg(2\bar F_6+\bar F_9
   - \bar F_{12} \bigg) \bigg], 
   \label{eq:GPelampl} \\
  P^{(M1,M1)0}(Q^2)&=-\sqrt{\frac{8}{3}} N_q \bar F_1, 
  \label{eq:GPmagnampl} \\
  P^{(L1,L1)1}(Q^2)&=- \frac{2}{3} N_q M_N \tau \bigg[ \bigg( \bar F_5 + \bar F_7 + 4\bar F_{11} \bigg) \nonumber \\
  &+ 4M_N \bar F_{12} \bigg], 
  \label{eq:GPselampl}  \\
  P^{(M1,M1)1}(Q^2)&=-\frac{2}{3}N_q\frac{M_N \tau}{1 + \tau}
   \bigg[ \bar F_5 - 2 M_N \tau \bar F_{12}\bigg], 
   \label{eq:GPsmagnampl} \\ 
  P^{(M1,L2)1}(Q^2)&=\frac{2\sqrt{2}}{3\sqrt{3}} N_q\bigg[ 
  - \frac{1}{2 (1 + \tau)} \bar F_5 \nonumber \\
  &+ \frac{1}{2} \bigg( \bar F_5 + \bar F_7 + 4 \bar F_{11} \bigg) + \frac{M_N \tau}{1 + \tau}\bar F_{12} \bigg], \quad 
  \label{eq:GPsm1l2ampl} \\
  P^{(L1,M2)1}(Q^2)&=-\frac{\sqrt{2}}{6} \frac{N_q}{1 + \tau}
  \bigg[ 4 M_N \bigg( 2 \bar F_6 + \bar F_9 \bigg)  \nonumber \\
  &+ \bigg( \bar F_7 + 4\bar F_{11}\bigg) + 2 M_N \tau \bar F_{12}\bigg], 
  \label{eq:GPsl1m2ampl}
\end{align}
where we introduced the shorthand notations 
\begin{align}
&\bar F_i(Q^2) \equiv F_i^{\mathrm{NB}}(Q^2,\nu = 0, t = -Q^2), 
\label{eq:vcsamplgp}\\
&N_q \equiv \left(\frac{1 + \tau}{1 + 2 \tau}\right)^{1/2},\quad \tau \equiv \frac{Q^2}{4 M_N^2}\,.\nonumber
\end{align}
In the GPs $P^{(\rho'L',\rho L)S}$  we used the multipole notation of Refs.~\cite{Guichon:1995pu,Drechsel:1997xv}: $\rho(\rho')$ denotes the longitudinal ($L$) or magnetic ($M$) nature of the initial (final) photon. The initial (final) orbital angular momentum is labeled by $L(L')$, and $S=0,1$ denotes non-spin-flip/spin-flip transition.

It is also convenient to introduce the $Q^2$ dependent generalizations of the electric dipole ($\alpha_{E1}$) and magnetic dipole ($\beta_{M1}$) polarizabilities as
\begin{eqnarray}
\alpha_{E1}(Q^2) &=& - \frac{e^2}{4 \pi} \sqrt{\frac{3}{2}}\, P^{(L1,L1)0}(Q^2), \\
\beta_{M1}(Q^2) &=& - \frac{e^2}{4 \pi} \sqrt{\frac{3}{8}}\, P^{(M1,M1)0}(Q^2). 
\end{eqnarray}

In the limit $Q^2 \to 0$ for the GPs, one finds the following  relations with the  static polarizabilities as measured in real Compton scattering (RCS)
\begin{eqnarray}
&&\alpha_{E1}(0)= \alpha_{E1} \;, \nonumber\\
&&\beta_{M1}(0)= \beta_{M1} \;, \nonumber \\
&&P^{(M1,L2)1}(0)=-\frac{4 \pi}{e^2}\, \frac{2 \sqrt{2}}{3 \sqrt{3}} \,
\gamma_{M1 E2} \;, \nonumber \\
&&P^{(L1,M2)1}(0)=-\frac{4 \pi}{e^2}\, \frac{\sqrt{2}}{3} \,
\gamma_{E1 M2} \;,  \nonumber\\ 
&&P^{(L1,L1)1}(0)= 0 \;, \nonumber\\
&&P^{(M1,M1)1}(0)= 0 \;,
\label{eq:polrcs}
\end{eqnarray}
where $\alpha_{E1}$ ($\beta_{M1}$) are the electric (magnetic) scalar dipole polarizabilities respectively, and where $\gamma_{M1 E2}$ and $\gamma_{E1 M2}$ are two of the four lowest-order spin polarizabilities of the nucleon~\cite{Drechsel:2002ar}.

\section{Dispersion relations for VCS amplitudes}

\begin{figure}[h]
	\centering
	 \includegraphics[width=0.45\textwidth]{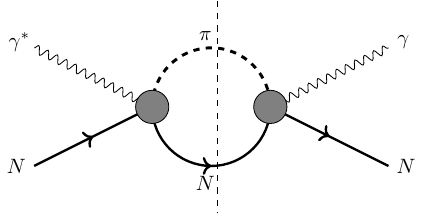}
	\caption{The $s$-channel $\pi N$ cut contribution in the dispersive formalism for VCS on the nucleon.}
	\label{fig:schannelcut}
\end{figure}

In~\cite{Pasquini:2001yy}, a fixed-$t$ dispersive framework was presented for the non-Born VCS amplitudes $F_i^{\mathrm{NB}}$, which are free of kinematical singularities and constraints. To obtain the corresponding dispersion relations, one needs to start from the analytic structure of the full amplitudes, which includes nucleon pole terms and the $s$-channel cut contributions starting at the $\pi N$ threshold (see Fig.~\ref{fig:schannelcut}).  

The nucleon poles are located at $s = M_N^2$ and $u = M_N^2$ and are expressed in terms of the crossing variable $\nu$ as
\begin{eqnarray}
s - M_N^2 &=& 2 M_N\,(\nu - \nu_N), \nonumber \\ 
u - M_N^2 &=& - 2 M_N\,(\nu + \nu_N), 
\end{eqnarray}
with the $s$-channel nucleon pole position 
\begin{eqnarray}
\nu_N = \frac{Q^2 + t}{4 M_N}.
\end{eqnarray}

The $s$-channel cut contributions in the dispersive formalism start at the $\pi N$ threshold position $\nu_0$, given by
\begin{eqnarray}
\nu_0 = \nu_N + m_\pi + \frac{m_\pi^2}{2 M_N},
\end{eqnarray}
with $m_\pi$ the pion mass. The convergence of the $s$-channel dispersion integrals depends on the high-energy behavior of the VCS amplitudes $F_i^{\mathrm{NB}}$. For amplitudes that decrease sufficiently fast at high energy, analyticity allows one to express their real parts, at fixed values of $t$ and $Q^2$, through an unsubtracted dispersion integral over the $s$-channel discontinuity as
\begin{align}
\label{eq:unsubtdr}
    F_i^{\mathrm{NB}}(Q^2,\nu, t) &= 
    F_i^{s\text{-pole}}(Q^2,\nu,t) - F_i^\mathrm{B}(Q^2,\nu,t)\nonumber \\
     & + \frac{2}{\pi} \int_{\nu_{0}}^{\infty} d\nu' \frac{\nu'\, {\rm Im}_s F_i(Q^2,\nu', t)}{ \nu'^2 - \nu^2 - i \varepsilon} , \quad
\end{align}
where $F_i^{s\text{-pole}}$ denotes the on-shell nucleon pole contribution in the $s$-channel (with pole positions located at $\nu = \pm \nu_N$), 
$F_i^\mathrm{B}$ is the Born contribution defined in gauge-invariant way for an off mass shell nucleon as in~\cite{Guichon:1995pu,Guichon:1998xv}, and where the dispersion integral starts from the pion production threshold $\nu_0$. In evaluating the non-Born VCS amplitudes in the dispersive framework, one thus needs the difference between the pole ($F_i^{s\text{-pole}}$) and Born ($F_i^{\mathrm{B}}$) amplitudes, which are given by

\begin{eqnarray}
&&F_1^{s\text{-pole}}(Q^2,\nu,t) - F_1^{\mathrm{B}}(Q^2,\nu,t) = - \frac{\kappa\,F_P(Q^2)}{4 M_N^3},  \nonumber \\
&&F_5^{s\text{-pole}}(Q^2,\nu,t) - F_5^{\mathrm{B}}(Q^2,\nu,t) = + \frac{\kappa\, F_P(Q^2)}{4 M_N^4},  \nonumber \\
&&F_{11}^{s\text{-pole}}(Q^2,\nu,t) - F_{11}^{\mathrm{B}}(Q^2,\nu,t) = - \frac{\kappa\, F_P(Q^2)}{16 M_N^4}, \nonumber \\
&&F_{i}^{s\text{-pole}} - F_{i}^{\mathrm{B}} = 0,  
\; \mathrm{for} \; i = 2, 3, 4, 6, 7, 8, 9, 10, 12, \nonumber \\
\label{eq:pmborn} 
\end{eqnarray}
where $\kappa$ is the anomalous magnetic moment of the nucleon and $F_P(Q^2)$ is its Pauli form factor, with $F_P(0) = \kappa $. One notices from Eq.~(\ref{eq:pmborn}) that for the combination $F_5 + 4 F_{11}$ the difference between pole and Born contribution vanishes, i.e. 
\begin{eqnarray}
(F_{5} + 4 F_{11})^{s\text{-pole}} - (F_{5} + 4 F_{11})^{\mathrm{B}} = 0.  
\end{eqnarray}
Therefore,  we only need to account for the difference between the pole and Born contributions for two combinations out of the twelve VCS amplitudes, which we take as $F_1$ and $F_5$. One furthermore notices from Eq.~(\ref{eq:pmborn}) that the difference between the nucleon pole and Born terms is energy independent. Consequently, for all twelve VCS amplitudes $F_i$ ($i = 1,...,12$), the following relation holds
\begin{eqnarray}
&&F_{i}^{s\text{-pole}}(Q^2,\nu,t) - F_{i}^{\mathrm{B}}(Q^2,\nu,t) \nonumber \\
&&= F_{i}^{s\text{-pole}}(Q^2,0,t) - F_{i}^{\mathrm{B}}(Q^2,0,t)\,. 
\label{eq:pmborn2}
\end{eqnarray}

From the high-energy Regge behavior of the VCS invariant amplitudes ($\nu \to \infty$ at fixed $t$ and $Q^2$), it was found in~\cite{Pasquini:2001yy} that the amplitudes $F_1^{\mathrm{NB}}$ and $F_5^{\mathrm{NB}}$ do not drop fast enough in $\nu$ for the unsubtracted dispersion integral in Eq.~(\ref{eq:unsubtdr}) to hold. The leading physical contributions responsible for this behavior are the $t$-channel $\pi \pi$ exchange with scalar and tensor quantum numbers in $F_1^{\mathrm{NB}}$ and the $t$-channel $\pi^0$ pole contribution in $F_5^{\mathrm{NB}}$. 
In the formalism developed in~\cite{Pasquini:2001yy}, the unsubtracted dispersion integrals for the VCS amplitudes $F_1^{\mathrm{NB}}$ and $F_5^{\mathrm{NB}}$ were implemented within a finite-energy sum-rule scheme: the unsubtracted integrals were evaluated up to an upper cutoff, and the remaining high-energy part was approximated by asymptotic contributions, as discussed below. Although this approach describes existing VCS data well from threshold up to the $\Delta(1232)$ region, it introduces a model dependence associated with the treatment of the high-energy contributions. In view of the precision goals of present and forthcoming VCS measurements, and to potentially extend the applicability range of the theory, we instead develop a subtracted dispersion-relation framework for $F_1^{\mathrm{NB}}$ and $F_5^{\mathrm{NB}}$, which improves convergence. If the subtraction functions can be constrained by data (or by dispersive input), the resulting framework enables a more transparent propagation of uncertainties.

We will next review the unsubtracted DR formalism with asymptotic contributions as used in~\cite{Pasquini:2001yy}, and subsequently develop  the subtracted DR formalism.

\subsection{Unsubtracted dispersion formalism with asymptotic contributions}

Since Ref.~\cite{Pasquini:2001yy} aimed to present a VCS DR formalism up to $\Delta(1232)$ resonance energies, the dispersion integrals of Eq.~(\ref{eq:unsubtdr}) were evaluated only up to a finite upper limit $\nu \leq \nu_{\mathrm{max}}$, with $\nu_{\mathrm{max}}$ corresponding to the limit $W_{\mathrm{max}} = 2$~GeV. The $s$-channel discontinuities were saturated by their $\pi N$ contributions, as shown in  Fig.~\ref{fig:schannelcut}. For most VCS amplitudes, this procedure provides a rapidly convergent representation. However, for the amplitudes $F_1^{\mathrm{NB}}$ and $F_5^{\mathrm{NB}}$, the unsubtracted dispersion integrals of Eq.~(\ref{eq:unsubtdr}) do not converge due to their high-energy behavior. In addition, for the amplitude $F_2^{\mathrm{NB}}$, contributions from intermediate states beyond $\pi N$ are non-negligible and must be accounted for. To handle these cases, Ref.~\cite{Pasquini:2001yy} approximated the remaining high-energy part of the $s$-channel integral ($\nu > \nu_{\mathrm{max}}$) by energy-independent asymptotic contributions. 

For the amplitudes $F_1^{\mathrm{NB}}$, the asymptotic contribution was parameterized through an effective $t$-channel $f_0(500)$-pole, with a $Q^2$-dependent residue adjusted to reproduce the $Q^2$ dependence of the GP $\beta_{M1}(Q^2)$. 

For the amplitude $F_5^{\mathrm{NB}}$, the asymptotic contribution was parameterized through the $t$-channel $\pi^0$-pole contribution in terms of the $\gamma^\ast \gamma \to \pi^0$ transition form factor.

For the amplitudes $F_2^{\mathrm{NB}}$, the asymptotic contribution was adjusted to reproduce the $Q^2$ dependence of the GP combination $\alpha_{E1}(Q^2) + \beta_{M1}(Q^2)$.

\subsection{Subtracted dispersion relation formalism}

To overcome the convergence limitations of the unsubtracted framework~\cite{Pasquini:2001yy}, we now develop a once-subtracted fixed-$t$ DR formalism.
For the non-Born VCS amplitudes $F_1^{\mathrm{NB}}$, $F_5^{\mathrm{NB}}$, and for a specific combination involving $F_2^{\mathrm{NB}}$ as defined below, the unsubtracted DR of Eq.~(\ref{eq:unsubtdr}) is either not valid or converges too slowly. We therefore write a once-subtracted fixed-$t$ dispersion relation
\begin{align}
\label{eq:subtdr}
 F_i^{\mathrm{NB}} (Q^2,\nu, t) &= 
    F_i^{\mathrm{NB}} (Q^2,0, t) \nonumber \\
   &+ \frac{2}{\pi} \nu^2\, \int_{\nu_0}^{\infty} d\nu' \frac{{\rm Im}_s F_i(Q^2,\nu', t)}{\nu' (\nu'^2 - \nu^2 - i \varepsilon)} , \quad 
\end{align}
where the subtraction is performed at $\nu = 0$. Owing to Eq.~(\ref{eq:pmborn2}), the difference between pole and Born contributions cancels in this relation. 

The subtraction function $F_i^{\mathrm{NB}} (Q^2,0, t)$ in Eq.~(\ref{eq:subtdr}) is then evaluated through a once-subtracted DR in the variable $t$ at fixed $Q^2$. Choosing the subtraction point at $t=-Q^2$, we obtain the following subtracted DRs in the variable $t$:
\begin{align}
\label{eq:subtraction}
 F_i^{\mathrm{NB}} (Q^2, 0, t) &= F_i^{\mathrm{NB}} (Q^2, 0, -Q^2)  \nonumber \\
 &+ F_i^{t\text{-pole}}(Q^2,0,t)- F_i^{t\text{-pole}}(Q^2,0,-Q^2) \nonumber\\
 &+ \frac{(t+Q^2)}{\pi}  \int_{t_R}^{\infty} dt'
 \frac{\textup{Im}_{t}F_{i}(Q^2,0,t')}{(t'+Q^2)(t'-t- i \varepsilon)} \nonumber \\
 &+ \frac{(t+Q^2)}{\pi}  \int_{-\infty}^{t_L} dt' \frac{\textup{Im}_{t}F_{i}(Q^2,0,t')}{(t'+Q^2)(t'-t - i \varepsilon)}.
\end{align}
We will subsequently discuss the four contributions on the right-hand side ({\it rhs})  of Eq.~(\ref{eq:subtraction}). 

The first contribution on the {\it rhs} of Eq.~(\ref{eq:subtraction}) is the non-Born VCS amplitudes evaluated at the subtraction point $t = -Q^2$. Using Eqs.~(\ref{eq:GPelampl}) - (\ref{eq:GPsl1m2ampl}), these can be expressed in terms of GPs for the following six combinations of VCS amplitudes
\begin{eqnarray}
&&F_1^{\mathrm{NB}}, \quad F_2^{\mathrm{NB}}, \quad F_5^{\mathrm{NB}}, 
\quad F_5^{\mathrm{NB}} + F_7^{\mathrm{NB}} + 4 F_{11}^{\mathrm{NB}},  \nonumber \\  
&& 2 F_6^{\mathrm{NB}} + F_9^{\mathrm{NB}}, \quad F_{12}^{\mathrm{NB}}.
\end{eqnarray}
In this work, we will evaluate the subtracted DRs of Eqs.~(\ref{eq:subtdr}) and (\ref{eq:subtraction}) for the amplitudes $F_1^{\mathrm{NB}}$, $F_5^{\mathrm{NB}}$, and for the combination
\begin{eqnarray}
\tilde F_2^{\mathrm{NB}} &\equiv& F_2^{\mathrm{NB}} - \frac{t}{4 M_N^2 - t} 
\biggl( 2 F_6^{\mathrm{NB}} + F_9^{\mathrm{NB}} - F_{12}^{\mathrm{NB}} \biggr). \quad
\label{eq:f2tilde}
\end{eqnarray}
For these three amplitudes,  their values at the subtraction point can be expressed in terms of GPs as
\begin{eqnarray}
F_1^{\mathrm{NB}} (Q^2, 0, -Q^2) &=& \frac{4 \pi}{e^2} \frac{1}{N_q} \beta_{M1}(Q^2), 
\label{eq:F1GP} \\
\tilde F_2^{\mathrm{NB}} (Q^2, 0, -Q^2) &=& - \frac{4 \pi}{e^2} \frac{1}{N_q} \frac{1}{4 M_N^2 (1 + \tau)} \nonumber \\ 
&\times& \bigg\{ \alpha_{E1}(Q^2) + \beta_{M1}(Q^2) \biggr\}, 
\label{eq:F2tildeGP} \\
F_5^{\mathrm{NB}} (Q^2, 0, -Q^2) &=& - \frac{1}{N_q} \frac{3}{4 M_N} 
\biggl\{ P^{(L1,L1)1}(Q^2)
\nonumber \\
&+& \left( \frac{2 + \tau}{\tau} \right) P^{(M1,M1)1}(Q^2) 
\nonumber \\
&+& 2 M_N \tau \sqrt{\frac{3}{2}} P^{(M1,L2)1}(Q^2) \bigg\} .
\label{eq:F5GP}
\end{eqnarray}

The second contribution on the {\it rhs} of Eq.~(\ref{eq:subtraction}) is due to the $\pi^0$ $t$-channel pole and is non-zero only for the amplitudes $F_5^{\mathrm{NB}}$ and $F_{11}^{\mathrm{NB}}$
\begin{eqnarray}
F_5^{t\text{-pole}}(Q^2,0,t) &=& - 4 F_{11}^{t\text{-pole}}(Q^2,0,t), \nonumber \\ 
&=& \frac{g_{\pi NN}}{M_N} \frac{1}{t - m_\pi^2} F_{\pi^0 \gamma^\ast \gamma}(Q^2 ),
\label{eq:pipole}
\end{eqnarray}
with $\pi NN$ coupling given by $g_{\pi NN} = 13.13$, and where $F_{\pi^0 \gamma^\ast \gamma}(Q^2)$ is the $\gamma^\ast \gamma \to \pi^0$ transition form factor parameterized through the monopole form
\begin{eqnarray}
F_{\pi^0 \gamma^\ast \gamma}(Q^2 ) = F_{\pi^0 \gamma^\ast \gamma}(0 ) \frac{1}{1 + Q^2/\Lambda_\pi^2}.
\end{eqnarray}
We use the fit to the space-like single virtual $\gamma^\ast \gamma \to \pi^0$ world data up to $Q^2 < 9$~GeV$^2$ with  values~\cite{Danilkin:2019mhd}
\begin{eqnarray}
\Lambda_\pi^2 &=& (0.611 \pm 0.005)~{\mathrm{GeV}^2}, \nonumber \\
F_{\pi^0 \gamma^\ast \gamma}(0 ) &=& (0.273 \pm 0.003)~{\mathrm{GeV}^{-1}}.
\end{eqnarray}
Note that Eq.~(\ref{eq:pipole}) implies that the $\pi^0$ pole contribution drops out of the combination $F_5^{\mathrm{NB}} + 4 F_{11}^{\mathrm{NB}}$. Furthermore, for the amplitude $F_5$, the difference $F_5^{\mathrm{NB}}(Q^2,0,-Q^2) - F_5^{t\text{-pole}}(Q^2,0,-Q^2)$ entering Eq.~(\ref{eq:subtraction}) can be expressed, using Eq.~(\ref{eq:F5GP}), in terms of the dispersive spin GP contribution, i.e. beyond the $\pi^0$ pole. 

The third and fourth contributions on the {\it rhs} of Eq.~(\ref{eq:subtraction}) arise from the discontinuities across the $t$-channel right-hand cut and left-hand cut, respectively. The right-hand cut starts at the $\pi \pi$ threshold, $t_{R}=4 m_{\pi}^2$, and extends along the positive real $t$-axis.  The left-hand cut extends along the negative real $t$-axis and starts at
\begin{eqnarray}
t_{L}=-2 m_{\pi}^2 -4M_N m_{\pi} -Q^2, 
\end{eqnarray}
which corresponds to the $\pi N$ threshold for $\nu = 0$. 
For $t<t_L$, the left-hand-cut term in Eq.~(\ref{eq:subtraction}) has to be understood with the $t \to t+ i \epsilon$ prescription. It then decomposes into a principal-value integral giving the real part, and an additional imaginary part $i\, \textup{Im}_{t}F_{i}(Q^2,0,t)$ on the cut. The principal-value part can be evaluated numerically as
\begin{eqnarray}
 &&\frac{(t+Q^2)}{\pi} \mathcal{P} \int_{-\infty}^{t_L} dt' \frac{\textup{Im}_{t}F_{i}(Q^2,0,t')}{(t'+Q^2)(t'-t)} \nonumber \\
 &&= \frac{(t+Q^2)}{\pi}  \int_{-\infty}^{t_L} dt' \frac{\textup{Im}_{t}F_{i}(Q^2,0,t') - \textup{Im}_{t}F_{i}(Q^2,0,t)}{(t'+Q^2)(t'-t)} \nonumber \\
 &&+ \frac{1}{\pi} \textup{Im}_{t}F_{i}(Q^2,0,t) \, 
 \ln \left| \frac{t_L - t}{t_L + Q^2} \right|.
\end{eqnarray}

In the next two subsections, we will discuss in detail how to estimate the $t$-channel right- and left-hand cuts. 

\begin{comment}
For the VCS amplitudes that do drop fast enough in $t$, 
as alternative to the subtracted DR of Eq.~(\ref{eq:subtraction}),  the subtraction function  $F_{i}^{\mathrm{NB}}(Q^2,0,t)$ may also be evaluated using an unsubtracted dispersion integral in the variable $t$ as
\bea
\label{subt}
    {\rm Re} \, F_i^{\mathrm{NB}}(Q^2,0, t) &=& F_i^{t\text{-pole}}(Q^2,0,t)\nonumber \\
    &+& \frac{1}{\pi} \mathcal{P} \int_{t_{R}}^{\infty} dt'  \frac{{\rm Im}_t \, F_i(Q^2,0, t')}{t' - t} \nonumber \\
    &+& \frac{1}{\pi} \mathcal{P} \int_{-\infty}^{t_L} dt' \frac{{\rm Im}_t \,F_i(Q^2,0, t')}{t' - t} .
\eea
\end{comment}

\subsection{$t$-channel right-hand cut contribution due to $\pi \pi$ intermediate state}
\label{tchannelsec}

The leading $t$-channel right-hand cut contribution to the VCS amplitudes at $\nu=0$ arises from the $\pi \pi$ intermediate state (see Fig.~\ref{fig:tchannel_rhc}). 
In Appendix~\ref{app:pipi}, we express the corresponding discontinuities of the VCS amplitudes in terms of the helicity amplitudes for the processes $\gamma^\ast \gamma \to \pi \pi$ and $\pi \pi \to N \bar N$. 

\begin{figure}
	\centering
	 \includegraphics[width=0.45\textwidth]{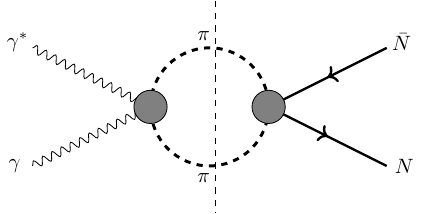}
	\caption{The $t$-channel $\pi \pi$ right-hand cut contribution in the subtracted dispersive formalism for VCS on the nucleon.}
	\label{fig:tchannel_rhc}
\end{figure}

For the $\gamma^\ast \gamma \to \pi \pi$ amplitudes, we use the $s$- and $d$-wave amplitudes of the dispersive formalism of~\cite{Danilkin:2018qfn}, which provides a good description of the available $\gamma \gamma \to \pi \pi$ data from threshold up to $\pi \pi$ invariant mass of 1.4~GeV, and which was extended to the processes with one or both photons having spacelike virtuality~\cite{Danilkin:2019opj}.   

For the $\pi \pi \to N \bar N$ amplitudes, two sets of $s$ and $d$-wave amplitudes are available. The older reconstruction of H\"ohler~\cite{hohler1983pion} is given up to $t \lesssim 0.78$~GeV$^2$ and was used in the RCS subtracted DR formalism of~\cite{Drechsel:1999rf}. In the more recent analysis of Hoferichter {\it et al.}~\cite{Hoferichter:2015hva}, which we employ here, the $s$- and $d$-wave amplitudes were reconstructed from Roy-Steiner equations for $\pi N$ scattering and provided up to $t \lesssim 4$~GeV$^2$.

Both the dispersive framework for $\gamma^\ast \gamma \to \pi \pi$ and the modern Roy–Steiner reconstruction for $\pi \pi \to N \bar N$ incorporate a coupled-channel description in the scalar-isoscalar channel ($I=0$, $s$-wave), and therefore cover both the $f_0(500)$ and $f_0(980)$ regions. In the present work, however, we restrict ourselves to the dominantly elastic $\pi\pi$ contribution in the $s$-wave. Therefore, when evaluating the $t$-channel right-hand cut integrals of~Eq.~(\ref{eq:subtraction}) for the $s$-wave, we truncate the dispersive integration at $0.78$ GeV$^2$, i.e., we keep only the elastic $f_0(500)$ region, below the onset of significant inelasticities. For the $d$-wave, both the $\gamma^\ast \gamma \to \pi \pi$ and $\pi \pi \to N \bar N$ inputs are based on single-channel analyses covering the $f_2(1270)$ region and are reliable up to at least $2$ GeV$^2$. Although the $\gamma^\ast \gamma \to \pi \pi$ and $\pi \pi \to N \bar N$ inputs are obtained from independent dispersive analyses, both rely on $\pi\pi$ phase-shift information constrained by Roy-type analyses \cite{Caprini:2011ky,Garcia-Martin:2011iqs}. As a consequence, Watson’s theorem holds in the elastic region, ensuring that the phases cancel in the unitarity relation below. 
Any residual mismatch in the phase input provides  an additional source of systematic uncertainty in the numerical evaluation.

For the three VCS amplitudes $F_1^{\mathrm{NB}}$, $\tilde F_2^{\mathrm{NB}}$, and $F_5^{\mathrm{NB}}$, which we evaluate in this work through the subtracted DR formalism, the $\pi \pi$ right-hand cut discontinuities (including $s$- and $d$-waves) are obtained from Eq.~(\ref{eq:imtpipiF1-12}) as
\begin{align}
&\textup{Im}_{t}F_{1}^{\pi \pi}(Q^2,0,t') = - \frac{1}{e^2} 
\frac{\beta_\pi(t')}{t' + Q^2} \nonumber \\
&
\hspace{0.5cm}
\times \biggl\{\frac{1}{4 M_N^2 - t'} h_0^0(t',Q^2) f_+^{0 \ast}(t') \nonumber \\
&
\hspace{0.5cm} 
+ \frac{5}{32} t' \beta^2_\pi(t') 
\left[ h_0^2(t',Q^2) - \sqrt{\frac{3}{2}} h_2^2(t',Q^2) \right] f_+^{2 \ast}(t') \biggr\}, \nonumber \\
& \label{eq:imtpipiF1}  \\
&\textup{Im}_{t}\tilde F_{2}^{\pi \pi}(Q^2,0,t') = - \frac{1}{e^2} 
 \frac{5}{16} \; \frac{\beta^3_\pi(t')}{4 M_N^2 - t'}  \nonumber \\
&
\hspace{0.5cm} 
\times \biggl\{  
\sqrt{\frac{3}{2}} \left(\frac{t'}{t' + Q^2}\right)  \, h_2^2(t',Q^2) \,  f_+^{2 \ast}(t') \nonumber \\
&
\hspace{0.5cm} 
- \frac{1}{2 M_N}\left(\frac{t'}{t' + Q^2}\right)^2 \biggl[  \sqrt{\frac{3}{2}} t' \, h_0^2(t',Q^2) 
\nonumber \\
&
\hspace{.5cm}
+  \sqrt{2 t' Q^2} \, h_1^2(t',Q^2)  - \frac{Q^2}{2} h_2^2(t',Q^2)  \biggr] f_-^{2 \ast}(t') \biggr\}, 
\label{eq:imtpipitildeF2} \\
&\textup{Im}_{t}F_{5}^{\pi \pi}(Q^2,0,t') = - \frac{1}{e^2} 
\left(\frac{t'}{t' + Q^2}\right) \frac{5}{32} \; \beta^3_\pi(t')  \nonumber \\
&
\hspace{0.5cm} 
\times \biggl\{ \biggl[ h_2^2(t',Q^2)- \sqrt{\frac{2 t'}{Q^2}} h_1^2(t',Q^2) \biggr] f_-^{2 \ast}(t') \biggr\}, 
\label{eq:imtpipiF5} 
\end{align}
where we defined the pion velocity in the $\pi \pi$ rest frame as
\begin{eqnarray}
\beta_\pi(t') = \sqrt{1 - \frac{4 m_\pi^2}{t'}}.
\label{eq:betapipi}
\end{eqnarray}
In Eqs.~(\ref{eq:imtpipiF1})-(\ref{eq:imtpipiF5}), $h^J_{\Lambda_\gamma}(t',Q^2)$ are the isospin $I = 0$ helicity amplitudes for the $\gamma^\ast \gamma \to \pi \pi$ process from Ref.~\cite{Danilkin:2018qfn}, where $J$ refers to the partial wave in the $\pi \pi$ system, and $\Lambda_\gamma$ is the difference of photon helicities in the $\gamma^\ast \gamma$ c.m. system.  Note that the $s$-wave $\pi \pi$ intermediate state only contributes to $\textup{Im}_{t}F_{1}$, whereas $\textup{Im}_{t} \tilde F_{2}$ and $\textup{Im}_{t}F_{5}$ only receive contributions from $d$-waves or higher partial waves.  Furthermore, $f_\pm^J(t')$ denote the isospin $I = 0$ $\pi \pi \to N \bar N$ helicity amplitudes for parallel ($+$) or anti-parallel ($-$) nucleon and anti-nucleon helicities~\cite{FrazerFulco}.

\subsection{$t$-channel left-hand cut contribution due to $\Delta(1232)$-exchange}
\label{tchannelseclhc}

\begin{figure}
	\centering
	 \includegraphics[width=0.45\textwidth]{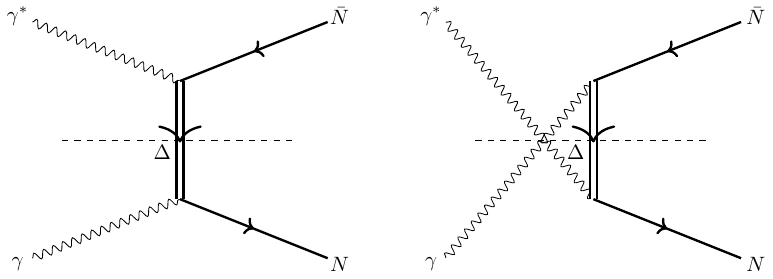}
	\caption{The $t$-channel $\Delta(1232)$-resonance exchange left-hand cut contribution in the subtracted dispersive formalism for VCS on the nucleon.}
	\label{fig:tchannel_lhc}
\end{figure}

The leading $t$-channel left-hand cut contribution to the VCS amplitudes at $\nu=0$ arises from the $\pi N$ intermediate state. In this work, we approximate the corresponding left-hand cut discontinuity by the spectral function associated with $\Delta(1232)$ excitation in the $s$- and $u$-channels (see Fig.~\ref{fig:tchannel_lhc}) including its finite width, $\Gamma_\Delta = 0.117$~GeV. Furthermore, we consider the dominant magnetic dipole excitations in the $\gamma^\ast N \Delta$ and $\gamma N \Delta$ vertices.  Details of the derivation are given in Appendix~\ref{app:delta}.

For the three VCS amplitudes $F_1^{\mathrm{NB}}$, $\tilde F_2^{\mathrm{NB}}$, and $F_5^{\mathrm{NB}}$, which we evaluate in this work through a subtracted DR, the left-hand cut discontinuities due to the $\Delta(1232)$-exchange are obtained from Eqs.(\ref{eq:f1delta})-(\ref{eq:f12delta}) as
\begin{align}
&\textup{Im}_{t}F_{1}^{\Delta}(Q^2,0,t') = 
\left[ \pi \rho^\Delta \left(M_N^2 - \frac{1}{2} (t' + Q^2); M_\Delta, \Gamma_\Delta \right) \right] \nonumber \\
&\hspace{1.cm} \times \frac{g_M(0)\, g_M(Q^2)}{M_N^2 Q_+^2} 
\biggl\{ - 2 M_N^2 (M_N + M_\Delta) (1 + \tau) \nonumber \\
&\hspace{3.5cm} + \left( M_\Delta + \frac{3}{4} M_N \right) (t' + Q^2) \biggr\}, \quad 
\label{eq:lhcF1delta} \\
&\textup{Im}_{t}\tilde F_{2}^{\Delta}(Q^2,0,t') = 
\left[ \pi \rho^\Delta \left(M_N^2 - \frac{1}{2} (t' + Q^2); M_\Delta, \Gamma_\Delta \right) \right] \nonumber \\
&\hspace{1.cm} \times \frac{g_M(0)\, g_M(Q^2)}{M_N^2 Q_+^2} 
\biggl\{ \frac{1}{2} (M_N + M_\Delta) \nonumber \\
&\hspace{3.75cm} + \frac{M_N}{4 (4 M_N^2 - t')}  (t' + Q^2) \biggr\}, \label{eq:lhctF2delta}\\
&\textup{Im}_{t}F_{5}^{\Delta}(Q^2,0,t') = 
\left[ \pi \rho^\Delta \left(M_N^2 - \frac{1}{2} (t' + Q^2); M_\Delta, \Gamma_\Delta \right) \right] \nonumber \\
&\hspace{1.cm} \times \frac{g_M(0)\, g_M(Q^2)}{M_N^2 Q_+^2} 
\biggl\{ - M_N (M_N + M_\Delta) \left(1 + \frac{\tau}{2} \right) \nonumber \\
&\hspace{3.75cm} + \left( M_\Delta + 7 M_N \right) \frac{(t' + Q^2)}{8 M_N} \biggr\}.
\label{eq:lhcF5delta}
\end{align}
where the normalized $\Delta(1232)$ spectral function $\rho^\Delta$  is defined in Eq.~(\ref{eq:spectraldel}). The variable $Q_+^2$ is defined in Eq.~(\ref{eq:defQ+}), and  $g_M(Q^2)$ denotes the covariant magnetic $\gamma^\ast N \Delta$ coupling of Eq.~(\ref{eq:gMgEgC_related_to_Ash}), for which we use the Ash form factors of the MAID2007 parameterization given in Eq.~(\ref{eq:GMGEGC}). 

Because the dispersion relation in $t$ is once-subtracted, the left-hand cut integrals exhibit rapid convergence. For VCS observables up to the $\Delta(1232)$ energy region, as considered in this work, contributions from higher-mass baryon resonances beyond the $\Delta(1232)$ are therefore expected to be suppressed.

\section{Results and discussion}

\begin{figure*}[!]
	\centering
\includegraphics[width=0.74\textwidth]{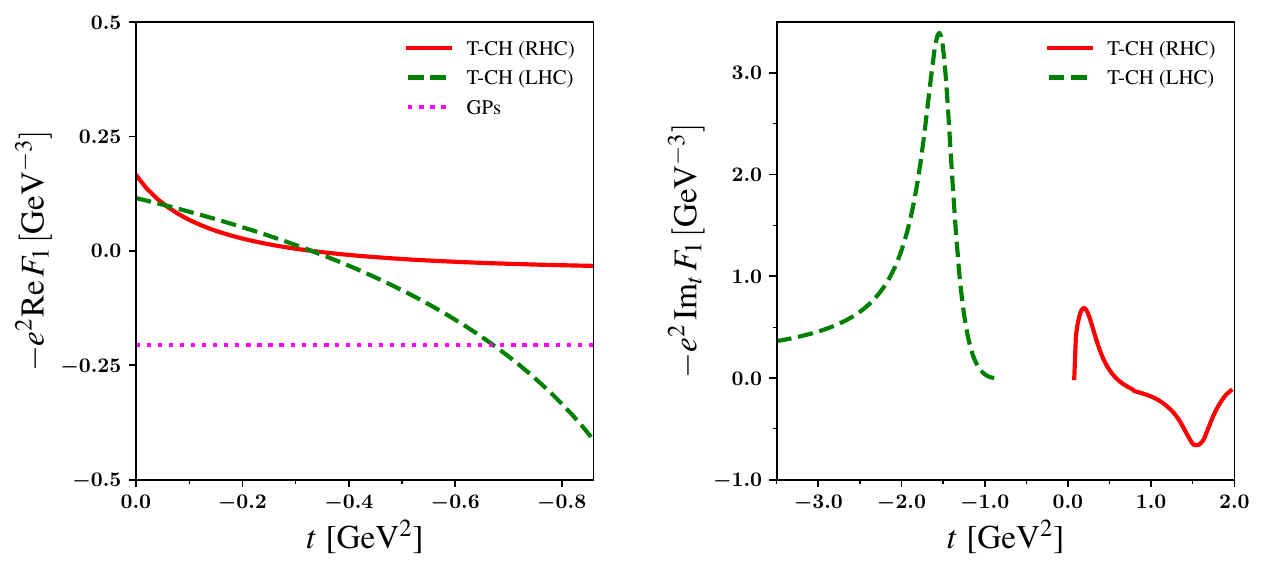}
	\caption{
    $t$-channel contributions to the real and imaginary parts of the amplitude $F_1^{\mathrm{NB}}(Q^2, 0, t)$ at $Q^2=0.33$ GeV$^2$ as function of $t$. 
    Left panel:
results from the $t$-channel subtracted DRs according to Eq.~(\ref{eq:subtraction}), with the separate contributions from the right-hand cut (RHC), left-hand cut (LHC), and the subtraction constant. The subtraction constant corresponds with a value of the GP $\beta_{M1}(Q^2)=1.20 \times 10^{-4}$ fm$^3$. Right panel: contributions to the imaginary part from the left-hand and right-hand cuts in the $t$-channel. }
   \label{fig:f1-ampl-tch-q2-0p33}
\end{figure*}

\begin{figure*}[!]
	\centering
\includegraphics[width=0.74\textwidth]{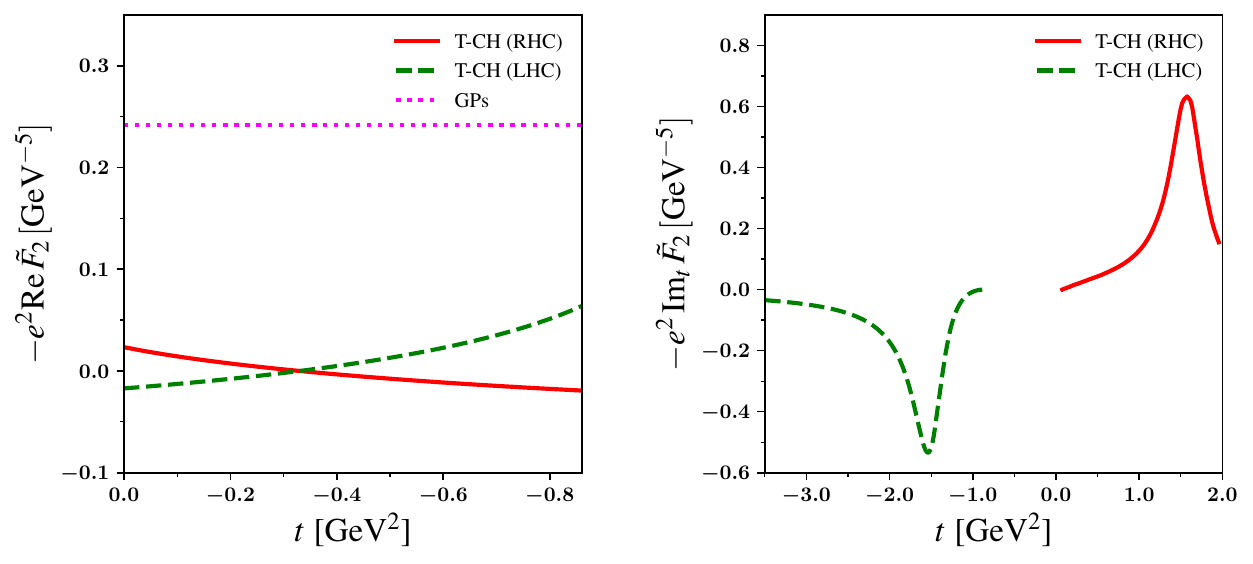}
\caption{The same as in Fig.~\ref{fig:f1-ampl-tch-q2-0p33} but for the amplitude $\tilde F_2^{\mathrm{NB}}(Q^2, 0, t)$ at $Q^2=0.33$ GeV$^2$. 
The subtraction constant corresponds with a value of the GP combination $\alpha_{E1}(Q^2) + \beta_{M1}(Q^2)= 5.44 \times 10^{-4}$ fm$^3$. }
   \label{fig:f2-ampl-tch-q2-0p33}
\end{figure*}

\begin{figure*}[!]
	\centering
\includegraphics[width=0.74\textwidth]{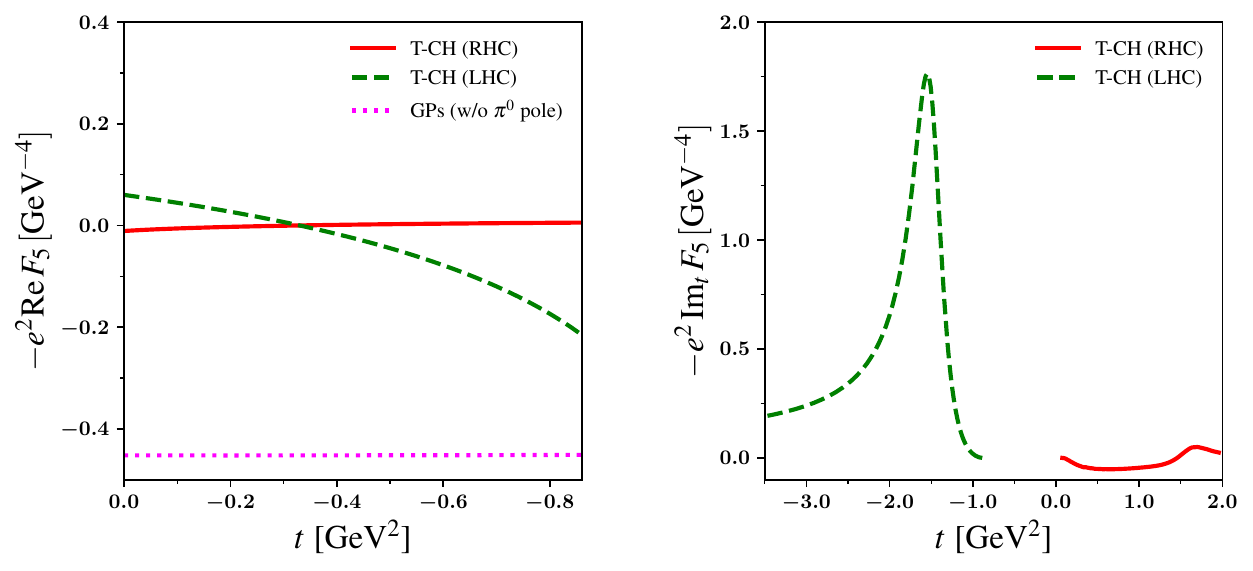}
\caption{The same as in Fig.~\ref{fig:f1-ampl-tch-q2-0p33} but for the amplitude $F_5^{\mathrm{NB}}(Q^2, 0, t)$. The GP value excludes the $\pi^0$ pole contribution and is calculated according to Eq.~(\ref{eq:unsubtdr}) for $t=-Q^2$ and $\nu=0$, with the upper limit of integration corresponding to $W_{\mathrm{max}}=2.0$ GeV.}
   \label{fig:f5-ampl-tch-q2-0p33}
\end{figure*}

We start our discussion with the $t$-channel dispersive input and the calculation of the subtraction functions at $\nu = 0$ for the non-Born VCS amplitudes $F_1^{\mathrm{NB}}$, $\tilde F_2^{\mathrm{NB}}$, and $F_5^{\mathrm{NB}}$, which are the main new elements of the present subtracted DR formalism for VCS. 
For illustration, the $t$-dependence of $F_1^{\mathrm{NB}}(Q^2, 0, t)$, $\tilde F_2^{\mathrm{NB}}(Q^2, 0, t)$, and $F_5^{\mathrm{NB}}(Q^2, 0, t)$ is shown in Figs.~\ref{fig:f1-ampl-tch-q2-0p33}, \ref{fig:f2-ampl-tch-q2-0p33}, and \ref{fig:f5-ampl-tch-q2-0p33} for $Q^2 = 0.33$~GeV$^2$. Several VCS experiments have taken data at this $Q^2$ value, including the most recent measurement of Ref.~\cite{Li:2022sqgnature}.

In the right panels of Figs.~\ref{fig:f1-ampl-tch-q2-0p33}-\ref{fig:f5-ampl-tch-q2-0p33}, we show the $t$-channel discontinuities $\textup{Im}_t F_i$ arising from the right-hand cut (RHC) due to the $\pi \pi$ intermediate state (see Sec.~\ref{tchannelsec}) and from the left-hand cut (LHC) due to the $\Delta(1232)$ resonance (see Sec.~\ref{tchannelseclhc}). The dominant feature of the RHC input is the $s$-wave $\pi \pi$ contribution corresponding to the $f_0(500)$ region, which contributes exclusively to $\textup{Im}_t F_1$ according to Eq.~\eqref{eq:imtpipiF1} (see the right panel of Fig.~\ref{fig:f1-ampl-tch-q2-0p33}). Furthermore, the $d$-wave $\pi \pi$ state contributes to all three discontinuities and shows a pronounced enhancement due to the tensor resonance $f_2(1270)$. As seen in Fig.~\ref{fig:f1-ampl-tch-q2-0p33}, the $f_0(500)$ and $f_2(1270)$ contributions to $\textup{Im}_t F_1$ enter with opposite sign, leading to a partial cancellation in the evaluating $\textup{Re} F_1$.

In this work, the numerical evaluation of both $s$- and $d$-wave $\pi \pi$ contributions to the $t$-channel RHC discontinuities in Eqs.~(\ref{eq:imtpipiF1})-(\ref{eq:imtpipiF5}) is performed using the $\gamma^\ast \gamma \to \pi \pi$ amplitudes of the dispersive formalism of~\cite{Danilkin:2018qfn} and the $\pi \pi \to N \bar N$ amplitudes of~\cite{Hoferichter:2015hva}, constructed from the Roy-Steiner equations for $\pi N$ scattering. In evaluating the corresponding real parts $\textup{Re} F_i$, the RHC subtracted dispersion integrals in Eq.~(\ref{eq:subtraction}) are computed up to an upper cutoff $\simeq 2$~GeV$^2$ in $d$-wave and $\simeq 0.78$ GeV$^2$ in $s$-wave~\footnote{In the present implementation we truncate the $s$-wave $\pi\pi$ contribution at $\simeq 0.78$ GeV$^2$ in order to isolate the $f_0(500)$ region. Extending the $s$-wave integration to higher cutoff would include the $f_0(980)$ region, where the $\gamma^\ast \gamma \to K\bar{K}$ and $K\bar{K}\to N \bar N$ inputs are needed in addition to the $\pi\pi$ channel. We have checked that the $\pi\pi$ contribution remains dominant above the $K\bar{K}$ threshold and omitting $K\bar{K}$ channel induces an imaginary part at the $\sim 15-20\%$ level in $\textup{Im}_t F_1$. Within our once-subtracted DR framework, the resulting impact of the $f_0(980)$ region on $\textup{Re} F_1$ is further suppressed.}. It was checked that the subtracted dispersion integrals exhibit  very fast convergence, and any contribution to $\textup{Im}_t F_i$ originating from the timelike region beyond $t \sim 2$~GeV$^2$ has a negligible impact when evaluating  $\textup{Re} F_i(Q^2, 0, t)$ in the spacelike region for $-t \lesssim 1$~GeV$^2$.

Figures~\ref{fig:f1-ampl-tch-q2-0p33}-\ref{fig:f5-ampl-tch-q2-0p33} also show the LHC discontinuities from the $\Delta(1232)$ resonance and the corresponding LHC subtracted dispersion integrals in Eq.~(\ref{eq:subtraction}). 
We checked that in evaluating the LHC contribution to $\textup{Re} F_i(Q^2, 0, t)$ in the spacelike region for $-t \lesssim 0.8$~GeV$^2$, the corresponding LHC subtracted dispersion integrals exhibit very fast convergence. They reach around 95\% or higher of their asymptotic value for the lower integration limit of $t_{lower} = - 3$~GeV$^2$, which corresponds to $W \simeq 1.5$~GeV for $\nu = 0$ and $Q^2 = 0.33$~GeV$^2$. Thus, further LHC discontinuities beyond $\Delta(1232)$ are expected to contribute at the few percent level or less to the subtracted dispersion integral for $\textup{Re} F_i(Q^2, 0, t)$.

The good convergence of the subtracted integrals entering $\textup{Re} F_i(Q^2, 0, t)$ in Eq.~(\ref{eq:subtraction}) in our subtracted DR formalism is achieved at the expense of introducing the subtraction constants 
$F_i^{\mathrm{NB}} (Q^2, 0, -Q^2) - F_i^{t\text{-pole}}(Q^2,0,-Q^2)$.  
For the amplitude $F_1$, the subtraction constant is related to the magnetic GP $\beta_{M1}(Q^2)$ according to Eq.~(\ref{eq:F1GP}), while for the amplitude $\tilde F_2$ it is related to the GP combination $\alpha_{E1}(Q^2) + \beta_{M1}(Q^2)$ according to Eq.~(\ref{eq:F2tildeGP}). In our formalism, both constants are to be extracted from a fit to the VCS data. In Figs.~\ref{fig:f1-ampl-tch-q2-0p33} and \ref{fig:f2-ampl-tch-q2-0p33}, we illustrate their contribution $\textup{Re} F_1$ and $\textup{Re} \tilde F_2$ using the GP values determined in the JLab experiment of~\cite{Li:2022sqgnature}. For the amplitude $F_5$, which contains the $\pi^0$ pole contribution, the subtraction constant involves the GP contribution beyond the $\pi^0$ pole. Its value is expressed in terms of a spin GP according to Eq.~(\ref{eq:F5GP}) and is calculated in the present work using Eq.~(\ref{eq:unsubtdr}) for $t=-Q^2$ and $\nu=0$, with the upper integration limit corresponding to $W_{\mathrm{max}}=2.0$ GeV.

\begin{figure*}[!]
	\centering
\includegraphics[width=0.92\textwidth]{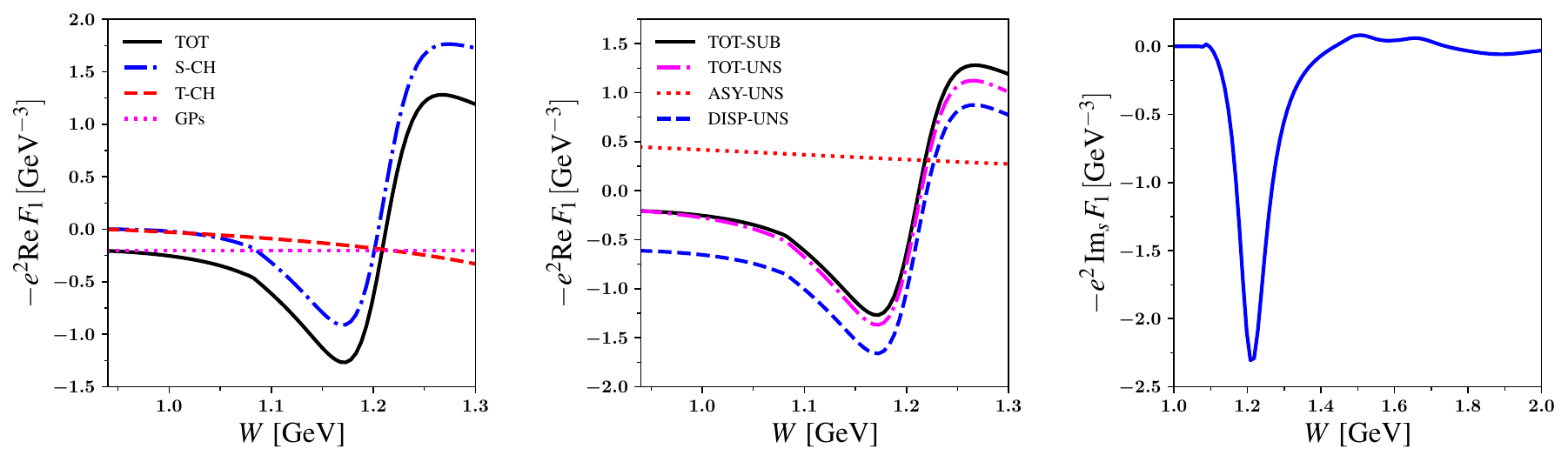}
	\caption{
    Real and imaginary parts of the amplitude $F_1^{\mathrm{NB}}(Q^2, \nu, t)$ at $Q^2=0.33$ GeV$^2$ and $\theta=150^o$  as function of $W$. 
    Left panel: results from the subtracted DRs with the separate contributions from the $s$- and $t$-channel, and for a value of the GP $\beta_{M1}(Q^2)=1.20 \times 10^{-4}$ fm$^3$.
    Middle panel: total results from subtracted DRs (solid black curve) compared with  the unsubtracted DR result of~\cite{Pasquini:2001yy} (dashed-dotted pink curve). For the unsubtracted DR result we also show the separate contributions of the dispersive $s$-channel contribution according to 
    Eq.~(\ref{eq:unsubtdr}) with the upper limit of integration corresponding to $W_{\mathrm{max}}=2.0$ GeV (dashed blue curve) and the asymptotic $t$-channel contribution (dotted red curve). 
    Right panel: contribution to the imaginary part from the $s$-channel cut. }
    \label{fig:f1-ampl-sch-tch-q2-0p33-theta150}
\end{figure*}

\begin{figure*}[!]
	\centering
\includegraphics[width=0.92\textwidth]{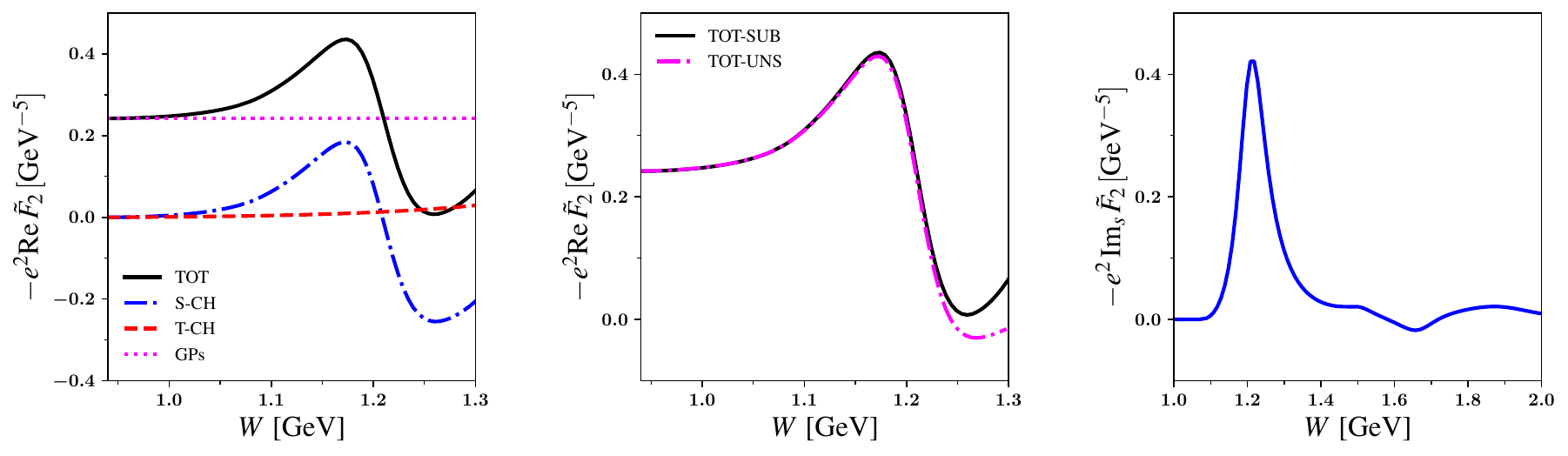}
	\caption{Real and imaginary parts of the amplitude $\tilde F_2^{\mathrm{NB}}(Q^2, \nu, t)$ at $Q^2=0.33$ GeV$^2$ and $\theta=150^o$  as function of $W$. 
    Left panel: results from the subtracted DRs   with the separate contributions from the $s$- and $t$-channel, and for a value of the GP combination $\alpha_{E1}(Q^2) + \beta_{M1}(Q^2) = 5.44 \times 10^{-4}$ fm$^3$. 
    Middle panel: total results from subtracted DRs (solid black curve) compared with  the unsubtracted DR result of~\cite{Pasquini:2001yy} (dashed-dotted pink curve). 
    Right panel: contribution to the imaginary part from the $s$-channel cut. }
    \label{fig:f2-ampl-sch-tch-q2-0p33-theta150}
\end{figure*}

\begin{figure*}[!]
	\centering
\includegraphics[width=0.92\textwidth]{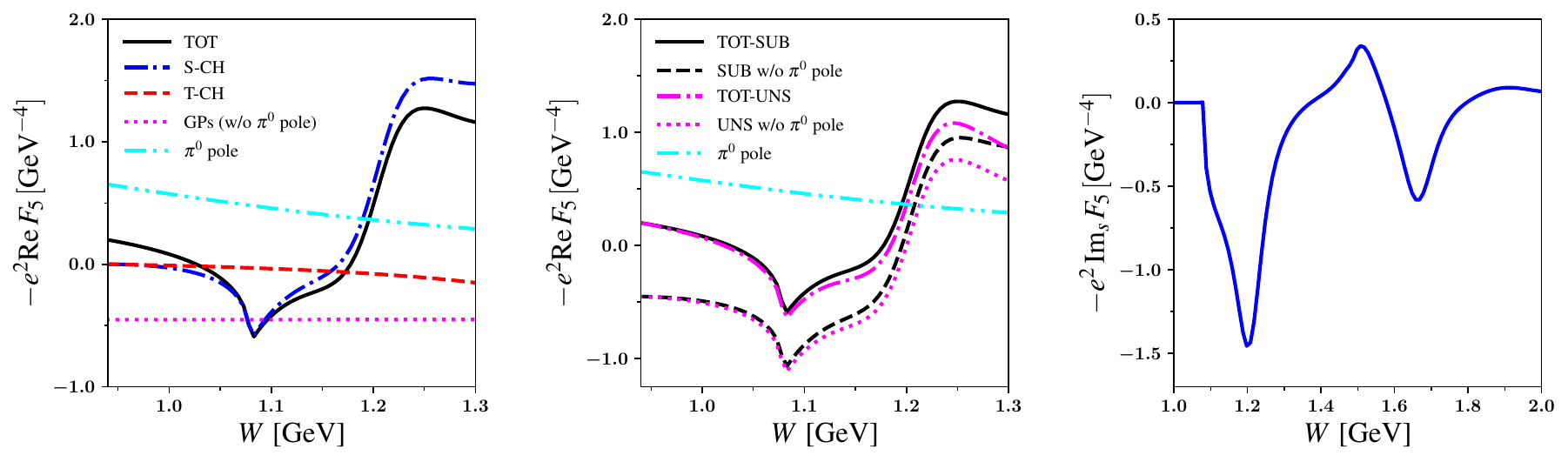}
	\caption{
    Real and imaginary parts of the amplitude $F_5^{\mathrm{NB}}(Q^2, \nu, t)$ at $Q^2=0.33$ GeV$^2$ and $\theta=150^o$  as function of $W$. 
    Left panel: results from the subtracted DRs, with the separate contributions from the $s$-channel, $t$-channel, $\pi^0$ pole, and GP contribution.  
    The GP value excludes the $\pi^0$ pole contribution and is calculated according to 
    Eq.~(\ref{eq:unsubtdr}) for $t=-Q^2$ and $\nu=0$, with the upper limit of integration corresponding to $W_{\mathrm{max}}=2.0$ GeV. 
   Middle panel: comparison of the results from subtracted DRs with the  unsubtracted DR result of~\cite{Pasquini:2001yy}, with and without the $\pi^0$ pole. Right panel: contribution to the imaginary part from the $s$-channel cut. }
   \label{fig:f5-ampl-sch-tch-q2-0p33-theta150}
\end{figure*}

Extracting the GPs from VCS measurements in the $\Delta(1232)$ energy region requires a reliable control of the $W$ dependence, which is provided by the dispersive calculation. The $s$-channel dispersive input which enters the subtracted DRs of Eq.~(\ref{eq:subtdr}), as well as the corresponding real parts 
$\textup{Re} F_i$, are shown in Figs.~\ref{fig:f1-ampl-sch-tch-q2-0p33-theta150}, \ref{fig:f2-ampl-sch-tch-q2-0p33-theta150}, and \ref{fig:f5-ampl-sch-tch-q2-0p33-theta150}  
as functions of the c.m. energy $W$ for $Q^2 = 0.33$~GeV$^2$ and for a fixed c.m. scattering angle $\theta = 150^o$. Along such a curve of fixed scattering angle, $W$ increases starting at $W = M$, where the amplitudes reduce to their GP values $F_i^{\mathrm{NB}}(Q^2, 0, -Q^2)$. The right panels of Figs.~\ref{fig:f1-ampl-sch-tch-q2-0p33-theta150}, \ref{fig:f2-ampl-sch-tch-q2-0p33-theta150}, and \ref{fig:f5-ampl-sch-tch-q2-0p33-theta150} show the $s$-channel discontinuities $\textup{Im}_s F_i$ entering the subtracted dispersion integrals. One sees that these $s$-channel discontinuities  are dominated by the low-$W$ region where the MAID2007 parameterization~\cite{Drechsel:2007if} provides a very good description of existing pion electroproduction data. 
For $\textup{Im}_s F_1$ and $\textup{Im}_s \tilde F_2$ the contributions beyond the $\Delta(1232)$ region are very small, while for $\textup{Im}_s F_5$ the contributions in the second and third nucleon resonance regions come with opposite sign. Thus, the corresponding subtracted dispersive integral for $\textup{Re} F_5$ is also nearly completely saturated by its contribution from the $\Delta(1232)$ region. One sees from the left panels of Figs.~\ref{fig:f1-ampl-sch-tch-q2-0p33-theta150} and \ref{fig:f2-ampl-sch-tch-q2-0p33-theta150} that the real parts of the VCS amplitudes $F_1$ and $\tilde F_2$ show the characteristic variation of an elastic resonance, with the real part of the $s$-channel dispersive contribution going through zero when crossing the $\Delta(1232)$ position. 
The real part of the amplitude $F_5$, shown in Fig.~\ref{fig:f5-ampl-sch-tch-q2-0p33-theta150}, displays in addition a pronounced cusp behavior at the $\pi N$ threshold. 
On the other hand, the $t$-channel subtracted DR contribution displays a smooth variation with $W$ for the VCS amplitudes, while $\textup{Re} F_5$ shows a prominent $\pi^0$ pole contribution. Furthermore, we also show the GP subtraction constant contribution, corresponding to $\textup{Re} F_i$ at $W = M$. In the middle panels of Figs.~\ref{fig:f1-ampl-sch-tch-q2-0p33-theta150}, \ref{fig:f2-ampl-sch-tch-q2-0p33-theta150}, and \ref{fig:f5-ampl-sch-tch-q2-0p33-theta150}, we compare the present subtracted DR formalism with the unsubtracted DR result of~\cite{Pasquini:2001yy}. While there is a good overall agreement between both frameworks in the $\Delta(1232)$ energy region, one notices that in particular on the higher energy side of the $\Delta(1232)$, where the experiment of \cite{Li:2022sqgnature} has taken data, there are clear differences that are significant when compared with the GP effects.

\begin{figure*}[!]
	\centering
\includegraphics[width=0.92\textwidth]{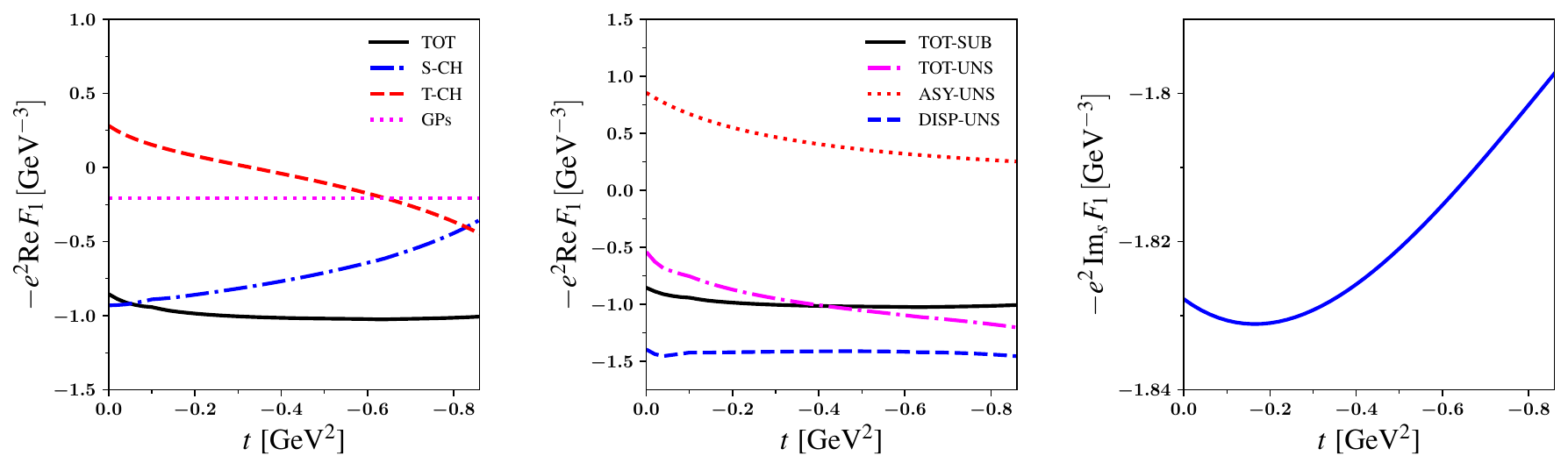}
	\caption{
    Real and imaginary parts of the amplitude $F_1^{\mathrm{NB}}(Q^2, \nu, t)$ at $Q^2=0.33$ GeV$^2$ and $W=1.19$ GeV as function of $t$. 
    Left panel: results from the subtracted DRs with the separate contributions from the $s$- and $t$-channel, and for a value of the GP $\beta_{M1}(Q^2)=1.20 \times 10^{-4}$ fm$^3$. 
        Middle panel: total results from subtracted DRs (solid black curve)  compared with the unsubtracted DR result of~\cite{Pasquini:2001yy} (dashed-dotted pink curve). For the unsubtracted DR result we also show the separate contributions of the dispersive $s$-channel contribution with the upper limit of integration corresponding to $W_{\mathrm{max}}=2.0$ GeV (dashed blue curve) and the asymptotic $t$-channel contribution (dotted red curve). 
        Right panel: contribution to the imaginary part from the $s$-channel cut. }
    \label{fig:f1-ampl-sch-tch}
\end{figure*}

\begin{figure*}[!]
	\centering
\includegraphics[width=0.92\textwidth]{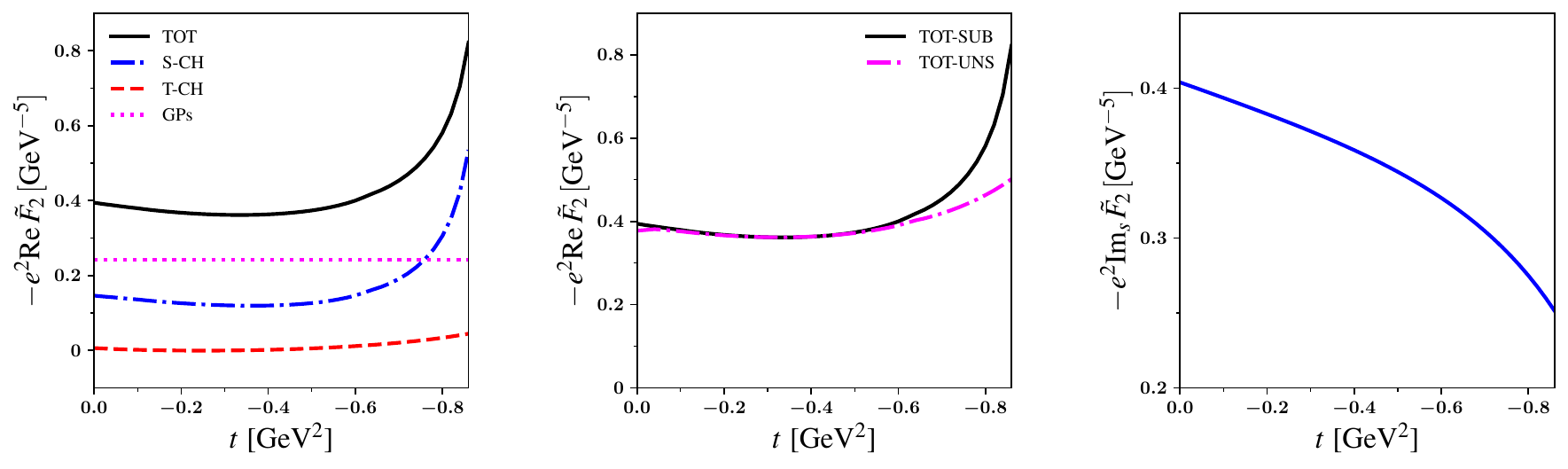}
	\caption{Real and imaginary parts of the amplitude $\tilde F_2^{\mathrm{NB}}(Q^2, \nu, t)$ at $Q^2=0.33$ GeV$^2$ and $W=1.19$ GeV as function of $t$. 
    Left panel: results from the subtracted DRs with the separate contributions from the $s$- and $t$-channel, and for a value of the GP combination $\alpha_{E1}(Q^2) + \beta_{M1}(Q^2) =5.44 \times 10^{-4}$ fm$^3$. 
        Middle panel: total results from subtracted DRs (solid black curve)  compared with the unsubtracted DR result of~\cite{Pasquini:2001yy} (dashed-dotted pink curve). 
        Right panel: contribution to the imaginary part from the $s$-channel cut. }
    \label{fig:f2-ampl-sch-tch}
\end{figure*}

\begin{figure*}[!]
	\centering
\includegraphics[width=0.92\textwidth]{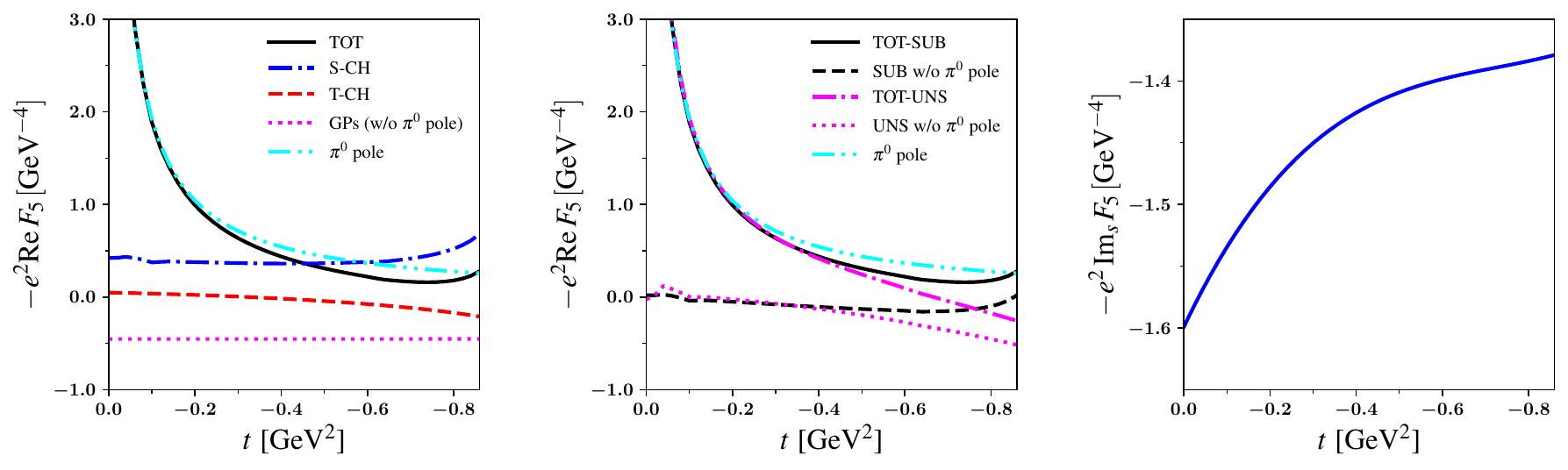}
	\caption{
    Real and imaginary parts of the amplitude $F_5^{\mathrm{NB}}(Q^2, \nu, t)$ at $Q^2=0.33$ GeV$^2$ and $W=1.19$ GeV as function of $t$. 
Left panel: results from the subtracted DRs, with the separate contributions from the $s$-channel, $t$-channel, $\pi^0$ pole, and GP contribution.  
    The GP value excludes the $\pi^0$ pole contribution and is calculated according to 
    Eq.~(\ref{eq:unsubtdr}) for $t=-Q^2$ and $\nu=0$, with the upper limit of integration corresponding to $W_{\mathrm{max}}=2.0$ GeV. 
   Middle panel: comparison of the results from subtracted DRs with the  unsubtracted DR result of~\cite{Pasquini:2001yy}, with and without the $\pi^0$ pole.
    Right panel: contribution to the imaginary part from the $s$-channel cut. }
    \label{fig:f5-ampl-sch-tch}
\end{figure*}

The above mentioned recent JLab VCS experiment of~\cite{Li:2022sqgnature} was performed at several fixed $W$ values in the $\Delta(1232)$ region. As illustration, we show in 
Figs.~\ref{fig:f1-ampl-sch-tch}, \ref{fig:f2-ampl-sch-tch}, and \ref{fig:f5-ampl-sch-tch} for a fixed $W = 1.19$~GeV, fixed $Q^2 = 0.33$~GeV$^2$ kinematics of \cite{Li:2022sqgnature}
the $t$-dependence of the VCS amplitudes $F_1$, $\tilde F_2$, and $F_5$ in the present subtracted DR formalism in comparison with their unsubtracted counterparts from \cite{Pasquini:2001yy}. One again notices clear differences between both formalisms. 
While for $\textup{Re} \tilde F_2$ and $\textup{Re} F_5$ the differences between the subtracted and unsubtracted results are mainly increasing for larger values of $-t$, for the amplitude $\textup{Re} F_1$ there is a clear difference over the whole $t$-region. This difference mainly originates from the different treatment of the $t$-channel discontinuities. In the unsubtracted formalism of \cite{Pasquini:2001yy}  the $t$-channel asymptotic contribution to the amplitude $F_1$ was modeled through a  $f_0(500)$-pole contribution, leading to a monopole behavior in $t$. The latter is responsible for the steeper behavior especially at smaller values of $-t$, as seen from the middle panel in Fig.~\ref{fig:f1-ampl-sch-tch}. In the present subtracted DR formalism on the other hand, the $t$-dependence is fully calculated and for $\textup{Re} F_1$ is given by the RHC and LHC contributions as shown in Fig.~\ref{fig:f1-ampl-tch-q2-0p33} and discussed above. Such different $t$-dependence is of relevance when extracting the GPs from the VCS data in the $\Delta(1232)$ region.

\begin{figure*}
	\centering
\includegraphics[width=0.8\textwidth]{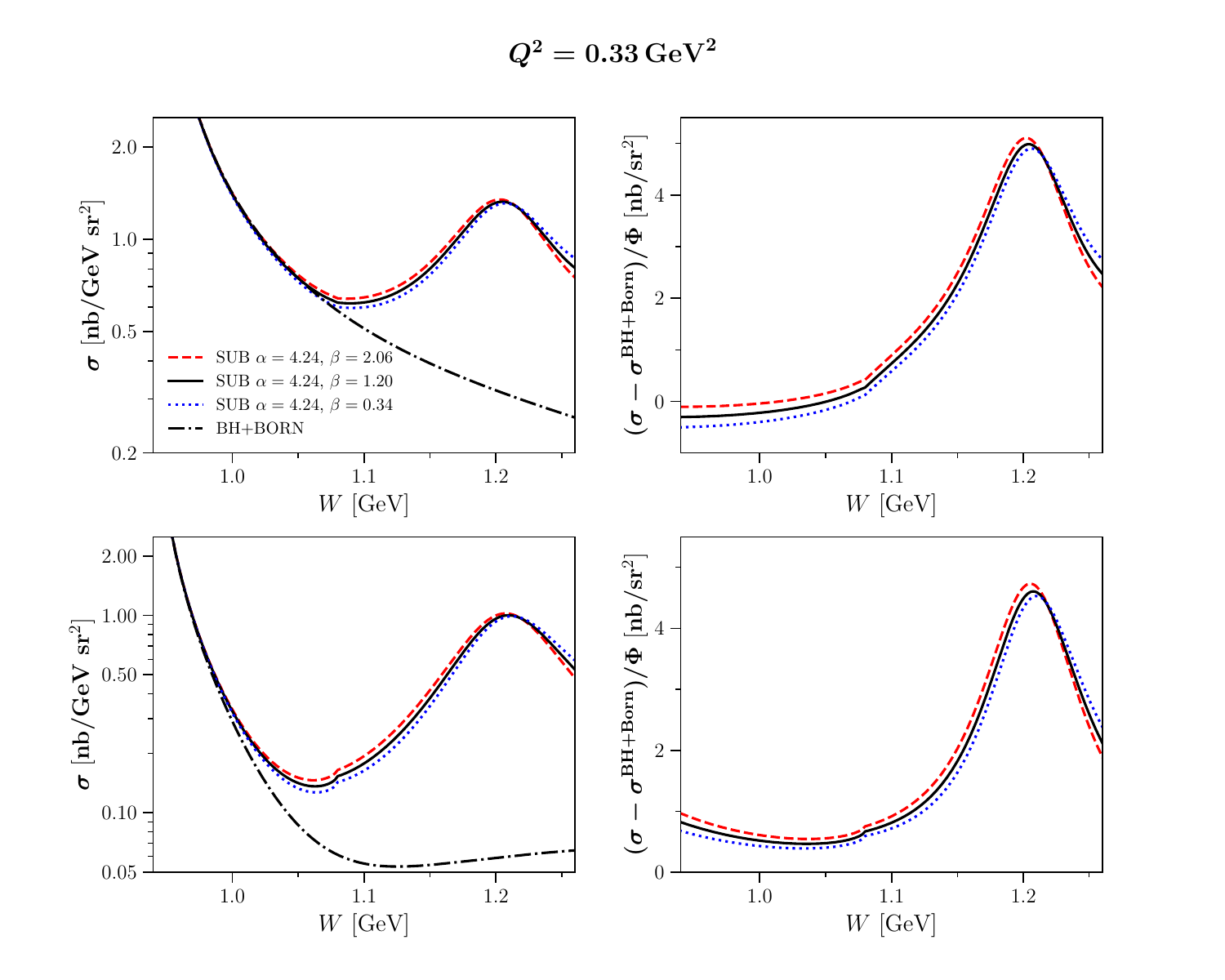}
	\caption{Left panels: $e^- p \to e^- \gamma p$ fivefold differential cross section of Eq.~(\ref{eq:cross}) as a function of the c.m. energy $W$ in the  kinematics of the upcoming VCS-II experiment at JLab~\cite{VCS-II:2023jhu}:
electron beam energy $E_e = 2.2$ GeV, initial photon virtuality $Q^2 = 0.33 $ GeV$^2$,  and photon c.m. scattering angle $\theta=150^o$. 
Right panels: results for $(\sigma-\sigma^{\mathrm{BH+Born}})/\Phi$ with $\Phi$ the phase space factor, given by Eq.~(\ref{eq:phasespace}).
Top and bottom panels correspond to azimuthal
angle $\phi=180^o$  and $\phi=0^o$, respectively. 
The dashed-dotted curves in the left panels show the result of the BH + Born cross section. 
 The other curves show the sensitivity in the subtracted DR formalism to the magnetic  GP $\beta_{M1}$ for fixed value of the electric GP $\alpha_{E1}$ (both in units $10^{-4}$ fm$^3$), as indicated
in the legend.
 }
	\label{fig:vcs-w-beta_fromthreshold}
\end{figure*}

\begin{figure*}
	\centering
    \includegraphics[width=0.65\textwidth]{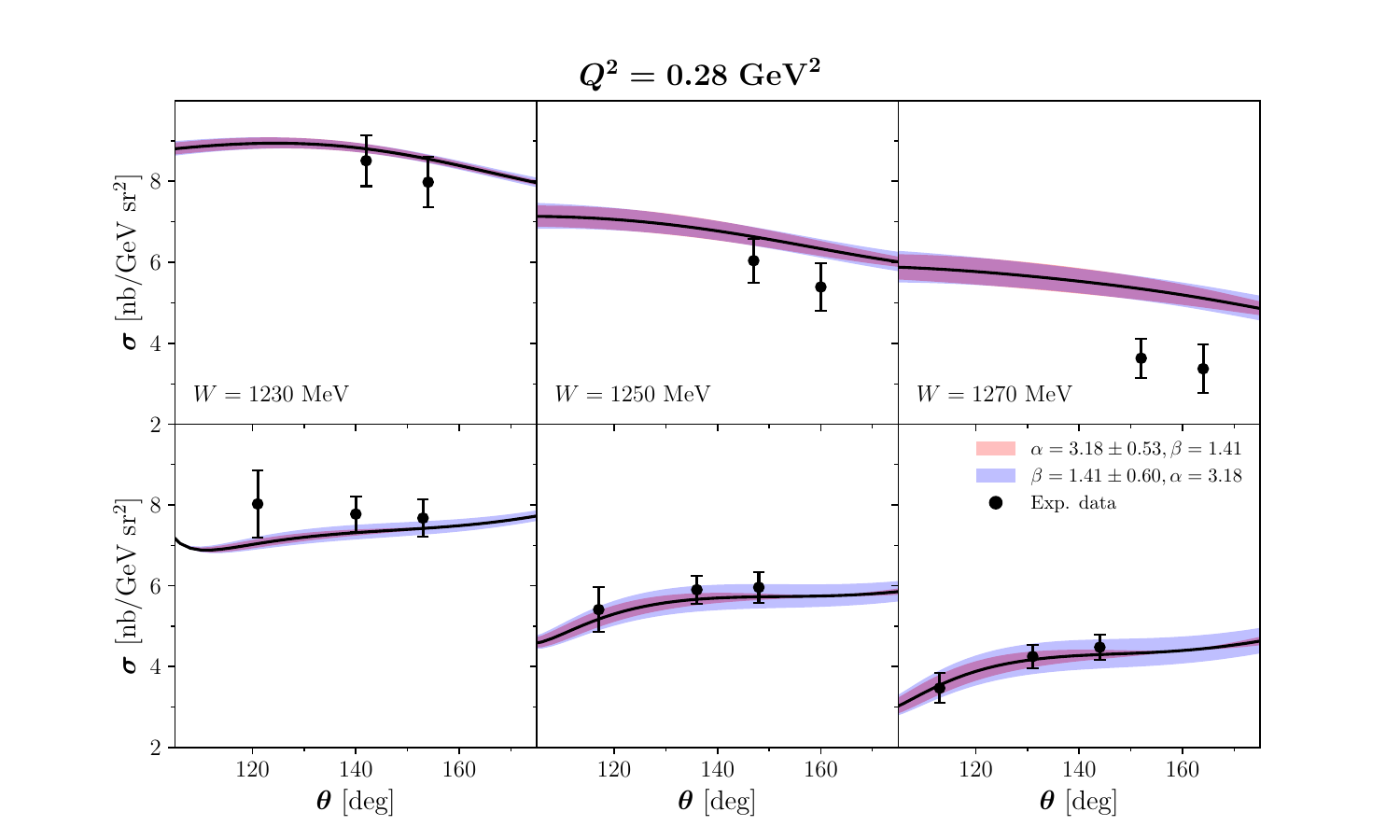}\\
    \includegraphics[width=0.65\textwidth]{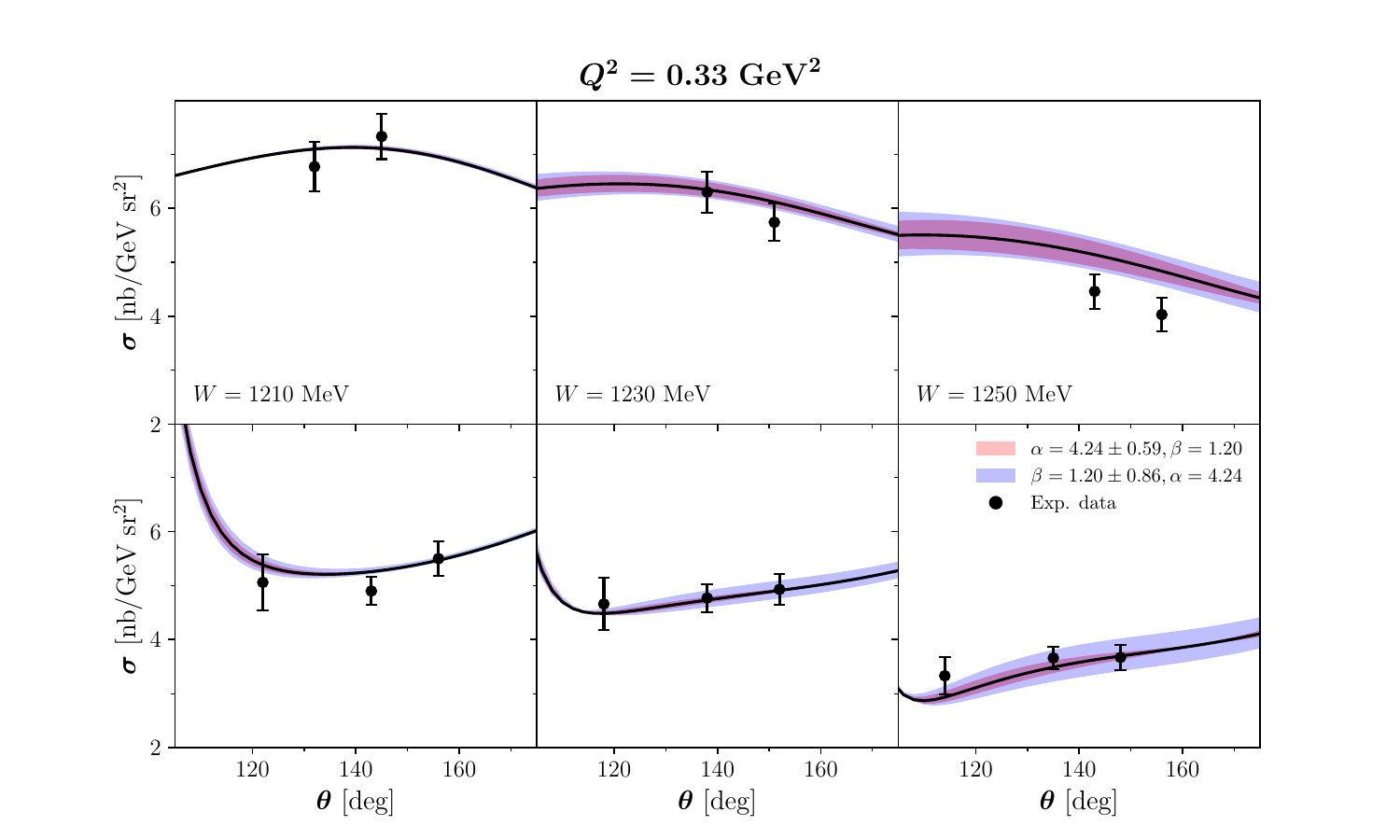}\\
\includegraphics[width=0.65\textwidth]{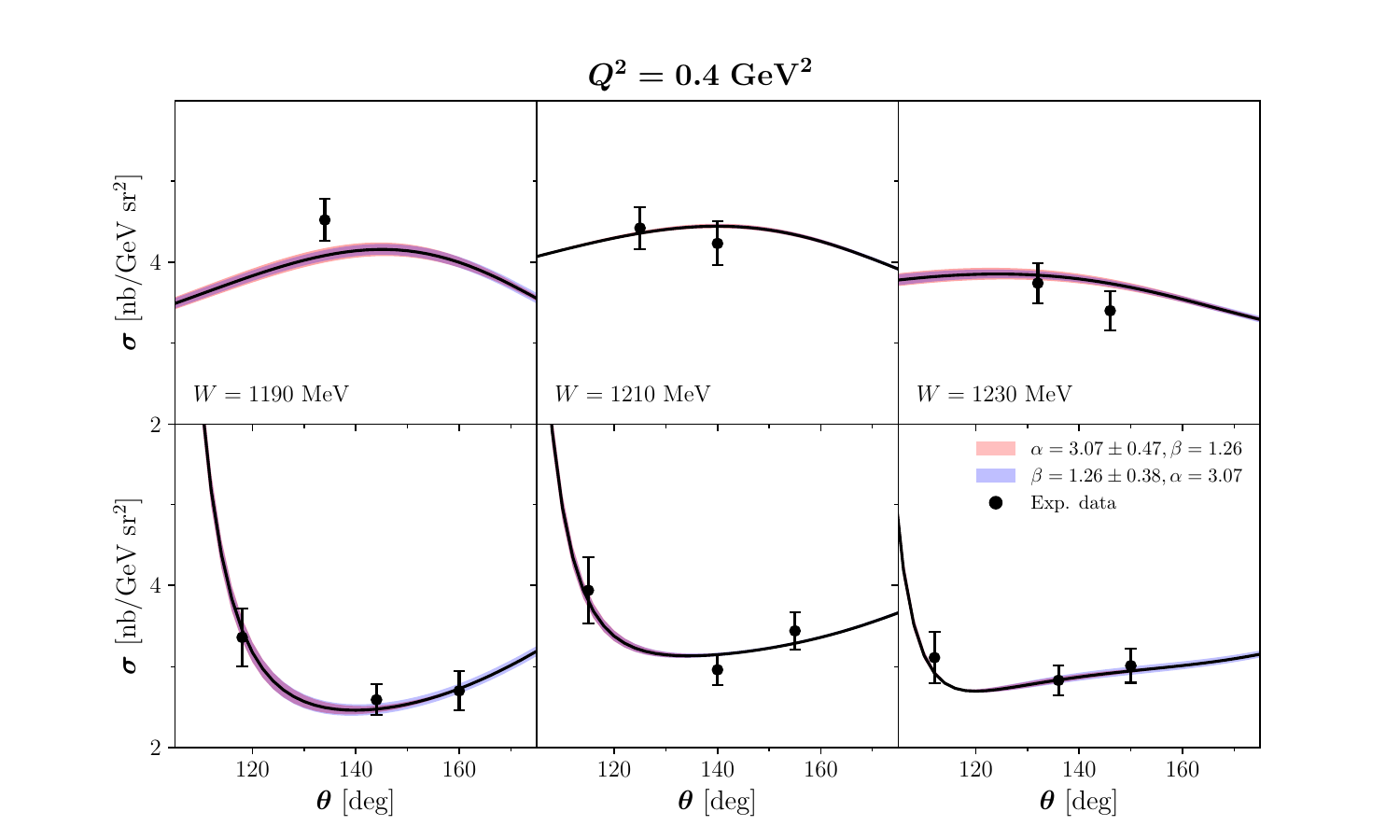}
	\caption{$e^- p \to e^- \gamma p$ fivefold differential cross section of Eq.~(\ref{eq:cross}) as a function of the photon c.m. scattering angle for the  kinematics of the JLab experiment~\cite{Li:2022sqgnature}:
electron beam energy $E_e = 4.56$ GeV, three values for photon virtuality $Q^2$, and several c.m. energy $W$ values as indicated on the panels. 
Top and bottom panels correspond to azimuthal angle $\phi=180^o$  and $\phi=0^o$, respectively.
The  bands show the $1 \sigma$ uncertainties associated with the values of the scalar GPs extracted in Ref.~\cite{Li:2022sqgnature}, as indicated in the legends (all in units $10^{-4}$ fm$^3$).
 }
	\label{fig:vcs-nature}
\end{figure*}

\begin{figure*}
	\centering
\includegraphics[width=0.75\textwidth]{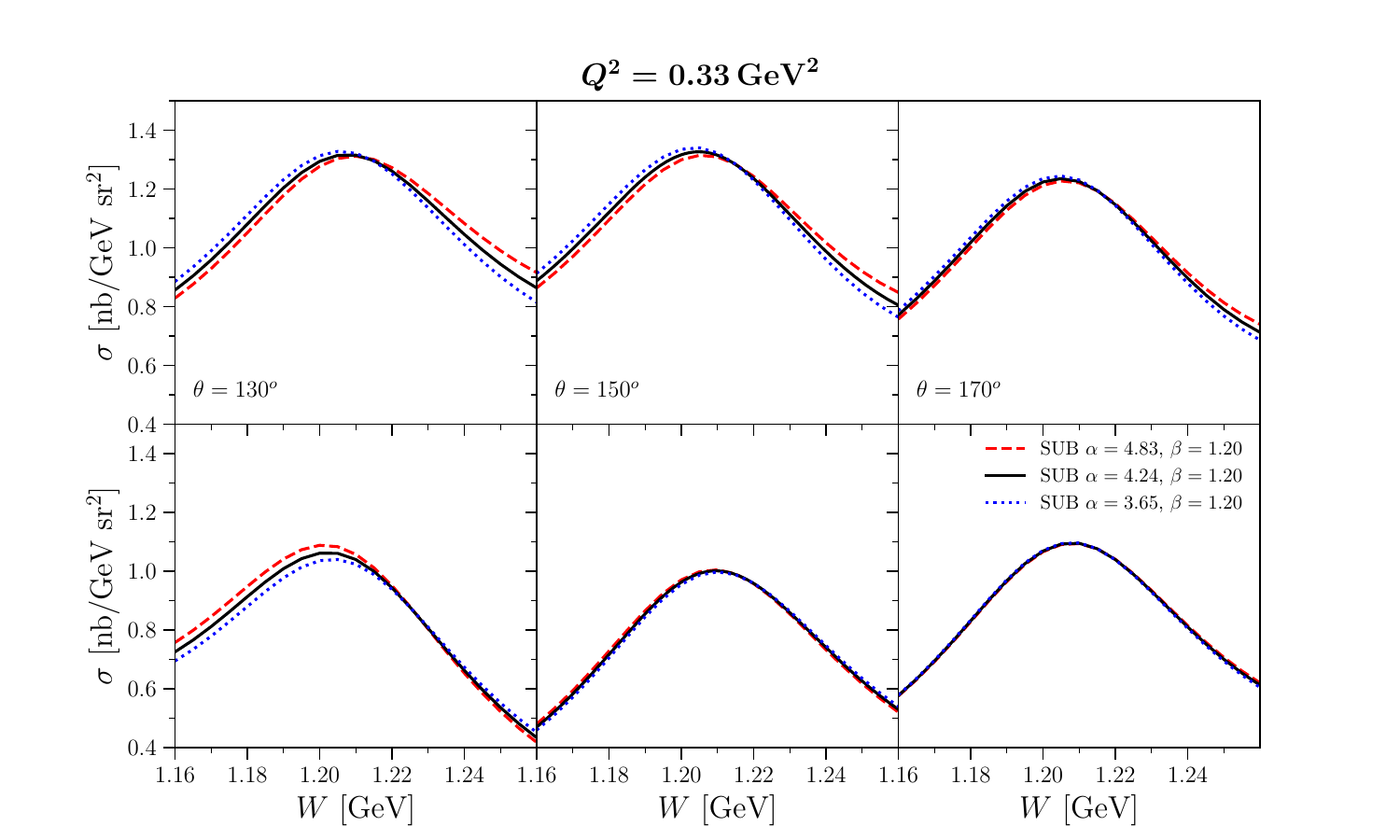}
	\caption{$e^- p \to e^- \gamma p$ fivefold differential cross section of Eq.~(\ref{eq:cross}) as a function of the c.m. energy $W$ in the  kinematics of the upcoming VCS-II experiment at JLab~\cite{VCS-II:2023jhu}:
electron beam energy $E_e = 2.2$ GeV, initial photon virtuality $Q^2 = 0.33 $ GeV$^2$, for several values of the photon c.m. scattering angle. Top and bottom panels correspond to azimuthal
angle $\phi=180^o$  and $\phi=0^o$, respectively.
 The curves show the sensitivity to the scalar electric GP $\alpha_{E1}$ for fixed value of the magnetic GP $\beta_{M1}$ (both in units $10^{-4}$ fm$^3$), as indicated
in the legend.
 }
	\label{fig:vcs-w-alpha}
\end{figure*}

\begin{figure*}
	\centering
\includegraphics[width=0.75\textwidth]{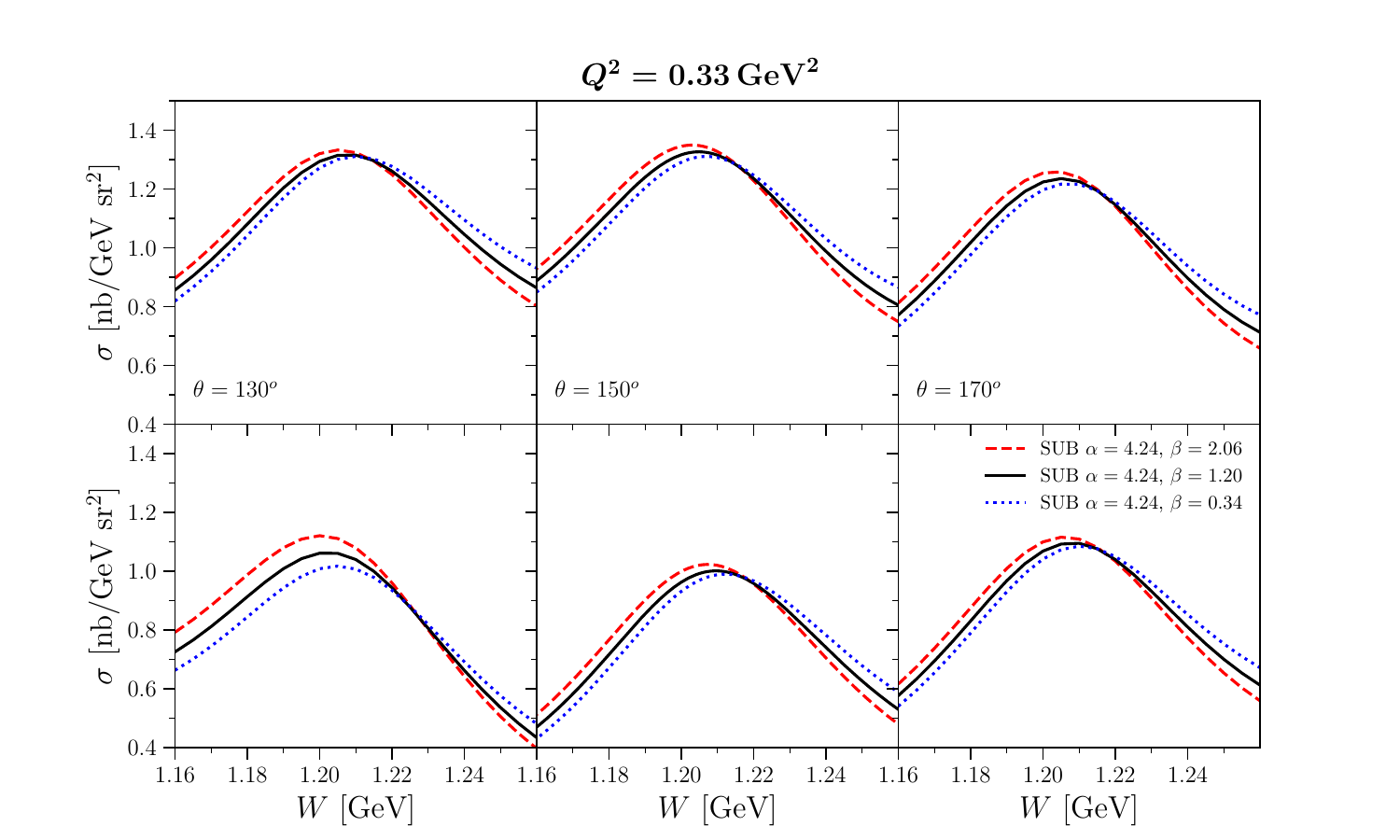}
	\caption{The same as Fig.~\ref{fig:vcs-w-alpha}, where 
 the curves show the sensitivity to the magnetic  GP $\beta_{M1}$ for fixed value of the electric GP $\alpha_{E1}$ (both in units $10^{-4}$ fm$^3$), as indicated
in the legend.
 }
	\label{fig:vcs-w-beta}
\end{figure*}

We next discuss the VCS observables and present our results in the subtracted DR formalism.  
The unpolarized fivefold differential cross section for the $e^- p \to e^-  \gamma p$ reaction is expressed as~\cite{Guichon:1995pu}
\begin{align}
\frac{d^5 \sigma}{(d \Omega^\prime_e)_\mathrm{L} d | \vec k^\prime|_\mathrm{L}  (d \Omega^\prime_p)_\mathrm{c.m.}} = 
\frac{\Phi}{4} \sum_{h, h', s, s',\lambda'} |{\mathcal M}_{\rm{BH}} + {\mathcal M}_{\rm{VCS}}|^2, 
\nonumber \\
\label{eq:cross}
\end{align}
where the scattered lepton solid angle ($d \Omega^\prime_e$) and momentum value 
($| \vec k^\prime|$) are expressed in the Lab (L) frame, in which the proton target is at rest (${\vec p} =0$), and where the scattered proton solid angle 
($d \Omega^\prime_p$) is expressed in the $\gamma^\ast p$ c.m. frame. 
The phase space factor $\Phi$ in Eq.~(\ref{eq:cross}) is given by~\cite{Guichon:1995pu}
\bea
\Phi =\frac{1}{(2 \pi)^5} \frac{1}{64 M_N} \frac{s - M_N^2}{s} \frac{E^\prime_e}{E_e},
\label{eq:phasespace}
\eea
with $E_e$ ($E^\prime_e$) the beam (scattered) electron energy in the Lab frame.  
Furthermore, the unpolarized cross section of Eq.~(\ref{eq:cross}) involves a helicity average over the squared invariant amplitude of the $e^- p \to e^-  \gamma p$ reaction. The $e^- p \to e^-  \gamma p$ invariant amplitude consists, at order ${\cal O}(e^3)$, of two terms: ${\cal M}_{\rm{VCS}}$ describing the VCS subprocess (lower diagram in Fig.~\ref{fig:epepgamma}), given by Eq.~(\ref{eq:vcs}), and ${\cal M}_{\rm{BH}}$ describing the BH subprocess (top diagrams in Fig.~\ref{fig:epepgamma}). The latter is given by 
\begin{align}
&{\cal M}_{\rm{BH}} = \frac{i e^3}{t} \bar u(k',h') \left\{ 
\gamma^\mu \frac{\gamma \cdot (k - q') + m}{- 2 k \cdot q'} \gamma^\nu 
\right. \nonumber \\
&+\left.   \gamma^\nu \frac{\gamma \cdot (k' + q') + m}{ 2 k' \cdot q'} \gamma^\mu \right\}
 u(k, h) \, \varepsilon^\ast_\nu(q',\lambda') \nonumber \\
&\times   \bar N(p', s') \left\{ F_D(t) \gamma_\mu + F_P(t) \frac{i \sigma_{\mu \alpha} (q - q')^\alpha}{2 M_N} \right\} N(p,s), \nonumber \\   
\label{eq:bh}
\end{align}
with $m$ the lepton mass. On the proton side, the BH amplitude  
is expressed in terms of the Dirac ($F_D$) and Pauli ($F_P$) proton form factors with spacelike argument $t$, for which we 
use the parameterization from \cite{A1:2010nsl}. 

The pioneering work of \cite{Guichon:1995pu} proposed to study the $e^- p \to e^- \gamma p$ reaction at low c.m. energy $W$ as a tool to extract the nucleon GPs. Specifically, it proposed to make a low-energy expansion (LEX) of the $e^- p \to e^- \gamma p$ amplitude in the outgoing photon c.m. energy
\bea
q^\prime_{c.m.} = \frac{W^2 - M_N^2}{2 W}.
\eea
In such LEX, the helicity averaged squared amplitude for the $e^- p \to e^- \gamma p$ reaction, entering Eq.~(\ref{eq:cross}), takes the form
\begin{align}
\frac{1}{4}\sum_{h, h', s, s',\lambda'} |{\mathcal M}_{\rm{BH}} + {\mathcal M}_{\rm{VCS}}|^2 &= 
\frac{a_{-2}}{q^{\prime \, 2}_{c.m.}} +  \frac{a_{-1}}{q^{\prime}_{c.m.}} + a_0 
\nonumber \\ 
&+ {\mathcal O}(q^\prime_{c.m.}).
\label{eq:lex}
\end{align}
The first two coefficients in this expansion, $a_{-2}$ and $a_{-1}$, are fully determined through a low-energy theorem (LET) in terms of the BH + Born amplitude, for details see \cite{Guichon:1995pu}. The nucleon structure information, beyond the nucleon elastic form factors in the BH + Born amplitude, appears to lowest order in $q^{\prime}_{c.m.}$ in the coefficient $a_0$, which can be expanded in terms of dipole GPs. The unpolarized cross section contains three GP combinations, which can be accessed by measuring the photon angular dependence and varying the beam energy. An experimental extraction of all six independent dipole GPs requires to perform also double polarization experiments involving a longitudinally polarized electron beam and either polarizing the nucleon target or measuring the nucleon recoil polarization along three orthogonal directions~\cite{Vanderhaeghen:1997bx}, see the review~\cite{Guichon:1998xv} for details of the formalism and the review of \cite{Fonvieille:2019eyf} for an overview of the low-energy VCS experiments performed so far.

We show in Fig.~\ref{fig:vcs-w-beta_fromthreshold} the sensitivity of the $e^- p \to e^- \gamma p$ cross section to the GPs from threshold into the $\Delta(1232)$-region for the kinematics of the upcoming VCS-II experiment at JLab~\cite{VCS-II:2023jhu}. We show the results for two values of the azimuthal angle $\phi$ between the lepton and hadron planes: $\phi = 0^o$ corresponding with the photon emitted in the same half-plane determined by the lepton beam and the scattered lepton, and $\phi = 180^o$ in which the photon is emitted in the opposite half-plane. One notices that in the low-energy region, below the pion production threshold, the cross section is largely dominated by the BH+Born amplitude, in accordance with the LET, and the GPs make an effect of order 10\% or less below pion production threshold. In the low-energy region, the sensitivity to the GPs arises through interference with the BH + Born amplitude. To enhance the sensitivity to the GPs, it has been proposed in \cite{Pasquini:2001yy} to measure the VCS process in the $\Delta(1232)$ region, requiring a theoretical formalism beyond the LEX of Eq.~(\ref{eq:lex}). For this purpose an unsubtracted dispersive formalism was developed in \cite{Pasquini:2001yy}. This formalism is improved in the present work through the largely data-driven subtracted DRs, in view of the precision goals of the present and future JLab experiments. In the subtracted DR formalism the scalar GPs directly appear as subtraction constants and are to be determined from a fit to the VCS cross sections. In the $\Delta(1232)$ region, one sees that the BH + Born process is only responsible for around 10 - 20\% of the cross section, which is predominantly driven by the $\Delta(1232)$-resonance excitation. The sensitivity to the GPs in this region therefore arises mainly through the interference with the real part of the $\Delta(1232)$ resonance amplitude. As the real part of the $\Delta(1232)$ production amplitude goes through a zero at the resonance position, where the amplitude is purely imaginary, one thus expects the  largest sensitivity to the GPs on either side of the resonance peak position.  As illustration, we show in Fig.~\ref{fig:vcs-w-beta_fromthreshold} the sensitivity of the $e^- p \to e^- \gamma p$ cross section to a $\pm 1\sigma$ (standard deviation) variation in the magnetic dipole GP $\beta_{M1}(Q^2)$ around its value at $Q^2 = 0.33$~GeV$^2$ measured at JLab~\cite{Li:2022sqgnature}, for the fixed value of the electric dipole GP $\alpha_{E1}(Q^2)$ measured in the same experiment.

In Fig.~\ref{fig:vcs-nature} a direct comparison between our subtracted DR formalism and the existing VCS JLab data from \cite{Li:2022sqgnature} is shown as function of the photon c.m. angle for three $Q^2$ values and for different $W$ values through the $\Delta(1232)$ energy range. As the GPs $\alpha_{E1}(Q^2)$ and $\beta_{M1}(Q^2)$ at each value of $Q^2$ are subtraction constants in our formalism, we show as a first step the comparison using the central values extracted in \cite{Li:2022sqgnature}, which were obtained using the unsubtracted DR formalism of \cite{Pasquini:2001yy}. Furthermore, in Fig.~\ref{fig:vcs-nature}  we also show the sensitivity bands in $\alpha$ and $\beta$ by considering $1 \sigma$ variations around their central values. For the data at $Q^2 = 0.4$~GeV$^2$, which are mostly centered near the peak of the $\Delta(1232)$, one notices that the sensitivity to the GP is reduced as the amplitude is mostly imaginary around the resonance position. This kinematics thus provides a good test of the dispersive formalism. One notices that the angular dependence is  well reproduced. For the lower $Q^2$ values, $Q^2 = 0.28$~GeV$^2$ and $Q^2 = 0.33$~GeV$^2$, shown in Fig.~\ref{fig:vcs-nature}, the existing data are mostly at and above the $\Delta(1232)$ peak position. Overall the subtracted DR formalism follows also well the angular distributions and the decrease of the cross section when going from the $\Delta(1232)$ peak to higher $W$ values. Merely at the highest $W$ bin, and for $\phi = 180^o$, one sees a deviation beyond the $1 \sigma$ sensitivity bands for the GPs. Before drawing conclusions as whether the origin of this deviation is caused by the GP values extracted in \cite{Li:2022sqgnature}, a global fit to all world data is required. Such a fit of the scalar GPs to the world VCS data using the subtracted DR formalism is planned for a future work.

As the real part of the $\Delta(1232)$ production amplitude changes sign when crossing the resonance position, also the interference effect with the GPs on the cross section changes sign when crossing the resonance position. Therefore, a simultaneous measurement both below and above the $\Delta(1232)$ resonance position will provide strong consistency checks on the extraction of the GPs $\alpha_{E1}$ and $\beta_{M1}$. Such a measurement is planned in the near-future at JLab~\cite{VCS-II:2023jhu} and will complement the existing JLab data from \cite{Li:2022sqgnature}, shown in Fig.~\ref{fig:vcs-nature}, which were mostly taken around or above the $\Delta(1232)$ peak position.  In Figs.~\ref{fig:vcs-w-alpha} and \ref{fig:vcs-w-beta} we show the predictions of our subtracted DR formalism for the expected sensitivity to $\alpha_{E1}$ and $\beta_{M1}$ in the kinematics of the forthcoming VCS-II experiment~\cite{VCS-II:2023jhu}. 
One clearly notices the sign change in the interference effect with the GPs when crossing the resonance position. One also notices in Fig.~\ref{fig:vcs-w-alpha}, especially for $\theta = 130^o$ below the resonance position, regarding the sensitivity to the GP $\alpha_{E1}$ there is a  change in interference for azimuthal angles $\phi = 180^o$ and $\phi = 0^o$. For $\phi = 180^o$ our results predict an anti-correlation between the GPs $\alpha_{E1}$ and $\beta_{M1}$: increasing the value of $\alpha_{E1}$ or decreasing the value of $\beta_{M1}$ affect the cross section in a similar way. On the other hand for $\phi = 0^o$ our results predict a correlation between the GPs $\alpha_{E1}$ and $\beta_{M1}$: increasing the value of $\alpha_{E1}$ or increasing the value of $\beta_{M1}$ affect the cross section in a similar way. Thus, the forthcoming JLab measurement~\cite{VCS-II:2023jhu} of the photon angular dependence of the $e^- p \to e^- \gamma p$ reaction both below and above the $\Delta(1232)$ position will allow for a strong consistency check on the extraction of the GPs using the subtracted DR formalism.

\begin{figure*}[!]
	\centering
\includegraphics[width=0.75\textwidth]{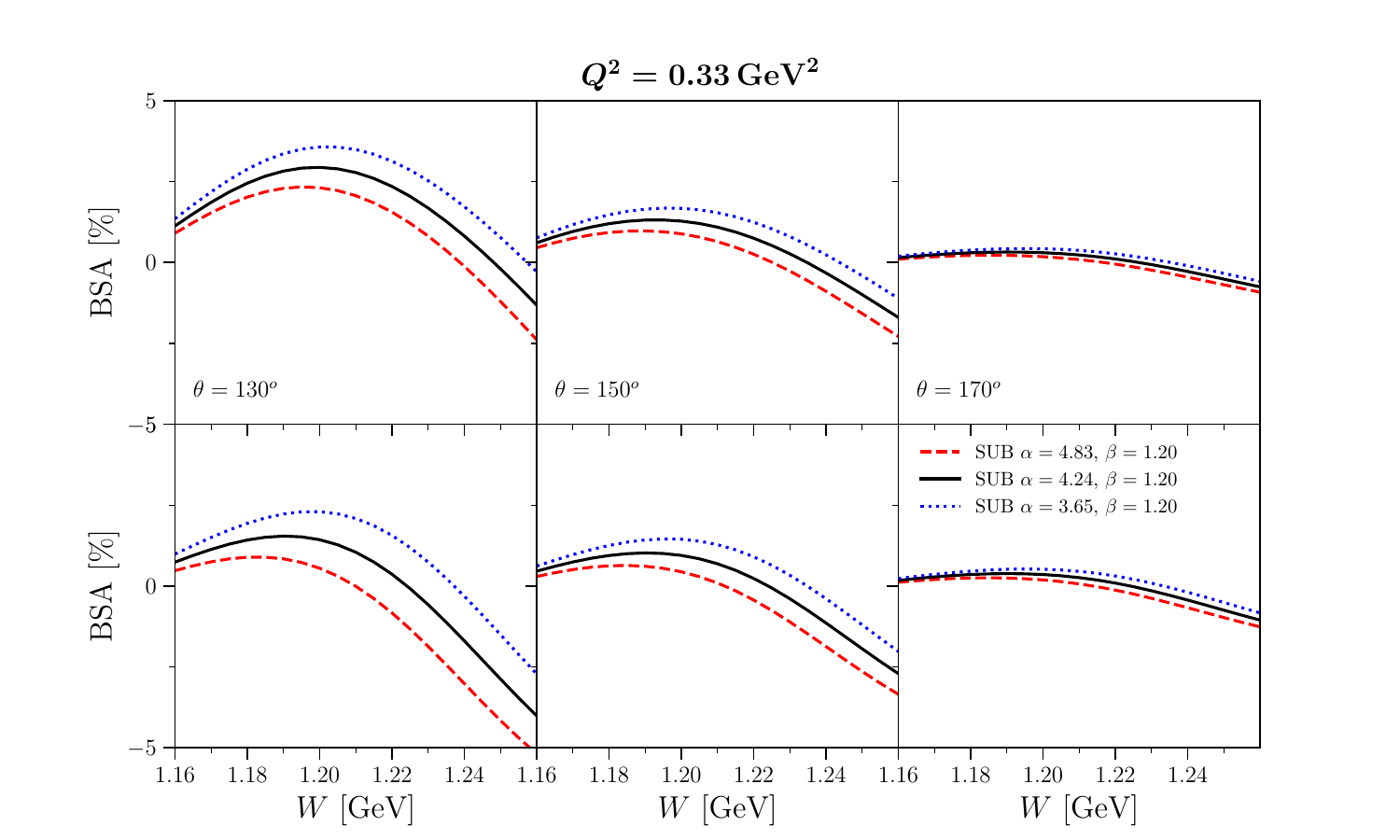}
	\caption{
   Beam spin asymmetry (BSA) for the $e^- p \to e^- \gamma p$ reaction as a function of the c.m. energy $W$ in the  kinematics of the planned VCS-IIIp experiment at JLab~\cite{VCS-BSA}:
electron beam energy $E_e = 2.2$ GeV, initial photon virtuality $Q^2 = 0.33 $ GeV$^2$, for several values of the photon c.m. scattering angle. Top and bottom panels correspond to azimuthal
angle $\phi=45^o$  and $\phi=90^o$, respectively.
 The curves show the sensitivity to the electric GP $\alpha_{E1}$ for fixed value of the magnetic  GP $\beta_{M1}$ (both in units $10^{-4}$ fm$^3$), as indicated
in the legend. 
 }
	\label{fig:bsa-w-alpha}
\end{figure*}

\begin{figure*}[!]
	\centering
\includegraphics[width=0.75\textwidth]{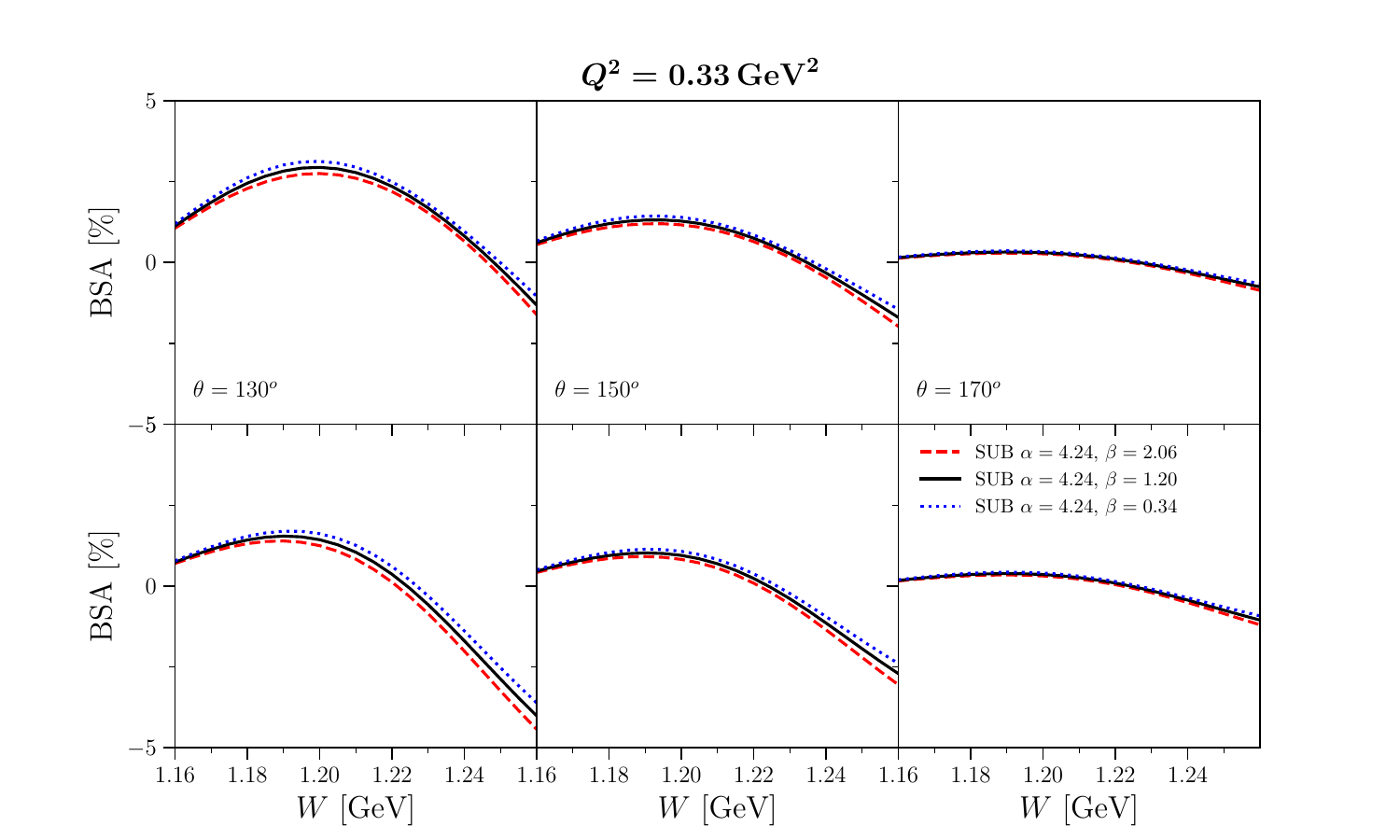}
	\caption{The same as Fig.~\ref{fig:bsa-w-alpha}, where 
 the curves show the sensitivity to the magnetic GP $\beta_{M1}$ for fixed value of the electric GP $\alpha_{E1}$ (both in units $10^{-4}$ fm$^3$), as indicated
in the legend. 
 }
	\label{fig:bsa-w-beta}
\end{figure*}

In order to test the imaginary part of the VCS DR formalism, 
a further planned JLab VCS-IIIp experiment~\cite{VCS-BSA} will measure the beam spin asymmetry (BSA) of the $e^- p \to e^- \gamma p$ reaction, which is the asymmetry for reversal of the electron beam helicity $h$, i.e., 
\begin{eqnarray}
\rm{BSA} = \frac{d^5 \sigma_{h = +1/2} - d^5 \sigma_{h = -1/2}}{d^5 \sigma_{h = +1/2} + d^5\sigma_{h = -1/2}},
\label{eq:bsa}
\end{eqnarray}
where $d^5 \sigma_h$ denotes the fivefold $e^- p \to e^- \gamma p$  differential cross section for an electron beam in a pure helicity state $h = \pm 1/2$. This observable is zero below the pion production threshold, as the $e^- p \to e^- p \gamma$ amplitude is purely real in that region. Above the pion production threshold, the BSA firstly tests the dispersive input as it is sensitive to the imaginary part of the VCS amplitude. Furthermore, the BSA numerator in Eq.~(\ref{eq:bsa}) has a $\sin \phi$ angular dependence. Its measurement thus requires out-of-plane kinematics. In Fig.~\ref{fig:bsa-w-alpha}, \ref{fig:bsa-w-beta}, we show our results for the BSA for $\phi = 45^o$ and $\phi = 90^o$. We see that the subtracted DR formalism predicts asymmetries in the range between -5\% to +3\% in the kinematics of the JLab VCS-IIIp experiment.  Fig.~\ref{fig:bsa-w-alpha} also illustrates that the BSA has a significant sensitivity to the GP $\alpha_{E1}$ especially below the $\Delta(1232)$ resonance position, while Fig.~\ref{fig:bsa-w-beta} shows that the BSA is nearly insensitive to the GP $\beta_{M1}$. This sensitivity to the GP $\alpha_{E1}$ in the BSA can be understood, as this observable results from the product of the imaginary part of the VCS amplitude for $\Delta(1232)$ production due to a transverse virtual photon (predominantly M1 excitation) with the real part of the VCS amplitude for a longitudinal virtual photon. The latter contains the dependence on $\alpha_{E1}$. The BSA thus provides a further strong cross-check on the extraction of $\alpha_{E1}$ from the unpolarized cross section.

In Figs.~\ref{fig:vcs-w-alpha}-\ref{fig:bsa-w-beta}, we show the sensitivity of the cross-section and beam-spin asymmetries to the scalar GPs. 
In the proposal of the upcoming JLab VCS-II experiment \cite{VCS-II:2023jhu}, it was estimated that the cross sections will be measured with a statistical uncertainty ranging from ±1\% to ±2\%, depending on the kinematics. The systematic uncertainties will be the dominant factor, ranging at $\approx$ ±4\%. 
A qualitative comparison of our results with these projected uncertainties shows that the differences between the curves due to varying the scalar GPs as shown in Figs.~\ref{fig:vcs-w-alpha}-\ref{fig:vcs-w-beta} are either broadly comparable or larger than the projected experimental uncertainties. For the BSA, the sensitivity to varying the GP $\alpha_{E1}$ as shown in Fig.~\ref{fig:bsa-w-alpha} is comparable to the  estimated experimental uncertainty, as reported in Ref.~\cite{VCS-BSA}, thus allowing an independent cross-check on the extraction of $\alpha_{E1}$.

\section{Conclusions and Outlook}

In this work\blue{,} we developed a once-subtracted dispersion-relation (DR) formalism for virtual Compton scattering (VCS) off the proton from threshold into the $\Delta(1232)$ region, with the aim of extracting the scalar generalized polarizabilities (GPs) with controlled input in view of the precision goals of the ongoing and forthcoming measurements at Jefferson Lab.

The central improvement over the previously used unsubtracted DR formalism is the replacement of the finite-energy sum-rule prescription and effective asymptotic parameterizations by a subtracted framework in which the relevant low-energy structure information enters through subtraction constants. In the present implementation, the VCS amplitudes for which the unsubtracted dispersion integrals are not convergent ($F_1^{\mathrm{NB}}$, $F_5^{\mathrm{NB}}$) or poorly convergent ($\widetilde F_2^{\mathrm{NB}}$) are reconstructed from fixed-$t$ once-subtracted dispersion integrals in $\nu$, while the subtraction functions are obtained from once-subtracted dispersion relations in $t$. This provides rapidly convergent dispersive integrals for both right-hand and left-hand cuts. The dominant $t$-channel dynamics are treated in a largely data-driven way through $\gamma^\ast\gamma \to \pi\pi$ and $\pi\pi \to N\bar N$ dispersive input, including the prominent $f_0(500)$ and $f_2(1270)$ regions, while the leading left-hand cut in the $\Delta(1232)$-energy region is estimated through $\Delta(1232)$ exchange with finite width.

We presented numerical results for the subtracted amplitudes and their decomposition into $s$- and $t$-channel contributions, highlighting the modified $t$-dependence relative to the unsubtracted formalism, in particular for $F_1^{\mathrm{NB}}$ where the earlier effective $f_0(500)$ pole approximation is replaced by an explicit dispersive evaluation. In the $s$-channel, the dispersive integrals entering the subtracted relations are largely saturated by the low-energy region up to and around the $\Delta(1232)$, which is essential for a quantitatively reliable description of the $W$ dependence in the resonance region.

Confronting the formalism with the latest VCS cross-section data in the $\Delta(1232)$ region, we find an overall good description of the angular distributions and of the observed decrease of the cross section with increasing $W$. A deviation appears only in the highest-$W$ bin for $\phi = 180^\circ$, where the data lie outside the sensitivity band obtained from the scalar GPs as extracted in the previous analysis. Establishing whether this indicates tension in the polarizability input or points to missing contributions requires a global analysis of the world data set within the present subtracted framework, which we plan to pursue in future work.

The subtracted DR formalism also yields clear expectations for near-future measurements designed to test both the real and imaginary parts of the VCS amplitude in the $\Delta$ region. Since the sensitivity to the scalar GPs $\alpha_{E1}(Q^2)$ and $\beta_{M1}(Q^2)$ is driven by interference with the real part of the $\Delta(1232)$ amplitude, the largest impact of the scalar GPs occurs on either side of the resonance peak, while it is reduced close to the peak where the amplitude is predominantly imaginary. For the planned kinematics of the VCS-II experiment at JLab, the predicted dependence on the azimuthal angle $\phi$ implies that measurements at $\phi = 0^\circ$ and $\phi = 180^\circ$, taken below and above the resonance, provide strong internal consistency checks and show different correlation patterns between $\alpha_{E1}$ and $\beta_{M1}$.

A complementary constraint will come from the polarization observables. In particular, the beam-spin asymmetry above pion threshold directly tests the dispersive input through its sensitivity to the imaginary part of the VCS amplitude. For the planned kinematics of the VCS-IIIp experiment at JLab we predict asymmetries at the few-percent level and a pronounced sensitivity to $\alpha_{E1}$ below the $\Delta(1232)$ peak position, while the dependence on $\beta_{M1}$ is much weaker, providing an independent cross-check on the extraction of $\alpha_{E1}$ from unpolarized cross sections.

Looking ahead, several extensions are natural. On the phenomenological side, a global fit of $\alpha_{E1}(Q^2)$ and $\beta_{M1}(Q^2)$ to the full set of VCS cross section data, including a consistent propagation of uncertainties from the pion electroproduction input and from the $t$-channel dispersive amplitudes, will be essential to fully exploit the precision of current and upcoming measurements. On the theory side, the treatment of the $t$-channel left-hand cut can be further refined by incorporating contributions beyond the leading $\Delta(1232)$ exchange, and the $t$-channel right-hand cut input can be improved by including the $f_0(980)$ in a coupled-channel framework. Finally, combining the present framework with dedicated double-polarization measurements offers a path toward a dispersive extraction of the spin GPs, thereby extending the reach of the subtracted DR approach beyond the scalar sector.

\section*{Acknowledgements}
This work was supported by the Deutsche Forschungsgemeinschaft (DFG, German Research Foundation), through the Collaborative Research Center "CRC1660: Hadrons and Nuclei as discovery tools" (Project ID 514321794).

\vspace{1cm}

\appendix

\section{$t$-channel helicity amplitudes for VCS}
\label{app:thelvcs}

\begin{figure}
	\centering
	 \includegraphics[width=0.45\textwidth]{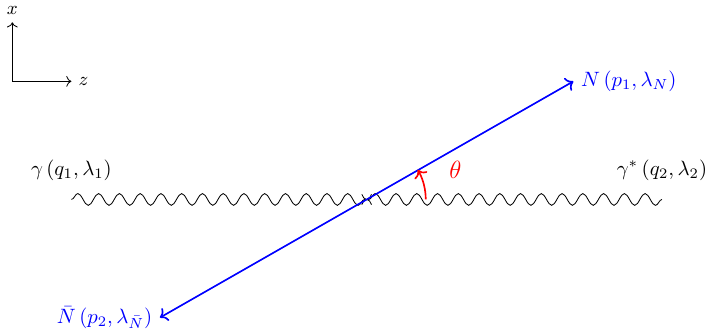}
	\caption{The kinematics of the $\gamma \gamma^\ast \to N \bar N$ process in the c.m. frame.}
	\label{fig:tchannel_kin}
\end{figure}

The $t$-channel helicity amplitudes for virtual Compton scattering, corresponding \blue{to} the $\gamma \gamma^\ast \to N \bar N$ process, can be written as
\begin{eqnarray}
\label{helicityamplt}
    T^{t}_{\lambda_1, \lambda_2, \lambda_{N}, \lambda_{\bar{N}}} &=& -e^2 \varepsilon_{2}^{\mu}(q_2,\lambda_2)\varepsilon_{1}^{\nu}(q_1,\lambda_1)
    \nonumber \\
    &\times& \Bar{u}(p_1,\lambda_N)\tilde{\mathcal{M}}_{\mu \nu}v(p_2,\lambda_{\bar{N}}),
\end{eqnarray}
where $\varepsilon_2$ and $\varepsilon_1$ are the polarization vectors of the virtual and real photon respectively, and $u$ and $v$ are the Dirac spinors of the nucleon and antinucleon in the final state. 
The tensor $\tilde{\mathcal{M}}_{\mu\nu}$ has the following definition:
\begin{equation}
\label{hadrtensorFs}
    \tilde{\mathcal{M}}_{\mu \nu}=\sum_{i=1}^{12} f_{i}(Q^2,\nu,t) \tilde{\rho}_{\mu \nu}^{\, i},
\end{equation}
where the gauge-invariant tensor basis $\tilde{\rho}_{\mu \nu}^{\,i}$ coincides with the basis introduced in Appendix A of \cite{Pasquini:2001yy}, but with the following crossing transformation (combined with a change of notation):
\begin{align}
    q \rightarrow q_{2}, \quad p \rightarrow -p_{2}, \quad q' \rightarrow -q_{1}, \quad p' \rightarrow p_{1}.
\end{align}
The $f_i$ are the same invariant amplitudes as in Eq.~\eqref{eq:vcsbasis}.
In this Appendix, we express the invariant amplitudes $F_{i}$ in terms of these $t$-channel helicity amplitudes.
As expected, the number of independent helicity amplitudes matches the number of twelve independent invariant amplitudes. We choose the following set of independent helicity amplitudes: the polarization of the antinucleon is fixed at $1/2$, while the polarizations of the other particles are varied, leading to the following independent amplitudes
\begin{align}
 \label{reducedhelampl}
    & T^{t}_{1,1,\frac{1}{2}, \frac{1}{2}}, \quad 
    T^{t}_{-1,-1,\frac{1}{2}, \frac{1}{2}},  \quad 
    T^{t}_{1,1,-\frac{1}{2} ,\frac{1}{2}},  \nonumber \\ 
    & T^{t}_{-1,-1,-\frac{1}{2}, \frac{1}{2}},  \quad 
    T^{t}_{1,0,\frac{1}{2}, \frac{1}{2}},  \quad 
    T^{t}_{-1,0,\frac{1}{2}, \frac{1}{2}}, \nonumber \\
    &T^{t}_{1,0,-\frac{1}{2} ,\frac{1}{2}},  \quad 
    T^{t}_{-1,0,-\frac{1}{2}, \frac{1}{2}},  \quad 
    T^{t}_{1,-1,\frac{1}{2}, \frac{1}{2}},  \nonumber \\ & T^{t}_{-1,1,\frac{1}{2}, \frac{1}{2}},  \quad 
    T^{t}_{1,-1,-\frac{1}{2}, \frac{1}{2}}, \quad 
    T^{t}_{-1,1,-\frac{1}{2}, \frac{1}{2}}.
 \end{align}
In order to find the relations with the invariant amplitudes $F_i$, as a first step, we implement the helicity amplitude for the $t$-channel of VCS of Eq.~\eqref{helicityamplt}. 
We work in the c.m. system of the $t$-channel process $\gamma \gamma^{*} \to N \bar{N}$ (see Fig.~\ref{fig:tchannel_kin} for the kinematics). In this frame, the four-momenta are defined as
\begin{align}
\label{fourmomenta1}
    &q_{1}^{\mu}=(q_{cm},0,0, q_{cm}), \\
    &q_{2}^{\mu}=(E_{q},0,0,-q_{cm}),\\
    &p_{1}^{\mu}=(E_{N},p_{N} \sin \theta_{t}, 0, p_{N} \cos \theta_{t}), \\
    &p_{2}^{\mu}=(E_{N},-p_{N} \sin \theta_{t}, 0,-p_{N} \cos \theta_{t}).\label{fourmomenta4}
\end{align}
In the following, for simplicity, $\theta_{t}$ will be denoted by $\theta$.
The energies and the magnitudes of the three-momenta in Eqs.~\eqref{fourmomenta1}-\eqref{fourmomenta4} can be expressed in terms of the Mandelstam variables as
\begin{align}
\label{whatisinsidemomenta}
    &q_{cm}=\frac{t+Q^2}{2 \sqrt{t}}, \, \, \, \, \,E_{q}=\frac{t-Q^2}{2\sqrt{t}}, \nonumber \\
    &E_{N}=\frac{\sqrt{t}}{2}, \, \, \,\quad \quad p_{N}=\sqrt{\frac{t}{4}-M_N^2}.
\end{align}
We use the Lorenz gauge for the photon polarization vectors. For the real photon they are given by
\begin{align}
\label{epsilonreal}
    &\varepsilon_{1}^{\mu}(q_{1},\lambda_1=1)=\frac{1}{\sqrt{2}}(0,-1,-i,0), \nonumber \\
    &\varepsilon_{1}^{\mu}(q_{1},\lambda_1=-1)=\frac{1}{\sqrt{2}}(0,1,-i,0),
\end{align}
whereas for the virtual photon, the polarization vectors read
\begin{align}
\label{epsilonvirtual}
    &\varepsilon_{2}^{\mu}(q_{2},\lambda_2=1)=\frac{1}{\sqrt{2}}(0,1,-i,0),\nonumber \\
    &\varepsilon_{2}^{\mu}(q_{2},\lambda_2=-1)=\frac{1}{\sqrt{2}}(0,-1,-i,0), \nonumber \\
    &\varepsilon_{2}^{\mu}(q_{2},\lambda_2=0)=\frac{1}{Q}(-q_{cm},0,0,E_{q}).
\end{align}
The spinor conventions for the nucleon  and the antinucleon in the final state characterized, respectively, by momenta $\vec{p}_1$ and $\vec{p}_2$ and by polarizations $\lambda_{N}$ and $\lambda_{\bar{N}}$ are
\begin{align}
\label{spinoru}
&u(\vec{p_1}, \lambda_N=1/2) = \sqrt{E_N + M_N} \begin{bmatrix} 
\chi_{\uparrow}(\theta) \\
\frac{p_N}{E_N + M_N} \, \chi_{\uparrow}(\theta)
\end{bmatrix}, \nonumber \\
&u(\vec{p_1}, \lambda_N=-1/2) = \sqrt{E_N + M_N} \begin{bmatrix} 
\chi_{\downarrow}(\theta) \\
-\frac{p_N}{E_N + M_N} \, \chi_{\downarrow}(\theta)
\end{bmatrix},
\end{align}
and 
\begin{align}
\label{spinorv}
&v(\vec{p}_{2}, \lambda_{\bar{N}}=1/2) = \sqrt{E_N + M_N} \begin{bmatrix} 
-\frac{p_N}{E_N + M_N} \, \chi'_{\downarrow}(\theta) \\
  \chi^{'}_{\downarrow}(\theta)
\end{bmatrix}, \nonumber \\
&v(\vec{p}_2, \lambda_{\bar{N}}=-1/2) = \sqrt{E_N + M_N} \begin{bmatrix} 
-\frac{p_N}{E_N + M_N} \, \chi'_{\uparrow}(\theta) \\
  -\chi'_{\uparrow}(\theta)
\end{bmatrix},
\end{align}
where
\begin{align}
\label{chiobjects}
&\chi_{\uparrow}(\theta) = \begin{bmatrix} \cos \theta/2 \\ \sin \theta/2 \end{bmatrix}, \quad
\chi_{\downarrow}(\theta) = \begin{bmatrix} -\sin \theta/2 \\ \cos \theta/2 \end{bmatrix}, \nonumber \\
&\chi'_{\uparrow}(\theta) = \begin{bmatrix} \sin \theta/2 \\ -\cos \theta/2 \end{bmatrix}, \quad
\chi'_{\downarrow}(\theta) = \begin{bmatrix} \cos \theta/2 \\ \sin \theta/2 \end{bmatrix}.
\end{align}
The relations between the independent $t$-channel helicity amplitudes and VCS invariant amplitudes can be written in matrix form as
\begin{equation}
\label{HAtoF}
    \vec{T^{t}}=\Tilde{A_{1}}\vec{F},
\end{equation}
where $\Tilde{A_{1}}$ is a $12 \times 12$ matrix, $\vec{F}$ is a vector whose components are the VCS invariant amplitudes $F_{i}$, and $\vec{T^{t}}$ is the vector containing the independent $t$-channel helicity  amplitudes listed in Eqs.~\eqref{reducedhelampl}. 
By inverting the matrix  $\Tilde{A_{1}}$, we can express the VCS invariant amplitudes in terms of the $t$-channel helicity amplitudes. 

Furthermore, it is useful to introduce reduced VCS $t$-channel helicity amplitudes $\tau_i$, $i=1,\dots,12$, defined by the following relations:
\begin{align}
T^t_{1,1,\frac{1}{2},\frac{1}{2}}&=\tau_{1}^t\nonumber\\
T^t_{-1,-1,\frac{1}{2},\frac{1}{2}}&=\tau_{2}^t\nonumber\\
T^{t}_{1,1,-\frac{1}{2},\frac{1}{2}}&=\sin(\tfrac{\theta}{2})\cos(\tfrac{\theta}{2})\tau_{3}^t\nonumber\\
T^t_{-1,-1,-\frac{1}{2},\frac{1}{2}}&=\sin(\tfrac{\theta}{2})\cos(\tfrac{\theta}{2})\tau_4^t\nonumber\\
T^t_{1,-1,\frac{1}{2},\frac{1}{2}}&=\sin^2(\tfrac{\theta}{2})\cos^2(\tfrac{\theta}{2})\tau_5^t\nonumber\\
T^t_{-1,1,\frac{1}{2},\frac{1}{2}}&=\sin^2(\tfrac{\theta}{2})\cos^2(\tfrac{\theta}{2})\tau_6^t\nonumber\\
T^t_{1,-1,-\frac{1}{2},\frac{1}{2}}&=\sin^3(\tfrac{\theta}{2})\cos(\tfrac{\theta}{2})\tau_7^t\nonumber\\
T^t_{-1,1,-\frac{1}{2},\frac{1}{2}}&=\sin(\tfrac{\theta}{2})\cos^3(\tfrac{\theta}{2})\tau_8^t\nonumber\\
T^t_{1,0,\frac{1}{2},\frac{1}{2}}&=\sin(\tfrac{\theta}{2})\cos(\tfrac{\theta}{2})\tau_{9}^t\nonumber\\
T^t_{-1,0,\frac{1}{2},\frac{1}{2}}&=\sin(\tfrac{\theta}{2})\cos(\tfrac{\theta}{2})\tau_{10}^t\nonumber\\
T^t_{1,0,-\frac{1}{2},\frac{1}{2}}&=\sin^2(\tfrac{\theta}{2})\tau_{11}^t\nonumber\\
T^t_{-1,0,-\frac{1}{2},\frac{1}{2}}&=\cos^2(\tfrac{\theta}{2})\tau_{12}^t.\label{tau1}
\end{align} 

In terms of the $t$-channel reduced helicity amplitudes $\tau_{i}^{t}$ of Eq.~\eqref{tau1}, the VCS invariant amplitudes $F_{i}(Q^2, \nu, t)$ read
\begin{widetext}

\begin{align}
\label{F1tau}
    -e^2F_1=& 
    \frac{1}{(t-4 M_N^2)^{1/2} \left(Q^2+t\right)} \biggl[ 
- \left(\tau_1^t+\tau_2^t\right)
- \frac{1}{4} \left(\tau_5^t+\tau_6^t\right) \biggr]  \\ \nonumber
&+\frac{4 \nu ^2 Q^2 M_N}{\left(Q^2+t\right)^3 \left(t-4
   M_N^2\right)^{3/2}} \biggl[ M_N \left(\tau_5^t +\tau_6^t\right)
   +\frac{\sqrt{t}}{2}  \left(\tau_7^t -\tau_8^t\right)\biggr] \\ \nonumber
   &-\frac{\nu}{(Q^2+t)^2 (t-4M_N^2)} \biggl[ t\left(\tau_3^t +\tau_4^t\right) +\frac{t}{2}(\tau_7^{t} +\tau_8^{t})+2\sqrt{2}Q M_N \left(\tau_9^t -\tau_{10}^t\right) +\sqrt{2}Q\sqrt{t}\left(\tau_{11}^t +\tau_{12}^t\right) \biggr],\\
   -e^2 F_{2} &= \frac{1}{(t-4 M_N^2)^{3/2} \left(Q^2+t\right)} \frac{1}{2} \biggl\{
-  \left(\tau_5^t+\tau_6^t\right)
    -\frac{\sqrt{t} }{2 M_N } \left(\tau_7^t-\tau_8^t \right) \biggr\},\\
    -e^2 F_{3} &= \frac{2 M_N}{(t-4 M_N^2)^{3/2} \left(Q^2+t\right)^2} \biggl\{
    \left(\tau_5^t+\tau_6^t \right) 
    %\nonumber \\ 
    %& 
    + \frac{\sqrt{t}}{2 M_N}  \left(\tau_7^t-\tau_8^t\right)  \biggr\} \\ \nonumber
   & - \frac{1}{\sqrt{2}\left(t-4M_N^2\right)\left(Q^2+t\right)Q \nu} \left[ \left(
    \tau_9^t -\tau_{10}^t \right)
    %\nonumber\\
    %&
    + \frac{\sqrt{t}}{2 M_N} \left(
    \tau_{11}^t+ \tau_{12}^t  \right) \right],
   \\ -e^2 F_{4} = & -\frac{1}{4 \nu  \left(t-4 M_N^2\right) \left(Q^2+t\right)}\left( \tau_7^t+\tau_8^t \right),\\
   -e^2 F_{5} =  &\frac{1}{ \left(Q^2+t\right)} \frac{1}{M_N \sqrt{t}} \biggl\{ 
- ( \tau_1^t-\tau_2^t) 
+ \frac{M_N^2 \left(Q^4+4 \nu ^2 Q^2+2 Q^2 t+t^2\right)}{\left(t-4M_N^2\right) \left(Q^2+t\right)^2}  \left(\tau_5^t-\tau_6^t\right) 
\nonumber\\
&- \frac{2 \sqrt{2}M_N Q \nu}{(t-4 M_N^2)^{1/2}(Q^2+t) } (\tau_9^t+\tau_{10}^t  ) -\frac{2 \nu ^2 M_N Q^2 }{(t-4 M_N^2)^{1/2}\left(Q^2+t\right)^2} \left( \tau_7^t - \tau_8^t \right)
\nonumber \\
& +\frac{1}{\left(t-4 M_N^2\right)^{1/2}} \biggl[ 
-M_N \biggl[ \left( \tau_3^t - \tau_4^t \right) +\frac{1}{2} \left( \tau_7^t - \tau_8^t \right) -\frac{\sqrt{2 t}}{Q} \left(\tau_{11}^t - \tau_{12}^t  \right) \biggr] \\ \nonumber
&+ \frac{\nu \left(t-4 M_N^2\right)^{1/2}}{\left(Q^2+t\right)} \left[  \sqrt{t}\left( \tau_3^t + \tau_4^t \right) -\sqrt{t} \frac{8M_N^{2}-t}{2(t-4M_N^2)}\left( \tau_7^t + \tau_8^t \right) +\sqrt{2}Q\left(\tau_{11}^t + \tau_{12}^t  \right)  \right] \biggr]  \biggr\},
\\
-e^2 F_{6} = &
\frac{1}{8 M_N \sqrt{t} (t-4 M_N^2)^{1/2} \left(Q^2+t\right)^3} 
\biggl\{ \biggl[ \left( (Q^2+t)^2 + Q^4 \right) \left( \tau_7^t - \tau_8^t \right) 
- \frac{4 M_N Q^2 (Q^2 + t)}{(t - 4 M_N^2)^{1/2}} \left(\tau_5^t-\tau_6^t
\right) \biggr] 
\nonumber \\
&- \frac{Q^2 (t - 4 M_N^2)^{1/2} (Q^2+t)}{4M_N \nu}\left[2 \sqrt{t} \left(\tau_3^t+\tau_4^t\right) 
+ \sqrt{t} \left( \tau_7^t + \tau_8^t \right) + 2 \sqrt{2} Q \left(\tau_{11}^t+\tau^t_{12}\right) \right]
\nonumber
\\
&
+\frac{\sqrt{2} Q \left(Q^2+t\right)^2}{\nu} 
\left(\tau_9^t+\tau_{10}^t\right)
 \biggr\},
\\
-e^2 F_{7} = &
+\frac{ \sqrt{t} }
{(t-4 M_N^2)^{1/2} \left(Q^2+t\right)^2}
\left(
\tau_3^t-\tau_4^t\right) 
%\nonumber \\
%&
+\frac{M_N }
{\sqrt{t} \left(t-4 M_N^2\right) \left(Q^2+t\right)}
\left(\tau_5^t-\tau_6^t\right)  
\nonumber \\  
&
-\frac{Q^2 }{ 2\sqrt{t} (t-4 M_N^2)^{1/2} \left(Q^2+t\right)^2}\left(\tau_7^t-\tau_8^t\right) -\frac{2M_N Q^2 \nu}{\left(t-4 M_N^2\right) \left(Q^2+t\right)^3}\left(\tau_7^t+\tau_8^t\right)
\nonumber\\
&
+\frac{\sqrt{2} Q}
{(t-4 M_N^2)^{1/2} \left(Q^2+t\right)^2}
\left(\tau_{11}^t-\tau_{12}^t \right),
\\
-e^2 F_{8} = &
-\frac{4 M_N }
{\sqrt{t} \left(t-4 M_N^2\right) \left(Q^2+t\right)^2}
\left(
\tau_5^t-\tau_6^t \right) 
%\nonumber \\ 
%&
+\frac{2 }
{\sqrt{t} (t-4 M_N^2)^{1/2} \left(Q^2+t\right)^2}(\tau_7^t-\tau_8^t) 
\nonumber \\ 
&
+\frac{1 }
{\sqrt{2}\sqrt{t} M_N Q \left(Q^2+t\right) \nu} \biggl[   \frac{2M_N}{(t-4 M_N^2)^{1/2}}\left(\tau_9^t+\tau_{10}^t\right)- \left( \tau_{11}^t+\tau_{12}^t\right)\biggr],
 \\
 -e^2 F_{9} = & 
%\nonumber \\ 
%&
- \frac{ \left(t-Q^2\right)}
{2\sqrt{t} \left(t-4 M_N^2\right) \left(Q^2+t\right)^2}
\left(\tau_5^t-\tau_6^t
 \right)- \frac{Q^4}
{4 M_N \sqrt{t} (t-4 M_N^2)^{1/2} \left(Q^2+t\right)^3} \left( \tau_7^t-\tau_8^t\right)
 \nonumber \\ 
 &
+ \frac{Q^2}{8  M_N^2 \left(Q^2+t\right)^2 \nu} \biggl[\left(\tau_3^t+\tau_4^t \right) +\frac{1}{2} \left( \tau_7^t+\tau_8^t\right)  +\frac{Q\sqrt{2}}{\sqrt{t}}\left(\tau_{11}^t+\tau_{12}^t \right) \nonumber\\
&-\frac{\sqrt{2}(Q^2+t)2M_N}{Q\sqrt{t}(t-4 M_N^2)^{1/2}}\left(\tau_9^t+\tau_{10}^t
\right) \biggr],\\
-e^2 F_{10} = & -\frac{1}
{8 M_N\left(Q^2+t\right) \nu} \biggl[\left(\tau_3^t+\tau_4^t \right) +\frac{1}{2}\left(\tau_7^t+\tau_8^t \right)\biggr] +\frac{\sqrt{t}}
{4(t-4 M_N^2)^{1/2} \left(Q^2+t\right)^2}  \left(\tau_7^t-\tau_8^t \right),
\\
-e^2 F_{11} = & +\frac{1}{4 M_N \sqrt{t} \left(Q^2+t\right)}\left( \tau_1^t-\tau_2^t\right) -\frac{\sqrt{t}}{8 (t-4 M_N^2)^{1/2} \left(Q^2+t\right)^2 } \left(\tau_7^t -\tau_8^t \right)  
%\nonumber\\
%& 
\nonumber \\
& 
+\frac{ Q^2 }{4 \sqrt{t} (t-4 M_N^2)^{1/2}\left(Q^2+t\right)^2} \left(\tau_3^t -\tau_4^t \right)-\frac{Q }{2 \sqrt{2} (t-4 M_N^2)^{1/2} \left(Q^2+t\right)^2} \left( \tau_{11}^t-\tau_{12}^t  \right)
%\nonumber \\
%& 
\nonumber\\
&  
+\frac{Q^2 \nu^2}{2\sqrt{t}\left(t-4M_N^2\right) \left(Q^2+t\right)^3} \biggl[ (t-4 M_N^2)^{1/2} \left(\tau_7^t -\tau_8^t \right) -2M_N \left(
\tau_5^t-\tau_6^t \right)  \biggr] 
\nonumber\\
&
-\frac{\nu}{8 M_N (t-4 M_N^2) \left(Q^2+t\right)^3} \biggl[ \left( t \left(Q^2+t\right)-4 M_N^2 Q^2\right) \left(\tau_7^t +\tau_8^t \right) -\frac{2\sqrt{2}Q (4M_N^2-t)\left(Q^2+t\right)}{\sqrt{t}}  \left( \tau_{11}^t+\tau_{12}^t  \right) \biggr]
%\nonumber \\
%& 
\nonumber \\
&
-\frac{\nu}{4 M_N (t-4 M_N^2)^{1/2} \left(Q^2+t\right)^2 }\biggl[(t-4 M_N^2)^{1/2} \left(\tau_3^t +\tau_4^t \right) -\frac{2 \sqrt{2} Q M_N}{\sqrt{t}} \left(  \tau_9^t+\tau_{10}^t  \right) \biggr],
\\
-e^2 F_{12} = &   
 %\nonumber\\
%&
+\frac{1}{2\sqrt{t} \left(t-4 M_N^2\right) \left(Q^2+t\right)} 
\left(\tau_5^t -\tau_6^t\right) -\frac{ Q^2 }{4M_N \sqrt{t} (t-4 M_N^2)^{1/2} \left(Q^2+t\right)^2}  \left(\tau_7^t - \tau_8^t \right)
\nonumber 
\\ \nonumber \label{F12tau}
&+\frac{1}{8 M_N^2 \left(Q^2+t\right) \nu} \biggl[ \left( \tau_3^t+\tau_4^t  \right) +\frac{1}{2}\left(\tau_7^t +\tau_8^t \right)+\frac{\sqrt{2}Q}{\sqrt{t}}\left( \tau_{11}^t+\tau_{12}^t\right) \biggr]   
%\nonumber\\
%&
\nonumber  \\
& -\frac{1}{2 \sqrt{2} \sqrt{t} M_N Q (t-4 M_N^2)^{1/2} \nu} \left( \tau_9^t+\tau_{10}^t \right).
\end{align}
\end{widetext}
As a consistency check, the RCS limit was considered. 
In this limit, the expressions for the invariant amplitudes given in
Eqs.~\eqref{F1tau}–\eqref{F12tau} were found to be consistent with the known relations between
the L’vov amplitudes $A_i$ and the $t$-channel helicity amplitudes for RCS (Eqs.~(B3) of Ref.~\cite{Drechsel:1999rf}).

\section{VCS $t$-channel discontinuities from the $\pi\pi$ intermediate state}
\label{app:pipi}

In the subtracted dispersion relation of Eq.~\eqref{eq:subtraction}, the $t$-channel integration is taken along the line $\nu=0$. Consequently, we need the $t$-channel discontinuities ${\rm Im}_{t} F_i(Q^2,\nu=0,t)$.
We evaluate the $\pi\pi$ contribution by inserting the $\pi\pi$ intermediate state into the $t$-channel unitarity relation and expressing the result in terms of helicity partial waves for $\gamma\gamma^\ast\to \pi\pi$ and $\pi\pi\to N\bar N$.

We first write the partial-wave expansion of the $t$-channel VCS helicity amplitudes
$\gamma(q_1,\lambda_1)\,\gamma^\ast(q_2,\lambda_2)\to N(\lambda_N)\,\bar N(\lambda_{\bar N})$ as

\begin{align}
T^t_{\lambda_1\lambda_2\lambda_N\lambda_{\bar{N}}}=\sum_{J}\frac{(2J+1)}{2}\, 
    T^{J}_{\lambda_1\lambda_2\lambda_N\lambda_{\Bar{N}}}(t, Q^2)\,      d^{J}_{\Lambda_{N} \Lambda_{\gamma}}(\theta),
    \label{pw}
\end{align}    
where $d^{J}_{m m'}(\theta)$ are Wigner $d$-functions, and the helicity differences $\Lambda_\gamma$ and $\Lambda_N$ are defined as
\begin{equation}
    \Lambda_\gamma\equiv \lambda_1-\lambda_2,\qquad \Lambda_N\equiv \lambda_N-\lambda_{\bar N}.
\end{equation}
The helicity amplitude for $\gamma\gamma^\ast\to\pi\pi$ satisfies the partial-wave expansion \cite{Danilkin:2018qfn,Danilkin:2019opj}
\begin{equation}
    T^{\gamma \gamma^{*} \to \pi \pi}_{\Lambda_{\gamma}}=\sum_{J\,even} (2J+1)\,h^{J}_{\lambda_1 \lambda_2} (t, Q^2)\, d^{J}_{\Lambda_{\gamma} 0} (\theta_{\pi \pi}),
\end{equation}
with $\theta_{\pi\pi}$ the c.m. angle of the $\gamma\gamma^\ast\to\pi\pi$ process. Finally, for $\pi\pi\to N\bar N$ we similarly write
\begin{equation}
    T^{\pi\pi\to N\bar N}_{\lambda_N\lambda_{\bar N}}
=\sum_{J}\frac{(2J+1)}{2}\,
T^{J}_{\lambda_N\lambda_{\bar N}}(t)\,d^{J}_{0,\Lambda_N}(\Theta),
\end{equation}
where $\Theta$ denotes the c.m. angle of the $\pi\pi\to N\bar N$ reaction. Isospin conservation and Bose symmetry restrict the $\pi\pi$ intermediate state to $I=0$ and even
$J$. The $\pi\pi$ contribution to the $t$-channel discontinuity is obtained from the unitarity relation, which for
fixed $J$ takes the form
\begin{align}
2\,\textup{Im}\,T^{J}_{\lambda_1\lambda_2,\lambda_N\lambda_{\bar N}}(t,Q^2)
&= S\,\frac{\beta_\pi(t)}{8\pi}\;
h^{J}_{\lambda_1\lambda_2}(t,Q^2) 
\nonumber \\
&\times \left(T^{J}_{\lambda_N\lambda_{\bar N}}(t)\right)^\ast,
\label{pweunit}
\end{align}
where $\beta_\pi(t)$ is defined in Eq.~(\ref{eq:betapipi}). For the isospin-$0$ $\pi\pi$ intermediate state considered here, one has the symmetry factor $S=1/2$. Parity invariance implies that
\begin{align}
    h^{J}_{\lambda_{1}\lambda_{2}}(t,Q^2)
   &= h^{J}_{-\lambda_{1},-\lambda_{2}}(t,Q^2),\nonumber\\
T^{J}_{-\lambda_{N},-\lambda_{\bar N}}(t)
   &= -\,T^{J}_{\lambda_{N},\lambda_{\bar N}}(t),
\end{align}
so that the independent amplitudes may be labeled by $\Lambda_{\gamma} = 0,1,2$ and $\Lambda_{N} = 0,1$. We therefore introduce the shorthand notations 
\begin{equation}
    h^{J}_{\Lambda_{\gamma}}(t,Q^2),\quad 
T^{J}_{\Lambda_{N}}(t)\,,
\end{equation}
with $T_{\Lambda_N=0}^J \equiv T_{\frac{1}{2}\frac{1}{2}}^J$. Using Eq.~\eqref{pw} together with Eq.~\eqref{pweunit}, and truncating the $\pi\pi$ system to $J=0,2$,
one obtains the following relations for the imaginary part of the reduced $t$-channel helicity amplitudes $\tau_i^t$ (defined in
Appendix~\ref{app:thelvcs}):
\begin{align}
{\rm Im} \tau_1^t &= {\rm Im} \tau_2^t 
=   \frac{\beta_{\pi}(t)}{64 \pi}
\biggl[ h^{0}_{0} (t, Q^2)\, \left(T^{0}_{0} (t) \right)^{*}  \nonumber\\ 
& \hspace{1cm} +\frac{5}{4}(1+3\cos 2\theta)\,
h^{2}_{0}(t,Q^2)\, \left( T^{2}_{0} (t)\right)^{*} \biggr], \nonumber\\
{\rm Im} \tau_3^t &= {\rm Im} \tau_4^t 
=  -\frac{5\sqrt{6}\,\beta_{\pi}(t)}{64 \pi}
\cos \theta \, 
h^{2}_{0}(t,Q^2)\, \left( T^{2}_{1} (t)\right)^{*},\nonumber\\
{\rm Im}\tau_5^t &= {\rm Im}\tau_6^t 
=   \frac{5\sqrt{6}\, \beta_{\pi}(t)}{64 \pi}
h^{2}_{2}(t,Q^2)\, \left( T^{2}_{0} (t)\right)^{*},\nonumber\\
{\rm Im}\tau_7^t &= -{\rm Im}\tau_8^t  
=   -\frac{5\, \beta_{\pi}(t)}{32 \pi}\,
h^{2}_{2}(t,Q^2)\, \left( T^{2}_{1} (t)\right)^{*},\nonumber\\
{\rm Im}\tau_9^t &= -{\rm Im}\tau_{10}^t =  \frac{5\sqrt{6}\,\beta_{\pi}(t)}{64 \pi}\,
\cos\theta \, 
h^{2}_{1}(t,Q^2)\, \left( T^{2}_{0} (t)\right)^{*},\nonumber\\
{\rm Im}\tau_{11}^t &=    -\frac{5\,\beta_{\pi}(t)}{64 \pi}\,
(1+2 \cos \theta)\,
h^{2}_{1}(t,Q^2)\, \left( T^{2}_{1} (t)\right)^{*},\nonumber\\
{\rm Im} \tau_{12}^t &=    -\frac{5\,\beta_{\pi}(t)}{64 \pi}\,
(-1+2\cos \theta)\,
h^{2}_{1}(t,Q^2) \left( T^{2}_{1} (t)\right)^{*}\,.
\end{align}
For the reaction $\pi\pi\to N\bar N$ we express the partial waves $T^J_{\Lambda_N}(t)$ through the Frazer-Fulco amplitudes $f_\pm^J(t)$~\cite{FrazerFulco} as
\begin{align}
T^{J}_{0} (t) &= \frac{16 \pi}{p_N} \left( p_N p_\pi \right)^J \,f_+^J (t), \nonumber\\
T^{J}_{1} (t) &= \frac{8 \pi \sqrt{t}}{p_N} \left( p_N p_\pi \right)^J \, f_-^J (t),
\end{align}
with the c.m. momenta
\begin{align}
    p_\pi = \frac{\sqrt{t}}{2} \beta_\pi, & \quad p_N =\sqrt{\frac{t}{4}-M_N^2}\,.
\end{align}
Combining the above relations with the definitions of the invariant amplitudes in Appendix~\ref{app:thelvcs}, we obtain
\begin{widetext}
\begin{align}
    -e^2 {\rm Im}_t F_1 (Q^2,0,t)&=
     \frac{\beta_\pi}{(t+Q^2)} 
    \left\{
    \frac{1}{(4 M_N^2- t)} h_{0}^{0}(t,Q^2)  f_{+}^{0\,*}(t)  
    +\frac{5}{32} t \beta_\pi^2 \left[ h_{0}^{2}(t,Q^2) -\sqrt{\frac{3}{2}}h_{2}^{2}(t,Q^2)\right] 
    f_{+}^{2\, *}(t) \right\},
    \nonumber \\
    -e^2 {\rm Im}_t F_2 (Q^2,0,t)&=
    \frac{t \beta_\pi^3}{(t+Q^2)(4 M_N^2 - t)} 
    \frac{5}{16} 
     h_{2}^{2}(t,Q^2) \left[ \sqrt{\frac{3}{2}} f_{+}^{2\,*}(t)  -\frac{t}{4M_N}
     f_{-}^{2\, *}(t) \right],\nonumber \\
    -e^2 {\rm Im}_t F_3 (Q^2,0,t)&=-
    \frac{M_N t \beta_\pi^3}{(t+Q^2)^2(4 M_N^2 - t)} 
    \frac{5}{4} 
    \left[h_{2}^{2}(t,Q^2) -  \sqrt{\frac{2t}{Q^2}}\,h_{1}^{2}(t,Q^2)
     \right]
     \left[ \sqrt{\frac{3}{2}}f_{+}^{2\,*}(t)  -\frac{t}{4M_N}
     f_{-}^{2\, *}(t) \right],
     \nonumber \\
    -e^2 {\rm Im}_t F_4 (Q^2,0,t)&=0, 
    \nonumber \\
    -e^2 {\rm Im}_t F_5 (Q^2,0,t)&=
 \frac{t \beta_\pi^3}{(t+Q^2)} 
    \frac{5}{32}
    \left[ h_{2}^{2}(t,Q^2) -  \sqrt{\frac{2t}{Q^2}} \,h_{1}^{2}(t,Q^2)
     \right]
     f_{-}^{2\, *}(t),
     \nonumber \\   
     -e^2 {\rm Im}_t F_6 (Q^2,0,t)&=
    \frac{ t \beta_\pi^3} {M_N (t+Q^2)^3} 
    \frac{5}{64} 
    \left[ \sqrt{\frac{3}{2}} Q^2 t \, h_{0}^{2}(t,Q^2) + \sqrt{2 Q^2 t} \,Q^2 \,h_{1}^{2}(t,Q^2)-
    \frac{Q^4+(t+Q^2)^2}{2}h_{2}^{2}(t,Q^2)
     \right]
     f_{-}^{2\, *}(t),
     \nonumber \\
      -e^2 {\rm Im}_t F_7 (Q^2,0,t)&=
    \frac{Q^2 \,  t \beta_\pi^3}{(t+Q^2)^2} 
    \frac{5}{32}
    \left[h_{2}^{2}(t,Q^2) -  \sqrt{\frac{2 t}{Q^2}}\,h_{1}^{2}(t,Q^2)
     \right]
     f_{-}^{2\, *}(t),
     \nonumber \\   
      -e^2 {\rm Im}_t F_8 (Q^2,0,t)&=-
 \frac{t \beta_\pi^3}{(t+Q^2)^2} 
    \frac{5}{8}
    \left[ h_{2}^{2}(t,Q^2) -  \sqrt{\frac{2t}{Q^2}}\,h_{1}^{2}(t,Q^2)
     \right]
     f_{-}^{2\, *}(t),
     \nonumber \\     
     -e^2 {\rm Im}_t F_9 (Q^2,0,t)&=-
    \frac{Q^2 t \beta_\pi^3 }{M_N (t+Q^2)^3} 
    \frac{5}{32} 
    \left[ \sqrt{\frac{3}{2}} t \, h_{0}^{2}(t,Q^2) +  \sqrt{2 Q^2 t}\,h_{1}^{2}(t,Q^2)-
    \frac{Q^2}{2}h_{2}^{2}(t,Q^2)
     \right]
     f_{-}^{2\, *}(t),
     \nonumber \\
       -e^2 {\rm Im}_t F_{10} (Q^2,0,t)&=
  \frac{t^2 \beta_\pi^3 }{(t+Q^2)^2} 
    \frac{5}{32}
    \left[ \sqrt{\frac{3}{2}} h_{0}^{2}(t,Q^2) -
    \frac{1}{2}h_{2}^{2}(t,Q^2)
     \right]
     f_{-}^{2\, *}(t),
    \nonumber \\
     -e^2 {\rm Im}_t F_{11} (Q^2,0,t)&=
    \frac{t^2 \beta_\pi^3}{(t+Q^2)^2} 
    \frac{5}{128} 
    \left[h_{2}^{2}(t,Q^2)+   \sqrt{\frac{2 Q^2}{t}}\,h_{1}^{2}(t,Q^2)
     \right]
     f_{-}^{2\, *}(t),
   \nonumber \\
       -e^2 {\rm Im}_t F_{12} (Q^2,0,t)&=-
     \frac{t \beta_\pi^3}{M_N(t+Q^2)^2} 
    \frac{5}{32}
    \left[ \sqrt{\frac{3}{2}} t \, h_{0}^{2}(t,Q^2) +  \sqrt{2 Q^2 t} \,h_{1}^{2}(t,Q^2)-
    \frac{Q^2}{2}h_{2}^{2}(t,Q^2)
     \right]
     f_{-}^{2\, *}(t)\,.\label{eq:imtpipiF1-12}
\end{align}
\end{widetext}
Note that the Born-subtracted $\gamma\gamma^\ast\to\pi\pi$ partial wave helicity amplitudes, $\bar h^J_{\Lambda_\gamma} \equiv h^J_{\Lambda_\gamma} - h^{J,\,\mathrm{Born}}_{\Lambda_\gamma}$, are subject to kinematic constraints at the soft-photon point and at two-pion threshold~\cite{Danilkin:2018qfn}. For the lowest partial waves relevant here one finds
\begin{align}\label{pwcontraints}
&\bar h_0^{0}(t,Q^2) \sim (t+Q^2)\,,\\
&\bar h_2^{2}(t,Q^2) \sim (t-4m_\pi^2)\,(t+Q^2)\,,\nonumber\\
&\bar h_2^{2}(t,Q^2)-\sqrt{\frac{2t}{Q^2}}\,\bar h_1^{2}(t,Q^2)
\sim (t-4m_\pi^2)\,(t+Q^2)^2\,,\nonumber\\
&\sqrt{\frac{3}{2}}\,t\,\bar h_0^{2}(t,Q^2)+\sqrt{2Q^2t}\,\bar h_1^{2}(t,Q^2)-\frac{Q^2}{2}\,\bar h_2^{2}(t,Q^2)\\
&\sim (t-4m_\pi^2)\,(t+Q^2)^3\,.\nonumber
\end{align}
As a consequence, in all combinations of the invariant amplitudes $F_i$ that enter the generalized polarizabilities (see Eqs.~\eqref{eq:GPelampl}-\eqref{eq:GPsl1m2ampl}) all denominators containing powers of $(t+Q^2)$ cancel against these zeros, which provides a consistency check of our evaluation of the $\pi\pi$ contribution. The Born term retains a soft-photon pole at $t=-Q^2$, which is of no consequence here since for spacelike $Q^2$ that point lies outside both integration ranges in Eq.~\eqref{eq:subtraction}.

\section{$\Delta(1232)$-exchange contribution to VCS amplitudes}
\label{app:delta}

\subsection{$\gamma^\ast N \Delta$ vertex function}
To describe the $\Delta(1232)$-exchange in the VCS process, we start from the Lagrangian describing the electromagnetic $N \to \Delta$ interaction:
\bea
{\cal L}_{em} &=& - e J_{em}^\mu A_\mu \nonumber \\
&=& -e \left\{ \bar \psi_\beta \, \Gamma^{\beta \mu}_{\gamma N \Delta} N + 
\bar N \, \tilde \Gamma^{\beta \mu}_{\gamma N \Delta}  \psi_\beta \right\} A_\mu,
\eea
where $N$ is the nucleon field and $\psi_\beta$ the $\Delta$ Rarita-Schwinger field.  Furthermore $\Gamma^{\beta \mu}_{\gamma N \Delta} $ 
and $\tilde \Gamma^{\beta \mu}_{\gamma N \Delta} $ stand for the vertex and its adjoint, which are related by:
\bea
\tilde \Gamma^{\beta \mu}_{\gamma N \Delta} \equiv \gamma^0  \left( \Gamma^{\beta \mu}_{\gamma N \Delta}  \right)^\dagger \gamma^0.
\eea  
The electromagnetic $N \to \Delta$ transition is  
described by three Lorentz structures corresponding with 
the magnetic dipole (M1), the electric quadrupole (E2), 
and the Coulomb quadrupole (C2) transitions. 
We follow the parameterization given in~\cite{Pascalutsa:2006up}:
\begin{eqnarray}
\Gamma_{\gamma N \Delta}^{\beta \mu} (p_\Delta, p)
&=& \sqrt{\frac{2}{3}} \biggl\{ g_M(q^2)  \Gamma_M^{\beta \mu}  
+ g_E(q^2)  \Gamma_E^{\beta \mu}  
\nonumber \\
&&  
\hspace{0.75cm}+  g_C(q^2)  \Gamma_C^{\beta \mu} \biggr\}, 
\label{eq:NDELem2}
\end{eqnarray}
with $p$ ($p_\Delta$) the nucleon ($\Delta$) four-momentum respectively, with four-momentum transfer $q = p_\Delta - p$. We have defined the vertex for the $p \to \Delta^+$ transition, which yields the isospin factor $\sqrt{2/3}$. 
The tensors in Eq.~(\ref{eq:NDELem2}) corresponding with magnetic dipole (M), electric quadrupole (E) and Coulomb quadrupole (C) $\gamma^\ast N \Delta$ excitations are given by the gauge-invariant expressions:
\begin{eqnarray}
\Gamma_M^{\beta \mu} &=& \frac{3}{2}\frac{(M_\Delta +M_N)}{M_N Q_+^2} \, 
\left[  i \varepsilon^{\beta\mu\kappa\lambda}(p_\Delta)_\kappa q_\lambda \right], 
\nonumber \\
\Gamma_E^{\beta \mu} &=& \frac{3}{2}\frac{(M_\Delta +M_N)}{M_N Q_+^2} \, 
\left[(q \cdot p_\Delta) \,  g^{\beta \mu} -  q^\beta (p_\Delta)^\mu \right] \,  \gamma_5, 
\nonumber \\
\Gamma_C^{\beta \mu} &=& \frac{3}{2}\frac{(M_\Delta +M_N)}{M_N M_\Delta Q_+^2} \, 
\left[- \gamma \cdot p_\Delta \, \left( q^\beta q^\mu- q^2 g^{\beta \mu} \right) \right. \nonumber \\
&&\left. \hspace{2.5cm} + \, \gamma^\beta \left( q \cdot p_\Delta \, q^\mu- q^2 (p_\Delta)^\mu \right) \right] \, \gamma_5,
    \nonumber \\
\label{eq:NDELem3}
\end{eqnarray}
where 
\bea
Q^2_\pm \equiv (M_\Delta \pm M_N)^2 + Q^2, 
\label{eq:defQ+}
\eea
and where we adopt the convention $\varepsilon_{0123} = +1$.  
The vertices defined in Eq.~(\ref{eq:NDELem3}) satisfy the property:
\begin{eqnarray}
(p_\Delta)_\beta \, \Gamma_{M, E, C}^{\beta \mu} = 0,
\end{eqnarray} 
which define them as ``consistent" couplings~\cite{Pascalutsa:2006up}, decoupling the unphysical spin-1/2 degrees of freedom present in the Rarita-Schwinger field. 
The corresponding form factors (FFs) $g_M$, $g_E$, and $g_C$ appearing in Eq.~(\ref{eq:NDELem2}) have spacelike virtuality. 
It is conventional to express these FFs 
%on 
at the resonance position, i.e. for $p_\Delta^2 = M_\Delta^2$, in terms of the magnetic dipole ($G_M^*$), electric quadrupole ($G_E^*$), and Coulomb quadrupole ($G_C^*$) transition FFs as:
\bea
    g_M&=&\frac{Q_+}{M_N+M_\Delta}(G_M^*-G_E^*),\nonumber\\
    g_E&=&-\frac{Q_+}{M_N+M_\Delta}\frac{2}{Q_-^2}\nonumber\\
    &&\times \{(M_\Delta^2-M_N^2 - Q^2) G_E^* + Q^2G_C^*\},\nonumber\\
    g_C&=&\frac{Q_+}{M_N+M_\Delta}\frac{1}{Q_-^2} \nonumber\\
    &&\times \{4M_\Delta^2 G_E^*-(M_\Delta^2-M_N^2 - Q^2)G_C^*\}, 
    \label{eq:gMgEgC_related_to_Ash}
\eea
with the so-called Ash FFs parameterized, for spacelike virtuality $q^2 = - Q^2 < 0$, through the MAID2007 analysis~\cite{Drechsel:2007if,Tiator:2011pw}:
\begin{eqnarray}
    G^{*}_{M}(Q^2) &=& 3.00 \, (1+ 0.01 \, Q^2)e^{- 0.23 Q^2}G_D(Q^2),\nonumber\\
    G^{*}_{E}(Q^2) &=& 0.064 \, (1- 0.021 \, Q^2)e^{- 0.16 Q^2}G_D(Q^2),\nonumber\\
    G^{*}_{C}(Q^2) &=& 0.124 \,  \frac{1 + 0.12 \, Q^2}{1 + 4.9 \frac{Q^2}{4M_N^2}} \frac{4M_\Delta^2 e^{- 0.23 Q^2}}{M_\Delta^2-M_N^2} G_D(Q^2), \;
    \nonumber \\
    \label{eq:GMGEGC}
\end{eqnarray}
with $Q^2$ in GeV$^2$ and the dipole FF:
\bea
G_D(Q^2) \equiv \frac{1}{(1 + Q^2/0.71)^2}. 
\eea

\subsection{VCS amplitudes for narrow $\Delta$-exchange in $s$- and $u$-channel}

As we are interested to model the $t$-channel left-hand cut for $\nu = 0$ through the $\Delta$-exchange, we calculate the narrow $\Delta(1232)$-exchange in $s$- and $u$-channel to the VCS amplitudes $f_i$ in Eq.~(\ref{eq:vcsbasis}) that are even in $\nu$ (i.e. for $i = 1, 2, 5, 6, 7, 9, 11, 12)$. For $\nu = 0$, the contribution to the odd amplitudes $f_3, f_4, f_8$, and $f_{10}$ vanishes. 

As the $\gamma^\ast N \Delta$ transition for low $Q^2$ is completely dominated by the magnetic dipole excitation in Eq.~(\ref{eq:NDELem2}), we will model the left-hand cut contribution to the VCS amplitudes keeping only the magnetic dipole excitation in the $\gamma^\ast N \Delta$ transition. 
The corresponding VCS tensors ${\cal M}^{\mu \nu}$ of Eq.~(\ref{eq:vcs}) 
due to $s$- and $u$-channel $\Delta$-exchange are given by:
\begin{widetext}
\bea
{\cal M}_{s\Delta}^{\mu \nu} &=& \tilde \Gamma_{\gamma N \Delta}^{\alpha \nu} (p', p_s) \frac{\left( \gamma \cdot p_s + M_\Delta \right)}{s - M_\Delta^2} \left[ - g_{\alpha \beta} + \frac{1}{3} \gamma_\alpha \gamma_\beta \right] 
\Gamma_{\gamma N \Delta}^{\beta \mu} (p_s, p), 
\label{eq:sdel} \\ 
{\cal M}_{u\Delta}^{\mu \nu} &=& \tilde \Gamma_{\gamma N \Delta}^{\alpha \mu} (p', p_u) \frac{\left( \gamma \cdot p_u + M_\Delta \right)}{u - M_\Delta^2} \left[ - g_{\alpha \beta} + \frac{1}{3} \gamma_\alpha \gamma_\beta \right] 
\Gamma_{\gamma N \Delta}^{\beta \nu} (p_u, p),  
\label{eq:udel}
\eea
\end{widetext}
with $p_s = p + q$ and $p_u = p - q'$.
By projecting Eqs.~(\ref{eq:sdel}) and (\ref{eq:udel}) on the VCS tensor basis of Eq.~(\ref{eq:vcsbasis}), the resulting expressions for the VCS amplitudes for $\nu = 0$ due to $s$- and $u$-channel $\Delta$ exchange are found as:
\begin{widetext}
\bea
f_1^\Delta(Q^2, 0, t) &=& F_\Delta(Q^2, t)
\bigg\{ - 2 M_N^2(M_N + M_\Delta) 
+ \left( M_\Delta + \frac{3 M_N}{4} \right) t 
+ \frac{1}{2} \left( M_\Delta + \frac{M_N}{2} \right) Q^2 \biggr\}, 
\label{eq:f1delta} \\
f_2^\Delta(Q^2, 0, t) &=& F_\Delta(Q^2, t)
\bigg\{ \frac{1}{2} (M_N + M_\Delta) 
+  \frac{1}{16 M_N} (t + Q^2)  
\biggr\}, 
\label{eq:f2delta} \\
f_5^\Delta(Q^2, 0, t) &=& F_\Delta(Q^2, t)
\bigg\{ - M_N (M_N + M_\Delta) 
+ \frac{1}{8 M_N} \left( M_\Delta + 7 M_N \right) t 
+ \frac{3}{4} Q^2 \biggr\}, 
\label{eq:f5delta} \\
f_6^\Delta(Q^2, 0, t) &=& F_\Delta(Q^2, t)
\bigg\{ \frac{1}{8} (M_N + M_\Delta) 
- \frac{1}{32 M_N} (t + Q^2)  \biggr\}, 
\label{eq:f6delta} \\
f_7^\Delta(Q^2, 0, t) &=& F_\Delta(Q^2, t)
\bigg\{ \frac{3}{4} Q^2 \biggr\}, 
\label{eq:f7delta} \\
f_9^\Delta(Q^2, 0, t) &=& 0, 
\label{eq:f9delta} \\
f_{11}^\Delta(Q^2, 0, t) &=& F_\Delta(Q^2, t)
\bigg\{ - \frac{1}{32 M_N} (M_N + M_\Delta) t - \frac{3}{16}  Q^2 \biggr\}, 
\label{eq:f11delta} \\
f_{12}^\Delta(Q^2, 0, t) &=& F_\Delta(Q^2, t)
\bigg\{ \frac{1}{4} (M_N + M_\Delta) \biggr\}, 
\label{eq:f12delta} 
\eea
\end{widetext}
where we have defined the shorthand for the common prefactor:
\bea
F_\Delta(Q^2, t) \equiv \frac{g_M(0)\, g_M(Q^2)}{M_N^2 Q_+^2  \left( M_N^2 - \frac{1}{2}(t + Q^2) - M_\Delta^2 \right)}. \label{eq:narrowdel}
\nonumber \\
\eea
In Ref.~\cite{Lensky:2016nui}, the nucleon GPs were calculated in Baryon ChPT. As part of that calculation, also the $\Delta$-pole contribution in the limit $t = -Q^2$ was evaluated in the $\delta$-counting scheme. We have checked that the expressions of Eqs.~(\ref{eq:f1delta})-(\ref{eq:f12delta}) exactly agree with the expressions of Appendix B of Ref.~\cite{Lensky:2016nui} in the limit $t = -Q^2$, when making the replacement 
$Q_+^2 \to (M_\Delta + M_N)^2$ in Eq.~(\ref{eq:narrowdel}) to be consistent with the power counting scheme of Ref.~\cite{Lensky:2016nui}.

\subsection{$\Delta(1232)$-resonance spectral function}

The expressions of Eqs.~(\ref{eq:f1delta})-(\ref{eq:f12delta}), which have been obtained for a narrow $\Delta$, see Eq.~(\ref{eq:narrowdel}), are of the pole form:
\bea
A(Q^2, t) = \frac{g(Q^2, t)}{s - M_\Delta^2},
\label{eq:sucut}
\eea
with $s = M_N^2 - (t+Q^2)/2$ for $\nu = 0$, and where $g(Q^2, t)$ is a smooth numerator function.  
The $t$-channel discontinuity of an amplitude of the form of Eq.~(\ref{eq:sucut}) for $\nu = 0$ is then obtained by the analytic continuation $t \to t + i \varepsilon$ resulting in:
\bea
{\rm Im}_t A(Q^2,t) = g(Q^2, t) \left[ \pi \delta(s - M_\Delta^2) \right].
\label{eq:discontinuitystable}
\eea
Note the opposite sign when calculating the $s$- and $t$-channel discontinuities for $\nu = 0$.  

Because in the realistic case the $\Delta$(1232) baryon has a width of $\Gamma_\Delta=0.117\,\text{GeV}$, we next discuss the discontinuity for the case of an unstable particle. 
For unstable particles, the standard Cutkosky rules have to be extended by also including  the resonance decay in the $\pi N$ channel, corresponding with a cut for $s \geq s_0 \equiv (M_N + m_\pi)^2$. It amounts to replace the discontinuity for the stable particle of Eq.~(\ref{eq:discontinuitystable}) by
\bea
{\rm Im}_t A(Q^2, t) &=& g(Q^2, t) 
 \nonumber \\
&\times& \left[ \pi \rho^\Delta(s = M_N^2 - \frac{1}{2}(t + Q^2); M_\Delta,\Gamma_\Delta) 
\right], \nonumber \\
\label{eq:discontinuityunstable}
\eea
with $\rho^\Delta(s; M_\Delta, \Gamma_\Delta)$ a spectral function, which describes the distribution of strength in the discontinuity function around the resonance position, with normalization
\bea
\int_{s_0}^\infty ds \, \rho^\Delta(s; M_\Delta, \Gamma_\Delta) = 1.
\label{eq:specnorm}
\eea
In the limit of a narrow resonance, the spectral function reduces to a $\delta$-function
\begin{align}
\rho^\Delta(s; M_\Delta, \Gamma_\Delta \to 0) \rightarrow \delta(s - M_\Delta^2), 
\end{align}
resulting in Eq.~(\ref{eq:discontinuitystable}).
For a finite width, we choose a Breit-Wigner lineshape for the spectral function 
\bea
\rho^\Delta(s; M_\Delta, \Gamma_\Delta) &=& \frac{N_\Delta}{\pi} 
\frac{M_\Delta \, \Gamma_\Delta(s)}{(s - M_\Delta^2)^2 + (M_\Delta  \Gamma_\Delta(s))^2}. \quad \quad\quad
\label{eq:spectraldel}
\eea
In the above equation, the energy dependence of the width function $\Gamma_\Delta(s)$ is given as in MAID2007~\cite{Drechsel:2007if}
\bea
\Gamma_\Delta(s) &=& \Gamma_\Delta \frac{|p_\pi(s)|^3}{|p_\pi(M_\Delta^2)|^3}
\frac{M_\Delta}{\sqrt{s}} \nonumber \\
&\times& \left( \frac{|p_\pi(M_\Delta^2)|^2 + X_\Delta^2}{|p_\pi(s)|^2 + X_\Delta^2} \right) \theta(s - s_0), 
\eea
where the pion momentum in the $\pi N$ c.m. frame is given by
\bea
|p_\pi(s)| = \frac{\lambda^{1/2}(s, M_N^2, m_\pi^2)}{2 \sqrt{s}},
\eea
with $\lambda(x,y,z)$ the K\"allén function
\begin{equation}
    \lambda(x,y,z) \equiv  x^2 + y^2 + z^2 - 2 x y - 2 x z - 2 y z.
    \label{eq:kinematics2}
\end{equation}
The $\Delta(1232)$ resonance in MAID2007 is described by the parameters
\bea
M_\Delta &=& 1.232~{\rm GeV},  \\
\Gamma_\Delta &=& 0.117~{\rm GeV}, \\ 
X_\Delta &=& 0.5~{\rm GeV}.  
\eea
The normalization of Eq.~(\ref{eq:specnorm}) of the spectral function 
implies for the normalization parameter $N_\Delta$ in Eq.~(\ref{eq:spectraldel})
\bea
N_\Delta = \frac{1}{1.1567}. 
\eea

\bibliography{bibliography}

%\printbibliography

\end{document}